\documentclass{aa}
\usepackage{natbib}
\usepackage{pdflscape} 
\usepackage{wasysym}
\usepackage[varg]{txfonts}
\usepackage{graphicx}
\usepackage{hyperref}
\bibpunct{(}{)}{;}{a}{}{,} 
\usepackage{longtable}
\usepackage{amssymb}

\usepackage{afterpage}

\def\ms{\hbox{\,m\,s$^{-1}$}}         
\def\m2s2{\hbox{\,m$^{2}$\,s$^{-2}$}} 
\def\kms{\hbox{\,km\,s$^{-1}$}}       

\def\logrhk{$\log$(R$^{\prime}_{HK}$)}

\begin{document}

\title{Measuring precise radial velocities on individual spectral lines}
\subtitle{I. Validation of the method and application to mitigate stellar activity
\thanks{Based on observations made with the {\footnotesize HARPS} instrument on the ESO 3.6 m telescope at La Silla Observatory under the GTO program 072.C-0488 and Large program 193.C-0972/193.C-1005/.}
\thanks{The {\footnotesize HARPS} RV measurements discussed in this paper are available in electronic form at the CDS via anonymous ftp to cdsarc.u-strasbg.fr (130.79.128.5) or via http://cdsweb.u-strasbg.fr/cgi-bin/qcat?J/A+A/.}
}
\author{X.~Dumusque\inst{1}
}
\institute{Observatoire astronomique de l'Universit\'e de Gen\`eve, 
               51 ch. des Maillettes, CH-1290 Versoix, Switzerland
}

\date{Received XXX; accepted XXX}

\abstract
{Stellar activity is the main limitation to the detection of an Earth-twin using the radial-velocity technique. Despite many efforts in trying to mitigate the effect of stellar activity using 
empirical and statistical techniques, it seems that we are facing an obstacle that will be extremely difficult to overcome using current techniques.}
{In this paper, we investigate a novel approach to derive precise RVs considering the wealth of information present in high-resolution spectra.}
{This new method consists in building a master spectrum from all available observations and measure the RVs of each individual spectral line in 
a spectrum relative to this master. When analysing several spectra, the final product of this approach is the RVs of each individual line as a function of time.}
{We demonstrate on three stars intensively observed with HARPS that our new method gives RVs that are extremely similar to the one 
derived from the HARPS data reduction software. 
 Our new approach to derive RVs demonstrates that the non-stability of daily HARPS wavelength solution
induces night-to-night RV offsets with an standard deviation of 0.4\ms, and we propose a solution to correct for this systematic.
Finally, and this is probably the most astrophysically relevant result of this paper, we demonstrate that some spectral lines are strongly affected by stellar activity while others are not. 
By measuring the RVs on two carefully selected subsample of spectral lines, we demonstrate that we can boost by a factor of 2 or mitigate by a factor of 1.6 the red noise 
induced by stellar activity in the 2010 RV measurements of $\alpha$\,Cen\,B.}
{ By measuring the RVs of each spectral line, we are able to reach the same RV precision as other approved techniques. In addition, this new approach allows us to
demonstrate that each spectral line is differently affected by stellar activity. Preliminary results show that studying in details the behaviour of each spectral line is probably the key to overcome the obstacle of stellar activity.}
   
\keywords{
Techniques: RVs -- Techniques: spectroscopy -- Stars: Activity -- Stars: individual: \object{HD10700} -- Stars: individual: \object{HD128621}-- Stars: individual: \object{HD10180}
}

\authorrunning{X. Dumusque}
\titlerunning{Measuring precise radial velocities on individual spectral lines}

\maketitle

\section{Introduction}
\label{sec:intro}

The radial-velocity (RV) technique was the first method that allowed the detection of exoplanets orbiting solar-type stars.
By measuring the stellar RV variations induced by the gravitational pull of orbiting planets, this method allows to indirectly measure the mass
of those companions. After the first detection of 51 Pegasi b reported in \citet{Mayor-1995}, the field of exoplanets boomed and new planetary detections, 
with smaller and smaller masses were announced \citep[e.g.][]{Butler-1999, Santos-2004e, Lovis-2006, Queloz-2009, Pepe-2011, Anglada-Escude-2016, Feng:2017ac}.

With the improvement of high-resolution spectrographs over time, and with more performant techniques to extract the RV information from raw frames, the RV precision reached a limit of 1\,\ms, 
or perhaps slightly better for spectrographs like HARPS \citep[][]{Mayor-2003} and HARPS-N \citep[][]{Cosentino-2012}. At this precision, we start to characterise
spurious RV stellar signals induced by physical phenomena happening on the surface of solar-like stars. 

On the timescale of a few minutes, stellar oscillations modify the velocity of the stellar surface and therefore induce a spurious RV signal \citep[e.g.][]{Kjeldsen-1995, Bouchy-2002, Arentoft-2008, Bazot-2012}.
On timescales from minutes to days, we can see the effect of granulation, which is driven by the variation of convection on the stellar surface \citep[e.g.][]{Del-Moro-2004a, Del-Moro-2004b, Lefebvre-2008, Dumusque-2011a, Cegla-2013, Meunier-2016}.
On a timescale similar to the stellar rotation period, stellar magnetic activity, responsible for the appearance of inhomogeneities like spots and faculae on the stellar surface induces a semi-periodic RV signal that can mimic a planetary signal \citep[e.g.][]{Saar-1997b, Queloz-2001, Desort-2007, Meunier-2010a, Dumusque-2011b, Dumusque-2014b, Borgniet-2015}.
Finally, on the timescale of several years, solar-like magnetic cycle induces long-term drift in RV measurements \citep[][]{Lindegren-2003, Dumusque-2011c, Lovis-2011b}.

Those stellar signals perturb the detection of tiny planetary signals. We recall here that the Earth induces a RV variation of 0.1\,\ms\,on our Sun, which is an order of magnitude smaller 
that the known sources of stellar signals described in the preceding paragraph. Over the years, different techniques have been developed to try mitigating the impact of those stellar signals, 
and we present here a list of the most common one in use:
\begin{itemize}
\item Observing stars for 15 minutes allows to average out the signal induced by stellar oscillations \citep[][]{Pepe-2011}.
\item Taking several measurements per night of the same target mitigates the impact of granulation \citep[][]{Dumusque-2011a}.
\item Probing the variation observed in activity indicators like the \logrhk\,\citep[][]{Noyes-1984}, the BIS SPAN \citep[][]{Queloz-2001} or H$\alpha$ \citep[e.g.][]{Bonfils-2007,Robertson-2014} can help in 
disentangling stellar activity from planetary signals. In addition, using statistical techniques like Gaussian Processes \citep[e.g.][]{Haywood-2014, Rajpaul-2015, Jones:2017aa, Delisle:2018aa} 
or Moving Average \citep[e.g.][]{Tuomi-2013a} can mitigate to a certain level the impact of the red noise induced by stellar activity.
\item Fitting the RVs with the long-term trend observed in the \logrhk\,activity index reduces the impact of solar-like magnetic cycle on the RVs \citep[][]{Dumusque-2011c, Dumusque-2012, Delisle:2018aa}.
\end{itemize}
Although these different techniques have been applied on several RV data set and have been able to detect interesting planetary systems, it seems that nowadays we are facing the obstacle of stellar activity, 
which is extremely difficult to overcome. We therefore need to rethink the way the RVs are derived if we want one day to mitigate stellar activity at the 0.1\,\ms precision level that should be reached 
by the ESPRESSO spectrograph \citep[][]{Pepe-2014}.

In this paper, we investigate a new way of deriving RVs. The idea behind this study is to use the wealth of information contained in high-resolution stellar spectra rather than averaging all the information into the 
few parameters derived from the cross-correlation functions \citep[][]{Baranne-1996}. Some preliminary works from \citet{Davis:2017aa}, \citet{Thompson-2017} and \citet{Wise:2018aa} show that stellar activity affects spectral lines in a different way.
This is not surprising as each spectral line have its how sensitivity to temperature, to magnetic field strength and to convection velocity, three physical parameters that are strongly modified in the presence of stellar activity.
We therefore investigate in this paper the possibility of measuring the RV on each individual spectral line. 

In Sec.~\ref{sec:precise_RVs}, we discuss
the different techniques that have been used so far to derive precise RVs and discuss their advantages and limitations. In Sec.~\ref{sec:rv_spectral_line} we describe the technique we used in this paper to measure the 
RV on each individual line, and in Sec.~\ref{sec:RV_comparison} we compare our new RVs with the ones derived using the HARPS Data Reduction Sofware (DRS), which is the gold standard for deriving precise RVs from HARPS spectra.
In Sec.~\ref{sec:night_to_night_offsets}, we use our new RV extraction procedure to investigate night-to-night RV offsets induced by the non-stability  of HARPS wavelength solutions. In Sec.~\ref{sec:mitigating_activity}, we demonstrate that stellar activity affects 
each spectral line in a different way, and that by using a smart selection of spectral lines when measuring the RVs, it is possible to either boost the RV stellar activity signal by a factor of 2, or mitigate it by a factor of 1.6. Finally, we discuss our results and conclude in Sec.~\ref{sec:discussion}.

\section{Measuring precise stellar radial velocities}
\label{sec:precise_RVs}

In the last 15 years, we could reach the radial-velocity precision of 1\ms. Two different types of high-resolution spectrographs
were developed and improved to reach this goal:
\begin{itemize}
\item The first type of instrument consists of slit spectrographs not well controlled in temperature and pressure, 
which implies that atmospheric changing conditions induce hundreds of \ms\,variations due to the modification of the instrument point spread function (PSF). 
To be able to reach the \ms\,precision, the stellar light passes through
an iodine cell, that imprint the absorption spectrum of iodine on top of the stellar spectrum. The iodine emission spectrum
is affected by PSF variations in the same way as the stellar spectrum and therefore serves as a reference scale to measure precise RVs. 
By using a complex reduction, fitting simultaneously i) a stellar spectrum deconvolved from the instrument PSF, ii) an extremely 
high-resolution spectrum of the iodine cell, normally obtained with an Fourier Transform Spectrograph, and iii) the PSF of the spectrograph, it is possible to measure on small 
chunks of the stellar spectrum, generally a few Angstr\"om, the local RV shift. Finally averaging the RV shifts of all the different chunks of the stellar spectrum
gives stellar RVs with precision close to the \ms \citep[e.g.][]{Butler-1996}.

\item The second type of instrument consists of fibre-fed spectrographs put in vacuum \citep[][]{Baranne-1996}, which allows for an extreme stability in temperature and pressure.
The PSF of the instrument is stabilised, and therefore very small RV variations are observed, generally bellow the \ms\,level. To correct for these small instrumental variations, the
spectrum of a calibration lamp, thorium-argon (Th-Ar) or Fabry-Perot (FP) \'etalon, is simultaneously recorded on the detector with the stellar spectrum. 
To reach a precise RV measurement, several techniques have been used like cross correlation \citep[][]{Baranne-1996, Pepe-2002}, maximum likelihood 
template matching \citep[e.g.][]{Anglada-Escude-2012, Astudillo-Defru:2015aa} or least square deconvolution \citep[][]{Donati:1997aa}.
Those methods are used to average out the RV signal of all the spectral lines together, which allows to reach the \ms\,RV precision.
\end{itemize}

\subsection{Deriving precise radial velocities using stabilised RV instruments}
\label{subsec:precise_RVs}

The first reference of using the numerical cross-correlation technique to measure precise RVs in the context of exoplanets can be found in the paper describing the ELODIE spectrograph \citep{Baranne-1996}. Once the stellar spectrum is recorded on the CCD, a cross-correlation is performed using a synthetic template \citep[][]{Baranne-1996}. The result of this transformation, namely the cross-correlation function (CCF), is an average line profile that looks like an inverse Gaussian. The RV of the star is derived by fitting an inverse Gaussian to the CCF and taking its mean. Note that this cross-correlation technique is inherited from the method used before CCDs exited. During those times, a physical template with holes at the position of every stellar spectral line was shifted in the focal plane of the spectrograph to construct physically the cross-correlation function that was then recorded using a photometer \citep[i.e. for the CORAVEL instrument,][]{Baranne:1979aa}

The numerical cross-correlation technique is still used nowadays on stabilised spectrograph like HARPS, HARPS-N and soon ESPRESSO. The technique was improved in \citet{Pepe-2002a}, were the authors show that rather than using a binary mask, 0 for the continuum and 1 for the centre of spectral lines, weighting the mask by the depth of the spectral lines improve the RV precision. HARPS and HARPS-N proved that the technique was able to deliver in a simple and robust way sub-\ms\,RV precision on stellar spectra. In addition to simplicity, the cross-correlation with a numerical mask has the advantage of directly giving a precise RV estimate for a given star without requiring any preceding observation.

It was however demonstrated that for M dwarfs, the template matching technique gives better results than the CCF technique using a weighted synthetic mask optimised for M dwarfs \citep[e.g.][]{Anglada-Escude-2012, Astudillo-Defru:2015aa}. In this technique, the RV is derived by performing a maximum likelihood estimation between each stellar spectrum and a high signal-to-noise ratio (SNR) master spectrum built from all the available stellar observations, in which the Doppler shift is a free parameter. For M dwarfs, the template matching technique gives more precise RVs because due to the low temperature of these stars, spectral lines are everywhere in the spectrum and therefore a master stellar spectrum can access to all the possible 
RV information present in the spectrum, what a synthetic mask with a limited number of lines cannot do. This approach gives however a worse RV precision when applied to G and K dwarfs. This is probably because in this case, the template matching technique is sensitive to the noise present in the continuum where no RV information is present. Note however that this could be solved by masking spectrum continuum regions in the master stellar spectrum.

An advantage of the cross-correlation technique is that we can measure, in addition to the RV, the full width at half maximum (FWHM) and the bisector of the CCF. Several studies have shown that those indicators are useful to assess the stellar or planetary nature of a RV signal \citep[e.g.][]{Queloz-2001, Desort-2007, Meunier-2010a, Dumusque-2014b}. Indeed, a planet only induces a pure Doppler-shift of the stellar spectra, without changing the shape of spectral lines, and therefore without changing the FWHM and bisector of the CCF, which is not the case for all known stellar signals.

By using a template that contains all the available lines in a stellar spectrum, the CCF or template matching techniques can reach an extreme RV precision. However, by averaging all the spectral lines together, any strong effect that specific spectral lines might have will be diluted in the final product, and thus impossible to observe. The problem is therefore not linked to the CCF or template matching techniques, but by the fact that the RV information of all lines are averaged out together. For example, in \citet{Dumusque-2015a} the authors discovered that several stars observed with HARPS were presenting a RV signal of a few \ms\,with a period of one year, in phase with the Earth barycentric RV. By looking closer at the RV of each spectral line, they were able to show that only a few spectral lines, the ones crossing the HARPS detector stitchings, were affected by a yearly signal with an amplitude up to a hundred\,\ms. By averaging out the signal of those lines with all the others not affected, the residual signals was of the order of a few \ms.

\section{Measuring the radial velocity of individual spectral lines}
\label{sec:rv_spectral_line}

The new approach presented in this paper to derive precise RVs is an intermediate solution between measuring locally the RV like in the iodine cell technique, to avoid diluting the RV signal of every spectral line, and using a stellar master spectrum without considering the stellar continuum, which might be the limitation of the template matching technique on G-K dwarfs. The idea is to measure the RV on each spectral line using the template matching technique, and then to perform a weighted average over all those lines to obtain a precise RV measurement for a given star. 

The process of measuring the RV on each spectral line, starting from HARPS reduced 2d stellar spectra, can be split in different steps that we will describe further:
\begin{itemize}
\item for a given star, correct each HARPS 2d stellar spectrum from known systematics,
\item build a high SNR 2d stellar master spectrum by co-adding these corrected 2d stellar spectra,
\item select the spectral lines for which we want to measure their RVs,
\item measure the RVs on those individual spectral lines,
\item and finally combine the RV information of all spectral lines to get a precise RV measurement.
\end{itemize}

\subsection{Spectra corrections from known systematics}
\label{subsec:spectra_corr}

To get a stellar continuum that is flat, we must correct the stellar spectrum from the blaze of the instrument.
On HARPS, the daily calibrations measure the blaze using a tungsten lamp
and the information is recorded in a HARPS.XXX\_blaze\_FIBRE.fits\footnote{XXX being the time of the calibration and 
FIBRE being A or B depending if the file is referring to the blaze of the science or reference fibres, respectively} file. The name of this file is recorded in the header of every stellar observation, under the keyword \emph{ESO DRS BLAZE FILE}.
To correct for the effect of the blaze, we simply have to divide the stellar spectrum from it.

After dividing by the blaze, the stellar continuum is much flatter, however, we still see small linear trends on each spectral order.
This is because the stellar energy distribution is not corrected for. Taking into account the effective temperature of the
star we are studying, we measure, using the black body radiation, the flux that is recorded by each pixel on the CCD and correct for this effect. Note that to perform this correction, we need to know the wavelength of each pixel on the CCD, which is given by the wavelength solution. This solution is measured by the HARPS afternoon calibrations and is recorded in the header of each stellar observation. A total of 288 polynomial coefficients are saved under the keywords \emph{ESO DRS CAL TH COEFF LLX}, where X goes from 0 to 287. The wavelength $\lambda_{i,j}$ of pixel $j$ in order $i$ can be obtained using the following formula:
\begin{eqnarray}
\lambda_{i,j} = P_{4*i} + P_{4*i+1} \times j + P_{4*i+2} \times j^2 + P_{4*i+3} \times j^3,
\end{eqnarray}
where $P_{X}$ corresponds to the value of the keyword \emph{ESO DRS CAL TH COEFF LLX}, $i$ varying from 0 to 71, and $j$ from 0 to 4095.

The relative flux between the blue and red parts of the spectrum of a given star, commonly referred to as a colour index, is correlated
with atmospheric extinction that varies with airmass and changing atmospheric conditions \citep[e.g.][]{Bourrier-2014}. The airmass effect is corrected 
for before taking the measurement using an atmospheric dispersion corrector, however the effect due to changing weather conditions cannot be accounted for \emph{a priori}.
This change in relative flux will put different weights as a function of time on the RV of each spectral order and thus can induce a correlation between the RV measured and colour index.
To avoid this colour dependence, the total flux in each order is divided by the flux of a 
reference spectrum of similar spectral type, and the residuals are fitted with a 5th order polynomial, which coefficients are saved in the header of the 
observed spectrum under the keywords \emph{DRS FLUX CORR COEFFX}, X going from 0 to 4.
To get a similar colour index for each spectrum of a given star, and therefore obtain RVs that are insensitive to colour, we divide each spectrum with the corresponding 5th order polynomial. 

\subsection{Build a high signal-to-noise-ratio master stellar  spectrum}
\label{subsec:master_spectrum}

Once each HARPS 2d spectrum is corrected from known systematics (see Sect. \ref{subsec:spectra_corr}), building a high SNR stellar master 2d spectrum
is a rather easy task, however some small details need to be taken into account.

HARPS 2d spectra are recorded in the reference frame of the observatory, which velocity with respect to a given star changes as a function of time due to Earth orbiting the Sun and rotating on its own axis. 
This implies that the stellar spectra of a given star moves on the HARPS CCD as a function of time.
To get the spectral lines of all spectra overlapping each other, we need to change the reference frame of the observation to the solar system barycentre.
This can be done by Doppler shifting every stellar spectrum using the barycentric Earth RV (BERV) measured by the HARPS data reduction software (DRS) and saved in the header of
each observation under the keyword \emph{ESO DRS BERV}. Spectra can also be Doppler shifted by the presence of companions orbiting a given star. Therefore, in addition to correct for the BERV, we also correct for the RV measured by the HARPS DRS. Note however that since the width of a pixel on the HARPS CCD is equivalent to 820 \ms\,only massive close-in binaries and hot Jupiters will have a significant impact on the high SNR stellar master spectrum.

By red or blue-shifting the spectra to correct for the BERV and the effect of potential companions, the minimum and maximum wavelength of each order will be different
for each observation. We therefore search for the maximum of the minimum wavelength and the minimum of the maximum wavelength for each order over all observations,
and set those values as being the minimum and maximum wavelength of each order for the high SNR stellar master spectrum.

Each stellar spectrum will have its own wavelength solution, however, only one wavelength solution can be used for the stellar master spectrum.
We therefore choose the wavelength solution of the spectrum having the minimum BERV, and then 
linearly interpolated all the other spectra to this fixed wavelength solution. Choosing the wavelength solution for which the BERV of
the star is minimum will assure that the median RV of all the spectral lines will be the closest to 0.

Even after correcting the spectra from known systematics (see \ref{subsec:spectra_corr}), some orders still show small drifts. To remove those drifts before co-adding all the spectra together to build the high SNR 
stellar master, we fit the continuum of each order with a linear drift and correct for it. For reasons that we did not investigate, but probably due to a bad cosmic ray 
correction, the slope over an order can reach unusual values in some cases. To prevent adding bad spectra or systematics in the master stellar spectrum, we reject all spectra for which at least one slope over the orders 30 to 71 is further away than 5-$\sigma$ from the mean of all slopes. We do not reject spectra with bad slopes for orders smaller than 30 as fitting for the stellar continuum in the blue part of the spectrum is challenging.

To have a high SNR stellar master spectrum that does not present significant instruments systematics, we also rejected spectra for which the SNR measured 
in order 10 ($\sim$4040\,$\AA$) is smaller than 10 and for which the SNR measured in order 60 ($\sim$6110\,$\AA$) is larger than 450. This prevents of 
considering spectra that could be affect by the non-linearity of the detector but also by inefficiency in the charge transfert when the SNR is very low. 
On HARPS, the atmospheric dispersion corrector corrects for airmass colour effect up to an airmass of 2. To be conservative and to prevent adding to the stellar master template spectra
that could be contaminated by atmospheric effects, we select only observations with an airmass smaller than 1.5. Finally, to get a master template that is not affected 
by stellar activity, we select only spectra for which the calcium activity index is smaller than \logrhk $<-4.95$. Those cuts in airmass and \logrhk\,are very
restrictive and we choose them here as we will analyse, in Sec.~\ref{sec:RV_comparison}, stars that have thousands of spectra available. We will see in Sec.~\ref{subsec:spectral_line_RVs}
that the noise in the master spectrum will be neglected compared to the noise in each individual spectrum. Therefore, as a rough estimate, at least 25 spectra must be selected to build the master
as this will imply that on average the noise of the master will be five times smaller than the noise in each individual spectra.
For some stars, the restrictive cutoff in airmass and activity level selected here will never be reached. In those cases, we would advise to select at least the 25 less active spectra with 
airmass smaller than 2. This is however something that should be studied further.

After doing all those corrections and rejecting spectra that could exhibits some systematics, we can build the high SNR stellar master spectrum by co-adding all the individual corrected spectra.
The obtained stellar master spectrum is oversampled by a factor 82 using a cubic spline to reach a resolution of 10\,\ms. We did not investigate what is the optimal oversampling factor to 
be used. But for computation performance, we preferred doing a dense interpolation at this stage using a cubic spline and then a linear interpolation for the final refinement
to obtain a master spectrum which is on the same wavelength scale as each stellar spectrum (see Sec.~\ref{subsec:spectral_line_RVs}).
Because this stellar master spectrum is at very high SNR, we will be able, in Sec.~\ref{subsec:spectral_line_RVs}, to neglect the noise added during this oversampling process.

\subsection{Select the spectral lines to derive RVs}
\label{subsec:select_spectral_lines}

Three different synthetic templates are used by the HARPS DRS to measure the RV with the CCF technique: a G2, a K5 and a M2 mask. 
Those templates have been optimised and do not consider spectral lines 
strongly affected by magnetic activity, like the Ca II H and K, H$\alpha$ or Mg II lines, but also spectral lines in the red part of the spectrum strongly 
contaminated by tellurics. When measuring the RVs of G or K dwarfs, we will always derive the RVs on all the individual
spectral lines present in the K5 synthetic template to account for the maximum number of possible lines. 
If no line is present at the position defined by the mask due to a higher stellar effective temperature, the RV precision measured on this line will 
be extremely bad and the line will be rejected at a later stage (see Sec.~\ref{subsec:stellar_RV}).

\subsection{Measure the RV on individual spectral lines}
\label{subsec:spectral_line_RVs}

Let's consider that we want to measure, $RV_i$, the RV of spectral line $i$ with central wavelength $\lambda_{i}$.
We first need to define a window around $\lambda_{i}$ on which we will measure the RV of this line. To prevent 
windows overlapping in the case of close spectral lines, mainly in the blue part of the spectrum, but also to select the maximum number of pixel
to reach the best precision in RV, we selected a window width of 16 pixels, which corresponds to 0.17 and 0.30 $\AA$ in the bluest and reddest 
parts of HARPS spectra, respectively. For simplicity, we selected only a fixed number of pixels for all spectral lines. Note however that the window 
width could be optimised further as spectral lines in the red are much less packed than in the blue.

The second step consists in selecting the high SNR stellar master spectrum chunk S$_{ref,\,i}(\lambda)$ centred on $\lambda_{i}$ and that present a width of 
16 pixels. We record then the minimum $\nu_{min}$ and maximum wavelength $\nu_{max}$ of S$_{ref,\,i}(\lambda)$.

Then for each spectrum $j$ with its own wavelength solution, we select the spectrum chunk S$_{j,\,i}(\lambda)$ between $\nu_{min}$ and $\nu_{max}$ and perform linear interpolation on the edges so that S$_{j,\,i}(\lambda)$ for all times $j$ present the same boundaries as S$_{ref,\,i}(\lambda)$.

To calculate the RV shift of S$_{j,\,i}(\lambda)$ with respect to the reference S$_{ref,\,i}(\lambda)$, we need an estimate of the noise present in S$_{j,\,i}(\lambda)$.
This noise has two components that should be added in quadrature. One that is due to photon noise and that is equal to the square root of the flux, and another one that comes from the CCD read-out and dark current, estimated to 12 photo-electrons on HARPS, which can become significant at low SNR. To calculate the photon noise in HARPS spectra, we need to be careful to use
the raw flux before correcting for the blaze and the stellar spectral energy distribution (see \ref{subsec:spectra_corr}). Otherwise,
we will put too much weight on spectral lines that are on the edges of the orders and that have lower flux due to the blaze. To have the correct flux and photon noise, we adjust the flux 
of S$_{j,\,i}(\lambda)$ so that it is equivalent to the raw uncorrected spectrum.

To compute $RV_{i,\,j}$, the RV of line $i$ at time $j$, we use the template matching method described in \citet{Bouchy-2001b}, which uses the change in flux 
between two spectra and the derivative of one of them to measure the RV shift. This method can be applied if the two spectra are on the same wavelength scale,
if the shift between the two spectra is smaller than the sampling, and if the reference spectrum on which we calculate the derivative has a high SNR so that we can 
neglect the noise in this spectrum and in its derivative.
In our case, we apply this method on S$_{j,\,i}(\lambda)$ and S$_{ref,\,i}(\lambda)$, the latter being used as the reference.
Following Eq. 2 of Sec. 2 in \citet{Bouchy-2001b}, we can show that:
\begin{eqnarray} \label{eq:bouchy}
\mathrm{S}_{j,\,i}(\lambda) &=& A \left[ \mathrm{S}_{ref,\,i}(\lambda) + \frac{\partial \mathrm{S}_{ref,\,i}(\lambda)}{\partial \lambda}\delta \lambda \right]
\end{eqnarray}
where $A$ accounts for the difference in flux between S$_{j,\,i}(\lambda)$ and S$_{ref,\,i}(\lambda)$. Note that here we use spectra as a function of wavelength and not pixel. 
Moreover, we could use a more complex model to fit a zero-point offset to account for possible background contaminations, and/or a slope. We tried with an offset and this was not significantly improving
our results, so we preferred to remove this additional parameter. This is however something that should be investigated when analysing the RVs of fainter targets that the ones we analyse here. Regarding the slope,
we do not expect any significant change from spectrum to spectrum on such short wavelength windows.

Due to the BERV and potential orbiting companions, the shift between S$_{j,\,i}(\lambda)$ and S$_{ref,\,i}(\lambda)$ is often more than the sampling, which is 820\,\ms\,in the case of HARPS.
To use the technique described above, we need to reduce this shift to a level smaller than the sampling, which is done by applying a Doppler shift on S$_{ref,\,i}(\lambda)$ to add to the reference 
the RV shift induced by the BERV and potential orbital companions.

The template matching method requires that the template S$_{ref,\,i}(\lambda)$ is on the same wavelength scale as the spectrum chunk S$_{j,\,i}(\lambda)$ of one observation.
This is done by estimating the template at the wavelength of the observation using linear interpolation\footnote{Note that linear interpolation is precise enough here as S$_{ref,\,i}$ 
was already oversampled to 10 \ms\,(see Sec. \ref{subsec:master_spectrum}). As evaluating the template to the wavelength of the observation needs to be done for each line in each spectrum,
the linear interpolation performed here allows the method to be more computationally efficient.}.
Once this is done, we fit Eq. \ref{eq:bouchy} with a linear least-square algorithm taking
into account the noise in S$_{j,\,i}(\lambda)$, to get the best solution for $A$ and $A\,\delta \lambda$. The RV shift between S$_{j,\,i}(\lambda)$ and S$_{ref,\,i}(\lambda)$, 
$RV_{i,\,j}$, is then derived using the Doppler shift formula:
\begin{eqnarray} \label{eq:RV}
\frac{\delta V}{c} = \frac{\delta \lambda}{\lambda} \quad \implies RV_{i,\,j} = \frac{c}{\lambda}\,\frac{A\,\delta \lambda}{A},
\end{eqnarray}
where c is the speed of light. The square root of the diagonal elements of the covariance matrix obtained by the linear least square 
gives us the errors for $A$ and $A\,\delta \lambda$, $\sigma_A$ and $\sigma_{A\,\delta \lambda}$, respectively. 
The error for $RV_{i,\,j}$ is obtained from propagating these errors:
\begin{eqnarray} \label{eq:sig_RV}
\sigma_{RV_{i,\,j}} = \frac{c}{\lambda}\,\sqrt{\left[\frac{1}{A}\right]^2\,\sigma_{A\,\delta \lambda}^2 + \left[-\frac{A\,\delta \lambda}{A^2}\right]^2\,\sigma_A^2}.
\end{eqnarray}
For all the stellar spectra, we estimate $RV_{i,\,j}$ and $\sigma_{RV_{i,\,j}}$, and therefore get the RV of spectral line $i$ with its error as a function of time. 
Because the high SNR template spectrum is used as a reference, and because this spectrum is an average of all the stellar spectra corrected from 
the BERV and from the RV contribution of orbiting companions (see Sec.~\ref{subsec:master_spectrum}), the RV 
of each individual line will be centred around zero.

\subsection{Combining the RV information of all spectral lines to get the stellar RV}
\label{subsec:stellar_RV}

In the preceding section, we explained how we could get the RV of each individual line. Those RVs can be useful to study stellar physics at the
level of each spectral line. The RV precision on each line is however not very good, and reaches in the best case $\sim$10\ms\,(see Fig.\ref{fig:RV_correlation_with_activity}) 
when analysing the spectra of the extremely bright target $\alpha$\,Cen\,B (V=1.33). To reach the \ms\,precision, we need to combine the RV of all the spectral lines together.

To combine the RV information of all spectral lines, we first must perform some data cleansing. Due to imperfections on the CCD, saturation from cosmic rays that are not always corrected
properly, spectral lines from our K5 line list (see Sec.~\ref{subsec:select_spectral_lines}) that do not appear in a given star due to a different spectral type and probably other unknown effects, 
the RV measured on each spectral line can show some significant outliers. To reject bad RV measurements and 
bad spectral lines, we first remove from the RV of each line the effect of the BERV and the effect of potential companions by subtracting the RV measured by the HARPS DRS.
We perform then the following cleansing:
\begin{itemize}
\item We perform a $6-\sigma$ clipping on the $\chi^2$ of all the spectral lines in one observation obtained when fitting Eq. \ref{eq:bouchy} using a linear least square (see Sec. \ref{subsec:spectral_line_RVs}). 
We repeat the process twice, which eliminated spectral lines badly fitted.
\item We perform a $6-\sigma$ clipping on the RVs of all the spectral lines in one observation. We however keep all the RVs for which the value plus six times its error bar is 
compatible with the median RV of all the lines, which is close to 0 by construction. This removes RV measurements that are too far from the median of all lines in one spectrum.
We repeat the process twice.
\item We fit on the RVs of each spectral a second order polynomial as a function of time, and perform a $6-\sigma$ clipping, without rejecting points for which the error bar is compatible with the polynomial fit at $6-\sigma$.
\item We reject all spectral lines for which more than 1\% of the data are rejected by the three preceding cleansing.
\item We reject the 0.3\% of spectral lines with the worse RV precision.
\item We remove all lines for which the ratio between the standard deviation of the RVs and the median RV error, what we call further the quality factor, is larger than two. Those lines have a significant signal, which is not expected as the RV precision on a line is poor and
we expect stellar signals to be at a lower level.
\item We finally reject spectral lines for which their median RV is not compatible at $3-\sigma$ with the distribution of the mode of all lines. We use the mode here as the right and left sides of the distribution can have different tails.
\end{itemize}

After curating the RVs of all spectral lines using the criteria defined above, we combine the RV information of all the remaining lines to get a precise measurement of the stellar RV
at time $j$. This is done by performing a weighted average of the RVs measured on each spectral line, putting as weight their respective RV errors:
\begin{eqnarray} \label{eq:sig_RV}
RV_j = \frac{\sum_i\left[\frac{1}{\sigma_{RV_{i,\,j}}^2}\,RV_{i,\,j}\right]}{\sum_i\left[\frac{1}{\sigma_{RV_{i,\,j}}^2}\right]}.
\end{eqnarray}

\section{Comparison of the RVs derived from individual lines with the HARPS DRS RVs}
\label{sec:RV_comparison}

In this section, we compare for three stars the RVs derived from our new method, i.e analysing the RVs of each spectral line and then combining their information 
to get the RV of the star, with the RVs derived from the HARPS DRS using the CCF technique.

\subsection{Radial velocities of $\tau$\,Ceti}
\label{subsec:RV_HD10700}


The star $\tau$\,Ceti (\object{HD10700}) has been observed since the beginning of HARPS in 2003. Over 15 years, more than 10'000 spectra have been obtained, with a very dense sampling.
This G8 dwarf has the particularity of being inactive, with an activity level that always stayed below \logrhk=$-4.94$ over 15 years. This dwarf is probably in a similar state that the Sun was
during its Maunder Minimum, i.e. the period around 1645 to 1715 during which sunspots on its surface became exceedingly rare. $\tau$\,Ceti is therefore the perfect target to compare our new RVs 
with the RVs derived from the HARPS DRS.

Before deriving the RVs with our new RV extraction procedure, we have to build the master stellar spectrum that will be used as a reference to measure the RV of each single spectral line as a function of time (see Sec.~\ref{subsec:master_spectrum}).
For $\tau$\,Ceti, we selected spectra for which the SNR in order 10 was larger than 10, for which the airmass was below 1.2 and for which the activity level was below \logrhk=$-4.95$. We therefore
considered 5743 spectra to build the master spectrum, representing 50.8\,\% of all available spectra. Note that out of all those spectra, 15\% of them are coming from the five-day asteroseismology run performed in October 2004 \citep[][]{Teixeira-2009}. Although not done here, we should investigate if selecting a lot of spectra taken at the same time and therefore at the same particular state of the star does not have an impact on the final RVs. After rejecting the spectra for which the slope of the continuum in some spectral orders was too 
high (see Sec.~\ref{subsec:master_spectrum}), we were left with 3769 spectra (33.4\,\%), that were used to build the master stellar spectrum of $\tau$\,Ceti.

After computing the RV for each spectral line using the formalism described in Sec.~\ref{subsec:spectral_line_RVs}, we cleaned the data from any outlier before
combining the RV of all the spectral lines as explained in Sec.~\ref{subsec:stellar_RV}. In total, out of the 7331 lines available in the K5 line list used, 701 lines were rejected in this cleansing process, thus slightly less than 10\%\footnote{The sigma clipping on the chi-square rejected a total of 563402 data points out of 81036874 (equals 7331 spectral lines times 11054 spectra), thus 0.70\%. The sigma clipping of the RV of all spectral lines per observation rejected 155979 data points out of 81036874, thus 0.19\%. The sigma clipping of the RV of each line as a function of time removed 46945 data points out of 81036874, thus 0.06\%. When rejecting lines for which more than 1\% of the RV data points were rejected, the precedent cuts discarded 632 spectral lines. The following cleansing, being rejecting the lines with bad RV errors, rejecting lines with a bad quality factor and finally doing a sigma clipping on lines with a bad median RV, discarded 21, 26 and 22 spectral lines, respectively.}. Rejecting only 10\% of the lines might seems rather small, however, using the HARPS K5 template line list as a starting point in our analysis allows to remove most of the lines that would be rejected at this level (see Sec.\ref{subsec:select_spectral_lines}).

Finally, before analysing the newly derived RVs, we have to correct for the drift of the instrument, as explained in Sec.~\ref{instrumental_drift} of the appendix. As we can see in this section, our new RV extraction procedure on calibration spectra gives extremely similar results than the HARPS DRS, therefore we used our new drift measurements to correct for instrumental systematics in the $\tau$\,Ceti RV data set.

\begin{figure*}[!t]
\center
 \includegraphics[angle=0,width=0.80\textwidth,origin=br]{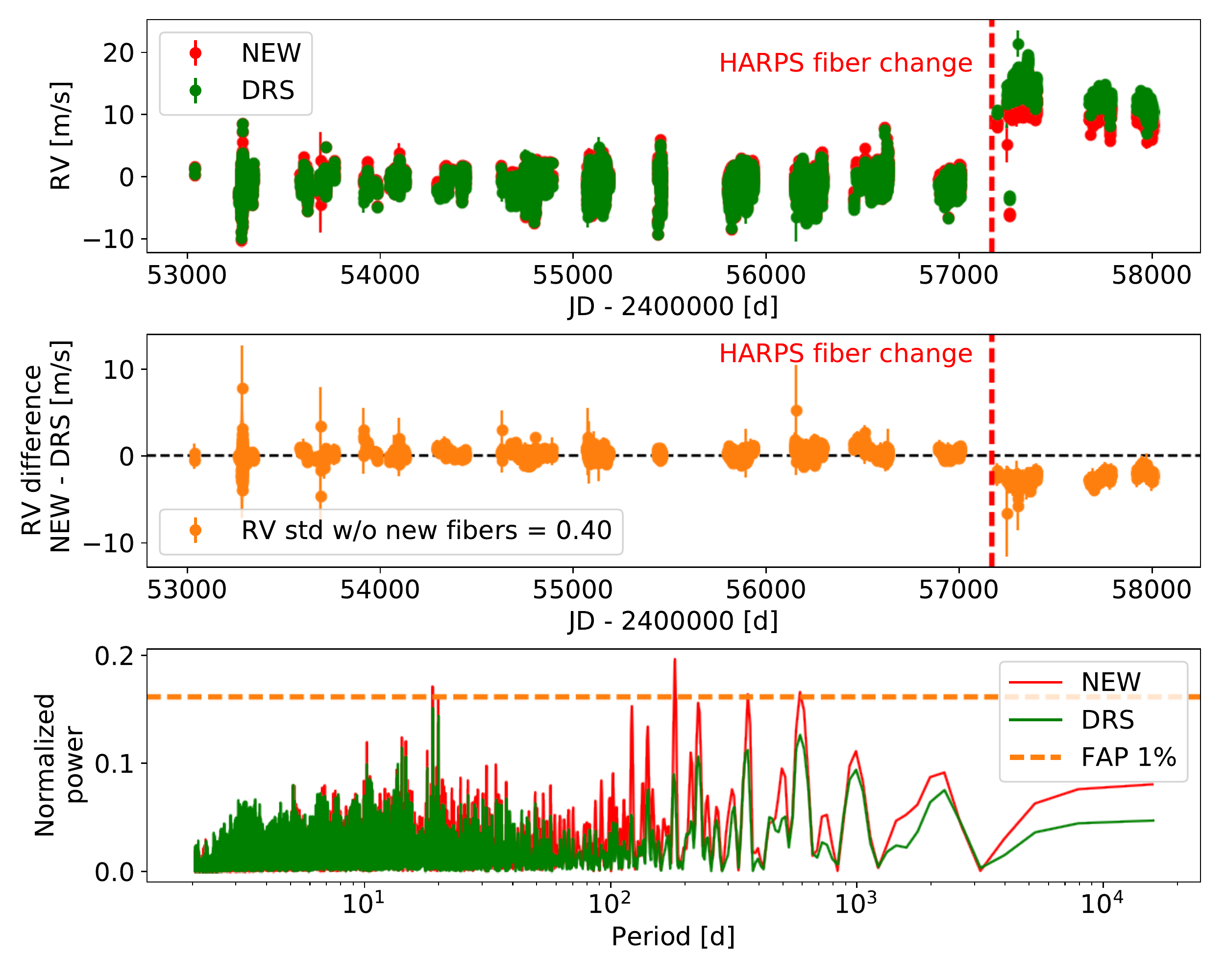}
 \caption[]{\emph{Top:} Comparison of the RVs of $\tau$\,Ceti derived using our new RV extraction procedure (red) with the RVs estimated from the HARPS DRS (green). The red vertical dashed line
  corresponds to the change from circular to octagonal fibres that was performed on HARPS on the 1st of June 2015 (JD=2457174). 
 \emph{Middle:} Difference between the RVs derived using our new RV extraction procedure and the RVs from the HARPS DRS. We show in the legend the standard deviation of the RV difference 
 for the measurements preceding the fibre change. 
 \emph{Bottom:} Periodograms of the RVs derived using our new RV extraction procedure (red) and the RVs derived by the HARPS DRS (green). To calculate those periodograms, we rejected the data after 
 the fibre change to prevent low frequency signals induced by the observed offset. The two periodograms have been normalised in such a way that their FAP of 1\% coincide. }
\label{fig:HD10700_comparison_with_DRS}
\end{figure*}

In the top plot of Fig.~\ref{fig:HD10700_comparison_with_DRS}, we compare the RVs for $\tau$\,Ceti measured using our new RV extraction procedure with the RVs estimated by the HARPS DRS.
The middle plot illustrate the difference between these two sets of RVs. The change of the 
HARPS fibres from circular to octagonal, which happened on JD=2457174 (1st June 2015), did not have the same impact on the two reductions. However, besides this offset, the RVs 
are extremely similar with a rms for the RV difference of 0.40\,\ms\,when considering the data before the fibre change\footnote{We note that we could remove this offset by building two master spectra, using either spectra taken before or after the fibre change, by measuring the RV of all lines in each spectrum using the corresponding master spectrum and finally by measuring the offset between the two masters.}.
Looking, in the bottom plot of Fig.~\ref{fig:HD10700_comparison_with_DRS}, at the periodograms of the two different sets of RVs for the data obtained before the fibre change, we see once again a similar behaviour. 
It seems however that there is slightly more power in our new RVs at periods between a hundred and a thousand days. Although we speak of signals with amplitudes of the order of 0.5\,\ms, we still investigate the cause of this difference in 
Sec.~\ref{Systematics in HD10700 RVs} of the Appendix. In resume, it seems that our new RV extraction procedure is slightly more sensitive to the 
effect induced by lines crossing CCD stitchings \citep[][]{Dumusque-2015a} and by micro-tellurics \citep[e.g.][]{Cunha-2014}.

\begin{figure*}[!t]
\center
 \includegraphics[angle=0,width=0.45\textwidth,origin=br]{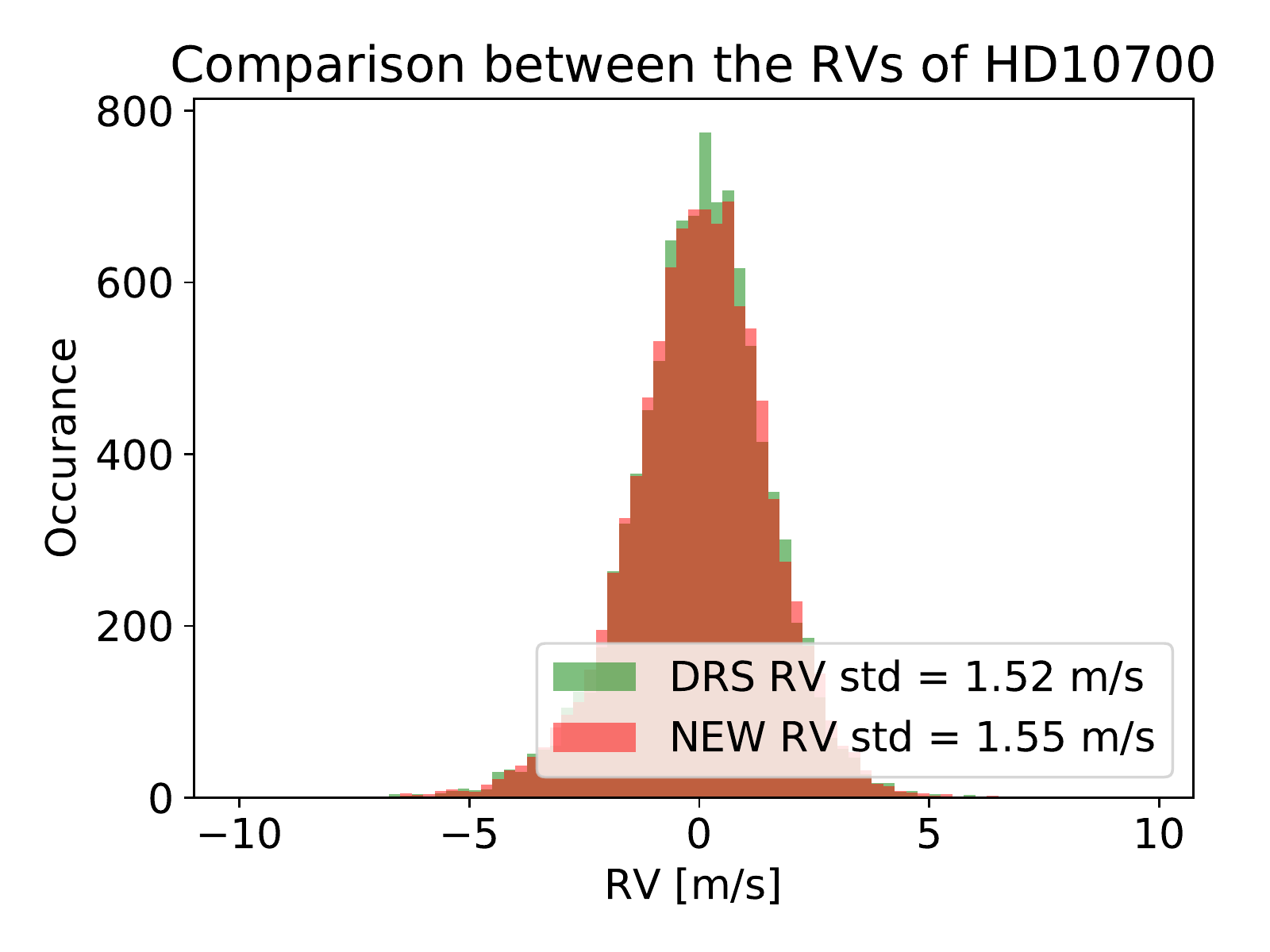}
 \includegraphics[angle=0,width=0.45\textwidth,origin=br]{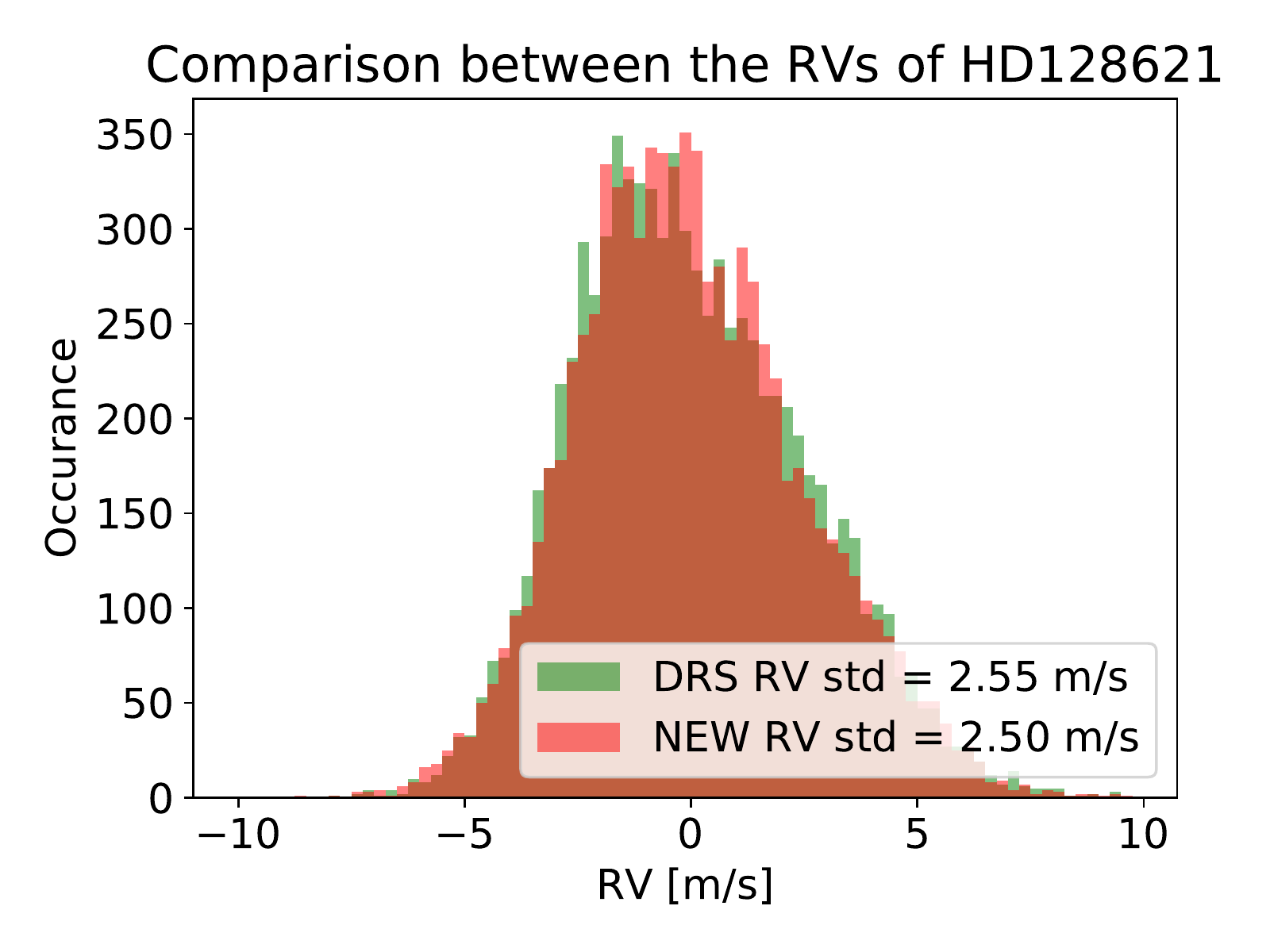}
 \includegraphics[angle=0,width=0.45\textwidth,origin=br]{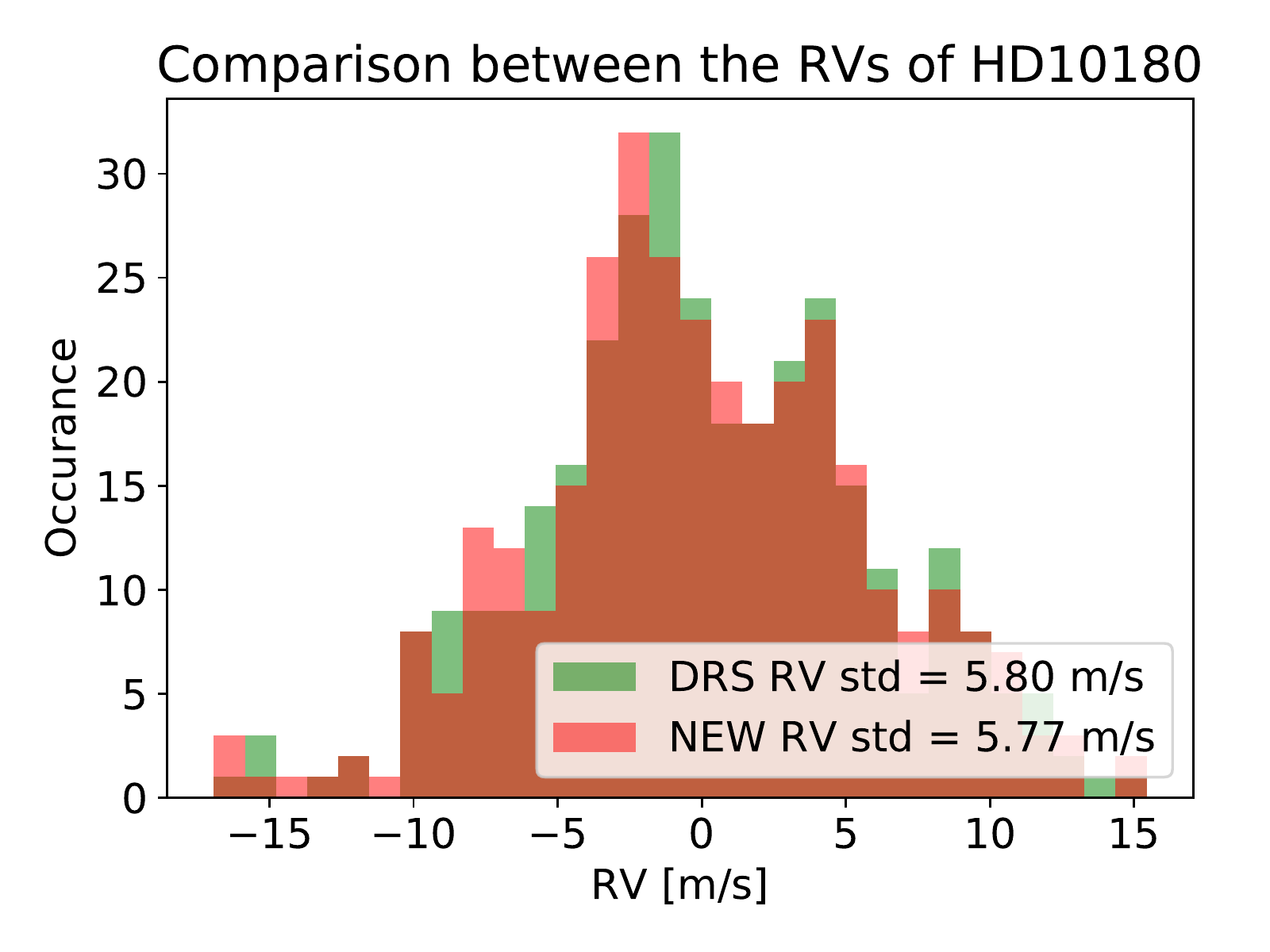}
  \caption[]{Histogram of the RVs of $\tau$\,Ceti, HD128621 and HD10180 derived using our new RV extraction procedure (red) and the RVs derived using the HARPS DRS (green). The legend shows the standard deviation obtained for both data sets.}
 \label{fig:comparison_STD_with_DRS}
\end{figure*}

Even though we see small differences in the periodogram of the RVs derived with our new RV extraction procedure and the HARPS DRS, when looking at the histogram of the RVs in Fig.~\ref{fig:comparison_STD_with_DRS}, 
we see very similar Gaussian-like distributions with a similar mean and standard deviation, confirming once again that the two set of RVs are extremely similar.

\subsection{Radial velocities of $\alpha$\,Cen\,B}
\label{subsec:RV_HD128621}

Another star that has been intensively observed with HARPS at extreme precision is $\alpha$\,Cen\,B (\object{HD128621}).
Contrary to $\tau$\,Ceti, the HARPS data of $\alpha$\,Cen\,B show a solar like magnetic cycle, with periods of minimum and maximum of
activity. Those data are therefore important to check how our reduction compares with the HARPS DRS in the case of RVs affected by stellar activity.

In \citet{Dumusque-2012}, the authors show that the data of $\alpha$\,Cen\,B are sometimes, depending on the observing conditions, contaminated by the light of $\alpha$\,Cen\,A
which is a few arcseconds away on the sky. We used here the same criterion as in \citet{Dumusque-2012} to reject contaminated measurements. In addition, it was also
shown in this paper that the data of $\alpha$\,Cen\,B exhibit a significant signal at one year and its different harmonics due to CCD stitching \citep[][]{Dumusque-2015a}.
To remove those perturbing signals, we rejected from our new RV extraction procedure spectral lines that were closer to CCD stitchings than 48 pixels. We did the same 
for the RVs derived by the HARPS DRS, by using a cross-correlation template that had those lines removed \citep[][]{Dumusque-2015a}. Note that in \citet{Bauer:2015aa} and Coffinet et al. 2018 (submitted to A\&A),
the authors develop a new recipe to derive HARPS wavelength solutions based on the physical size of the HARPS CCD pixels, which solves this problem of stitchings.

To build the master of $\alpha$\,Cen\,B, we selected spectra for which the SNR in order 10 was larger than 100, for which the airmass was below 1.5 and for which the activity level was below \logrhk=$-4.95$.
In addition, we rejected also spectra contaminated by $\alpha$\,Cen\,A \citep[see][]{Dumusque-2012}, which gave us in the end 1200 spectra, i.e. 5,7\% of all available spectra, that were used to build the master stellar spectrum.

During the data cleaning process described in Sec.~\ref{subsec:stellar_RV}, a total of 310 spectral lines were rejected out of 7317. In addition, 1071 lines were not considered when deriving the
final RVs because they were too close to CCD stitchings.

\begin{figure*}[!t]
\center
 \includegraphics[angle=0,width=0.80\textwidth,origin=br]{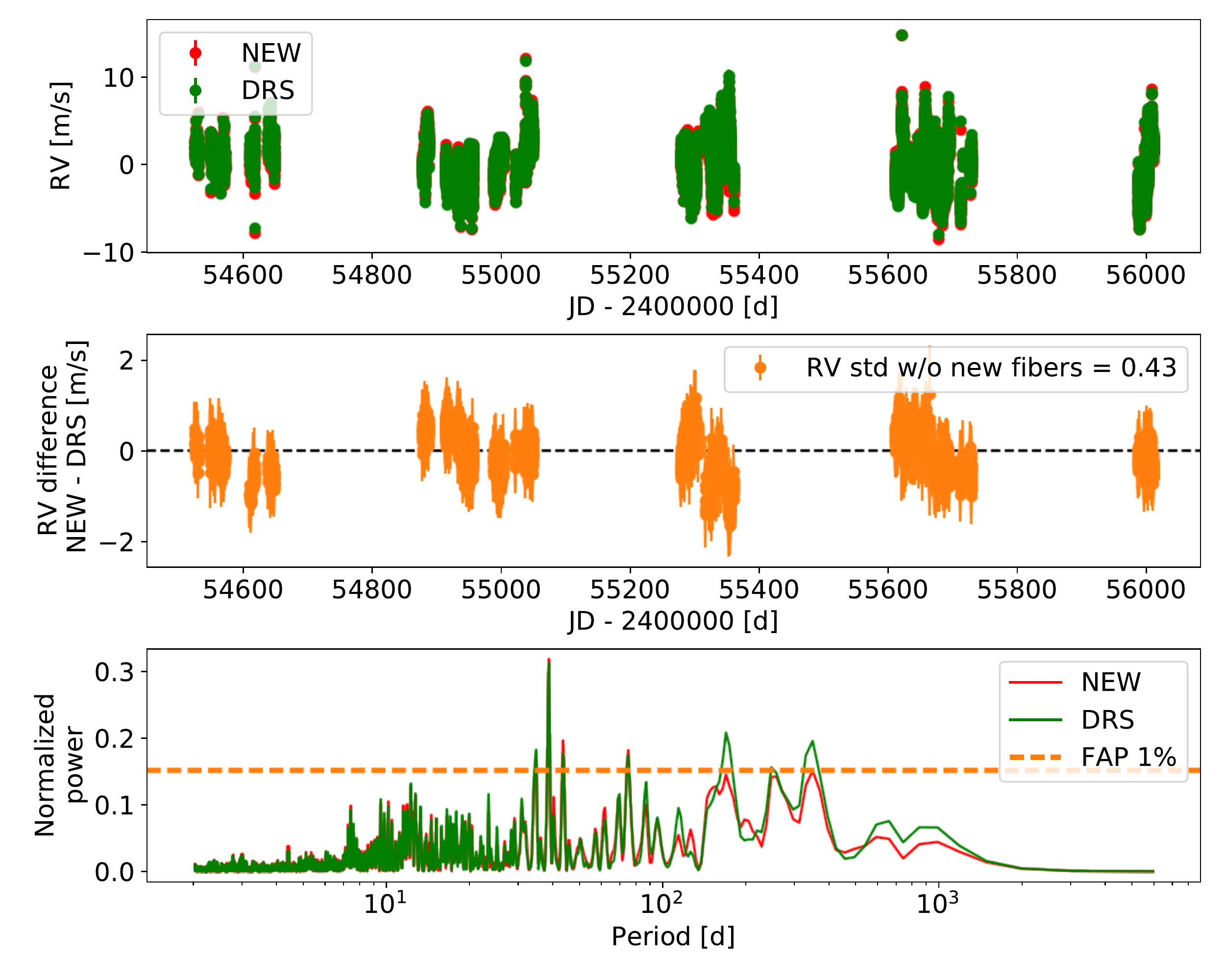}
\caption[]{Same as Fig.~\ref{fig:HD10700_comparison_with_DRS} but for the RVs of $\alpha$\,Cen\,B.}
\label{fig:HD128621_comparison_with_DRS}
\end{figure*}

We compare in Fig.~\ref{fig:HD128621_comparison_with_DRS}, as for $\tau$\,Ceti, the RVs obtained with our new RV extraction procedure and with the HARPS DRS. Once again, we see extremely similar RVs, with a RV difference standard deviation of
0.43\,\ms. The periodogram of both reduction are extremely similar, with this time signals at one year and its first harmonic that are slightly less significant when using our new RV extraction procedure. The histogram of the RVs, shown in Fig.~\ref{fig:comparison_STD_with_DRS}, show once again very similar distributions. Therefore, in the case of stellar activity, our new RV extraction procedure gives extremely similar results than the HARPS DRS.

\subsection{Radial velocities of HD10180}
\label{subsec:RV_HD10180}

After testing our new RV extraction procedure on $\alpha$\,Cen\,B, a star that shows solar-like activity, and on $\tau$\,Ceti, a very inactive star, we test here if 
our reduction can be used to look for planetary signals. Although 5 planets have been announced orbiting $\tau$\,Ceti \citep[][]{Feng:2017ac}, most of the signals found 
are really at the detection limit, and thus two different reductions of the data might give different results. We therefore analyse here a simpler case, the \object{HD10180} planetary system which 
is composed of 6 Neptune-like planets \citep[][]{Lovis-2011a}.

To build the stellar master spectrum used as reference to measure the RVs of each spectral line, we selected spectra that have a SNR in order 10 larger than 10, for which 
the airmass is smaller than 1.5 and for which the activity level is smaller than \logrhk=$-4.95$. After rejecting bad spectra as explained in Sec.~\ref{subsec:master_spectrum}, we are left with 210 spectra that are used to build the master stellar spectrum, which represents 62.5\% of all spectra available.

During the data cleaning process described in Sec.~\ref{subsec:stellar_RV}, a total of 497 spectral lines were rejected out of 7317.

\begin{figure*}[!t]
\center
 \includegraphics[angle=0,width=0.80\textwidth,origin=br]{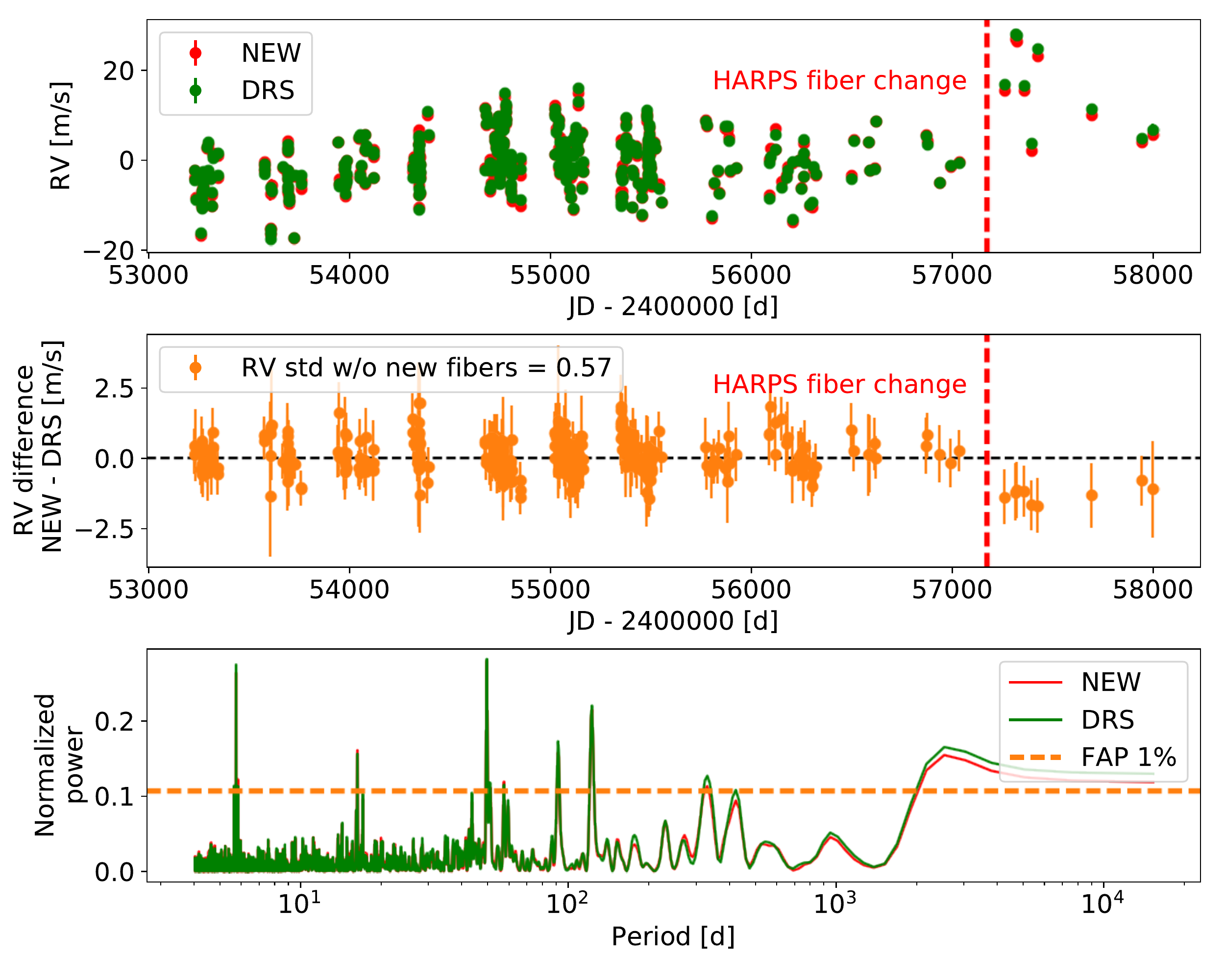}
 \caption[]{Same as Fig.~\ref{fig:HD10700_comparison_with_DRS} but for the RVs of HD10180.}
\label{fig:HD10180_comparison_with_DRS}
\end{figure*}

In Fig.~\ref{fig:HD10180_comparison_with_DRS}, we compare the RVs derived using our new RV extraction procedure with the RVs from the HARPS DRS. Like in the case of $\alpha$\,Cen\,B and $\tau$\,Ceti, we find very similar RVs, with a standard deviation of the
RV difference equals to 0.57\,\ms. The periodograms of the two set of RVs looks very similar, likewise the histogram of the RVs shown in Fig.~\ref{fig:comparison_STD_with_DRS}.

To check if our new RV extraction procedure gives as precise orbital parameters as the HARPS DRS for the planetary signals present in the data, we first rejected measurements obtained after the fibre change, as we estimated that nine data points is not enough 
to robustly fit for the offset induced by this intervention. We also did not consider, when fitting for the different signals, the Earth-mass planet announced in \citet{Lovis-2011a} at 1.2 days, as this discovery is at the limit 
of detection. After selecting the same data and model for both the RVs derived with our new RV extraction procedure and with the HARPS DRS, we used the MCMC tool available on the DACE platform\footnote{The DACE platform 
is available at \url{https://dace.unige.ch} while the online tools to analyse radial-velocity data can be found in the section Observations$=>$RVs} to derive precise orbital parameters for the planets orbiting HD10180.
We used the orbital parameters derived in \citet{Lovis-2011a} as initial guess for our MCMC fit.

The results of our orbital parameter estimation are summarised in Table~\ref{tab:HD10180_orbital_para}. As we can see, all the orbital parameters derived using both RV data sets are compatible within 1--$\sigma$. The log Posterior is also compatible within 1--$\sigma$,
and the RV residuals after removing the best fitting model only differs by 0.02\,\ms. Therefore, our new RV extraction procedure gives RVs that are as good as the HARPS DRS RVs to look for planetary signals.
\begin{table*}
\caption{Best-fitted solution for the planetary system orbiting HD10180, using either the RVs derived from our new RV extraction procedure (\emph{top}) or the ones derived by the HARPS DRS (\emph{bottom}). For each parameter, the median of the posterior is considered, with error bars computed from the MCMC chains using a 68.3\% confidence interval. $\sigma_{O-C}$ corresponds to the weighted standard deviation of the RV residuals around this best solutions. The orbital parameters derived using our new RV extraction procedure are all compatible within 1--$\sigma$ with the parameters derived using the HARPS DRS RVs.}  \label{tab:HD10180_orbital_para}
\def\arraystretch{1.5}
\footnotesize
\begin{center}
\begin{tabular}{lccccccc}
\hline
\hline
Param. & Units & HD10180b & HD10180c & HD10180d & HD10180e & HD10180f & HD10180g \\
\hline
\multicolumn{8}{c}{\large{NEW REDUCTION}}\\
\hline
$P$ & [d] & 5.7595$_{-0.0002}^{+0.0002}$  & 16.3528$_{-0.0029}^{+0.0027}$  & 49.7580$_{-0.0166}^{+0.0165}$  & 122.7190$_{-0.1505}^{+0.1429}$  & 605.2576$_{-8.0929}^{+8.4617}$  & 2358.5466$_{-60.4040}^{+68.1815}$  \\
$K$ & [m\,s$^{-1}$] & 4.61$_{-0.14}^{+0.14}$  & 2.71$_{-0.14}^{+0.14}$  & 4.29$_{-0.14}^{+0.15}$  & 2.85$_{-0.14}^{+0.17}$  & 1.37$_{-0.22}^{+0.25}$  & 2.68$_{-0.17}^{+0.18}$  \\
$e$ &   & 0.07$_{-0.03}^{+0.03}$  & 0.08$_{-0.05}^{+0.06}$  & 0.06$_{-0.04}^{+0.04}$  & 0.10$_{-0.05}^{+0.06}$  & 0.24$_{-0.13}^{+0.12}$  & 0.14$_{-0.06}^{+0.06}$  \\
$\omega$ & [deg] & -22.63$_{-23.99}^{+28.28}$  & -79.32$_{-41.99}^{+35.58}$  & 168.66$_{-33.50}^{+33.20}$  & 6.36$_{-35.39}^{+30.71}$  & -18.12$_{-36.54}^{+40.36}$  & -153.78$_{-30.57}^{+31.69}$  \\
$T_P$ & [d] & 55499.0677$_{-0.3838}^{+0.4527}$  & 55491.9314$_{-1.8815}^{+1.6317}$  & 55501.6305$_{-4.6578}^{+4.5708}$  & 55511.2088$_{-11.9208}^{+10.2449}$  & 55246.9659$_{-56.8119}^{+65.5251}$  & 53768.2714$_{-183.7064}^{+189.1611}$  \\
\hline
$Ar$ & [AU] & 0.0641$_{-0.0010}^{+0.0010}$  & 0.1285$_{-0.0021}^{+0.0020}$  & 0.2699$_{-0.0044}^{+0.0042}$  & 0.4926$_{-0.0080}^{+0.0077}$  & 1.4269$_{-0.0260}^{+0.0266}$  & 3.5351$_{-0.0838}^{+0.0869}$  \\
M.$\sin{i}$ & [M$_{\rm Earth}$] & 13.37$_{-0.57}^{+0.63}$  & 11.11$_{-0.69}^{+0.69}$  & 25.57$_{-1.15}^{+1.19}$  & 22.92$_{-1.46}^{+1.50}$  & 18.16$_{-2.89}^{+2.99}$  & 57.27$_{-4.04}^{+4.12}$  \\
\hline 
$\sigma_{(O-C)}$ & [m\,s$^{-1}$] & \multicolumn{6}{c}{1.60}\\
$\log{(\rm Post})$ &   & \multicolumn{6}{c}{-631.4464$_{-4.5282}^{+4.1029}$}\\
\hline
\multicolumn{8}{c}{\large{HARPS DRS}}\\
\hline
$P$ & [d] & 5.7595$_{-0.0002}^{+0.0002}$  & 16.3502$_{-0.0029}^{+0.0030}$  & 49.7573$_{-0.0162}^{+0.0140}$  & 122.7351$_{-0.1316}^{+0.1286}$  & 601.0734$_{-6.9850}^{+7.4069}$  & 2343.9809$_{-60.1932}^{+63.6823}$  \\
$K$ & [m\,s$^{-1}$] & 4.64$_{-0.14}^{+0.14}$  & 2.68$_{-0.14}^{+0.14}$  & 4.34$_{-0.14}^{+0.15}$  & 2.97$_{-0.16}^{+0.16}$  & 1.43$_{-0.22}^{+0.26}$  & 2.73$_{-0.16}^{+0.17}$  \\
$e$ &   & 0.06$_{-0.03}^{+0.03}$  & 0.07$_{-0.05}^{+0.06}$  & 0.04$_{-0.02}^{+0.04}$  & 0.09$_{-0.05}^{+0.05}$  & 0.26$_{-0.14}^{+0.12}$  & 0.11$_{-0.06}^{+0.06}$  \\
$\omega$ & [deg] & -31.00$_{-27.99}^{+30.84}$  & 280.35$_{-50.69}^{+38.49}$  & 162.70$_{-69.59}^{+55.20}$  & 0.11$_{-34.10}^{+34.88}$  & -30.91$_{-29.94}^{+34.59}$  & -148.51$_{-35.49}^{+32.56}$  \\
$T_P$ & [d] & 55498.9335$_{-0.4470}^{+0.4853}$  & 55508.3341$_{-2.0367}^{+1.8072}$  & 55500.9717$_{-9.3216}^{+7.7660}$  & 55385.5918$_{-11.7552}^{+11.7503}$  & 55215.6075$_{-45.2675}^{+54.7871}$  & 53807.7475$_{-212.7090}^{+199.9718}$  \\
\hline
$Ar$ & [AU] & 0.0641$_{-0.0010}^{+0.0010}$  & 0.1285$_{-0.0020}^{+0.0020}$  & 0.2699$_{-0.0042}^{+0.0041}$  & 0.4926$_{-0.0076}^{+0.0075}$  & 1.4208$_{-0.0252}^{+0.0242}$  & 3.5211$_{-0.0819}^{+0.0843}$  \\
M.$\sin{i}$ & [M$_{\rm Earth}$] & 13.47$_{-0.60}^{+0.60}$  & 10.99$_{-0.67}^{+0.68}$  & 25.91$_{-1.19}^{+1.16}$  & 23.78$_{-1.42}^{+1.62}$  & 18.93$_{-2.80}^{+2.98}$  & 58.28$_{-3.86}^{+4.06}$  \\
\hline 
$\sigma_{(O-C)}$ & [m\,s$^{-1}$] & \multicolumn{6}{c}{1.58}\\
$\log{(\rm Post})$ &   & \multicolumn{6}{c}{-630.8563$_{-4.5542}^{+3.8770}$}\\
\hline
\end{tabular}
\end{center}
\end{table*}

%
%

\section{Night-to-night RV offsets due to the wavelength solution}
\label{sec:night_to_night_offsets}

As explained in Sec. \ref{subsec:spectra_corr}, the HARPS DRS derives the
wavelength solution by fitting on each spectral order a 3rd order polynomial on the known wavelength of the Th-Ar emission lines \citep[][]{Redman:2014aa}.
On the edges of each order, the flux on the Th-Ar emission spectrum is low due to the blaze, and thus the polynomial is not well
constrained at those locations and can vary significantly from night-to-night. Therefore, some stellar spectral lines that appear on the edges of the spectrograph orders
can have a wavelength solution that varies of several dozens of \ms\,from night-to-night. Still due to the blaze, those spectral lines will have a small flux, and 
will therefore not contribute strongly to the HARPS DRS RVs, which combines the RV information of all spectral lines together. Nevertheless, the HARPS DRS RVs should still
be affected by night-to-night RV offsets below the \ms\,level.

In Sec.~\ref{sec:correcting_night_to_night_offsets} of the Appendix we use our new RV extraction procedure to demonstrate that the RVs of spectral lines falling on the edges of the HARPS detector are indeed affected by dozens of m/s night-to-night offsets. To mitigate this observed night-to-night RV offsets induced by the non-stability of the wavelength solutions, a possible solution is to use in combinaison to the Th-Ar spectrum, the much denser spectrum of the FP \'etalon to stabilise the wavelength solution on the edges of the HARPS CCD \citep[][]{Bauer:2015aa}. This work will be soon implemented in the HARPS DRS (Cersullo et al. 2018, submittted to A\&A) and will be available to the community. Performing a wavelength solution every night simplify a lot the data analysis as this fix the zero velocity point of the instrument every night, and it is therefore not necessary to estimate the drift of the instrument from night-to-night. In Sec.~\ref{sec:correcting_night_to_night_offsets} of the Appendix, we test another solution to mitigate the night-to-night offsets. This solution consists in building a master Th-Ar spectrum from all the Th-Ar calibrations, then applying the wavelength solution of the master to all Th-Ar spectra and measure the drift between them and the master. Doing so, only a unique wavelength solution is used, which solve for the problem of wavelength solution unstability. By using this technique in combination with our new RV extraction approach, we demonstrate that for the 4 years of $\alpha$\,Cen\,B RV data, the standard deviation of the night-to-night systematics can be estimated to 0.4\,\ms. Readers that want more details are invited to have look at  Sec.~\ref{sec:correcting_night_to_night_offsets} of the Appendix.

\section{Mitigating stellar activity in radial-velocity measurements}
\label{sec:mitigating_activity}

Now that we are confident that our new way of extracting the RV information from HARPS spectra
gives RVs that are as precise as the ones derived by the HARPS DRS, we can study the RV variation of each individual spectral line.

We want to investigate here the effect of stellar activity on each individual spectral line.
Stellar activity, due to the presence of strong magnetic fields in active 
regions, modifies locally the effective temperature inside spots, and inhibit convection in spots and faculae.
We know that each spectral line has a different sensitivity to temperature variation, and therefore, the appearance of a spot on the stellar surface 
should modify the stellar spectrum. In addition, spectral lines are formed at different depth in the stellar photosphere, and are thus affected differently
by convection, as the velocity of convection also varies with depth. In \citet{Gray-2009}, the author shows that the convective blueshift
is of a few hundred of \ms\,for shallow lines, formed deep inside the stellar photosphere, while it is negligible for very deep lines, formed close to the stellar surface. 
This effect is also well illustrated in the Fig.~3 appearing in \citet{Reiners-2015}. Therefore, when the convective blueshift is reduced due to the presence of strong magnetic fields in active regions,
shallow lines should be significantly affected, while deep lines should not. 

Some preliminary works have shown that each spectral line are affected differently by stellar activity. In \citet{Davis:2017aa}, the authors show, using PCA on data simulated
with SOAP 2.0 \citep[][]{Dumusque-2014b}, that spectral lines seems to be affected differently by activity. A similar result is also shown in \citet{Thompson-2017}and \citet{Wise:2018aa}, were the authors
demonstrate that stellar activity modifies the equivalent width of spectral lines. However, although variation in spectral line is seen in those analyses, it is not clear if those variations
induce a change in RV. Indeed, a spectral line could change in width and/or depth, without changing in asymmetry and therefore without inducing a RV effect.

To analyse the effect of stellar activity on each spectral line, we need RV observations significantly affected by activity, and with a dense enough sampling to detect stellar rotation.
We therefore used, like in \citet{Thompson-2017}, the 2010 data of $\alpha$\,Cen\,B. On those data, we clearly see 
a $\sim$5\,\ms\,sinusoidal variation due to stellar activity over two rotational periods of the star \citep[][]{Dumusque-2014c}. A similar signal is also seen in the \logrhk\,activity index \citep[][]{Dumusque-2012}.

When measuring RVs, the activity signal that we observe is coming from the spectral lines formed inside magnetic regions, faculae and spots, that are formed at the level of the stellar photosphere. The magnetic field that are at the origin of these regions goes up to the chromosphere and create regions that are called plage. The \logrhk\,activity index is known to probe plage, and although faculae, spots and plages are related to each other as they originate from the same magnetic fields, it is not necessarily expected that a one-to-one correlation exists between the RV variation induced by stellar activity and the \logrhk\,activity index.
We therefore decided here to select the RV
measured on all the spectral lines as our proxy for the activity signal, and we compared the RV of each individual line relative to it. To mitigate the night-to-night offsets observed in the RVs of spectral lines 
falling on the edges of the HARPS CCD described in Sec.~\ref{sec:night_to_night_offsets} and in Sec.~\ref{sec:correcting_night_to_night_offsets} of the Appendix, we used the RV data derived with our new RV extraction procedure using a unique wavelength solution. 
We can see in the right bottom plot of Fig.~\ref{fig:compa_wavesol} that mitigating those night-to-night offsets is extremely important in order to improve the correlation between the RV
of each spectral line and the RV measured on all the lines. We finally rejected from this analysis spectral lines falling close to CCD stitchings, to prevent being perturbed by the inaccuracy of the wavelength solution
close to those regions \citep[][]{Dumusque-2015a}.

\begin{figure*}[!h]
\center
 \includegraphics[angle=0,width=0.48\textwidth]{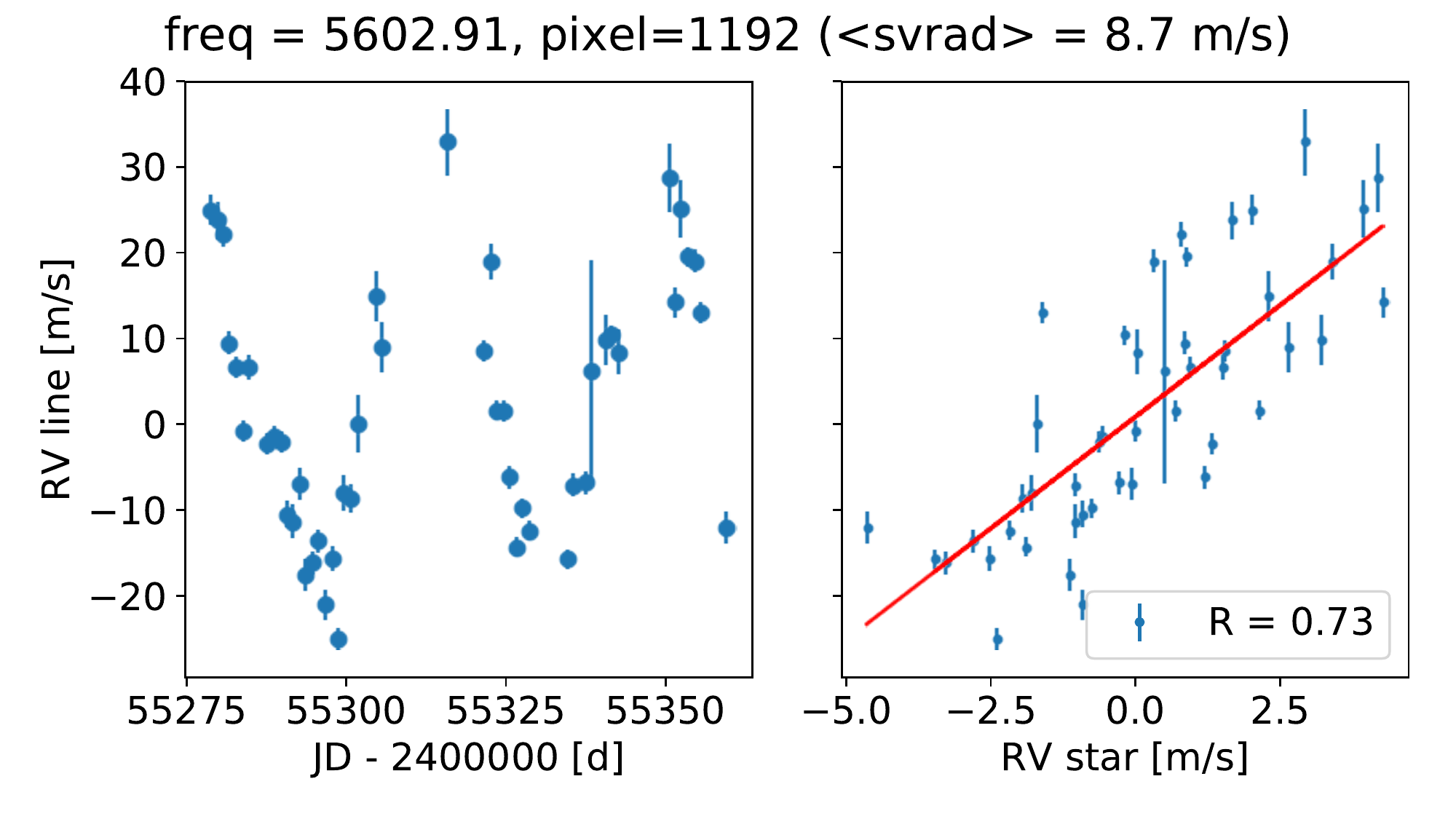}
  \includegraphics[angle=0,width=0.48\textwidth]{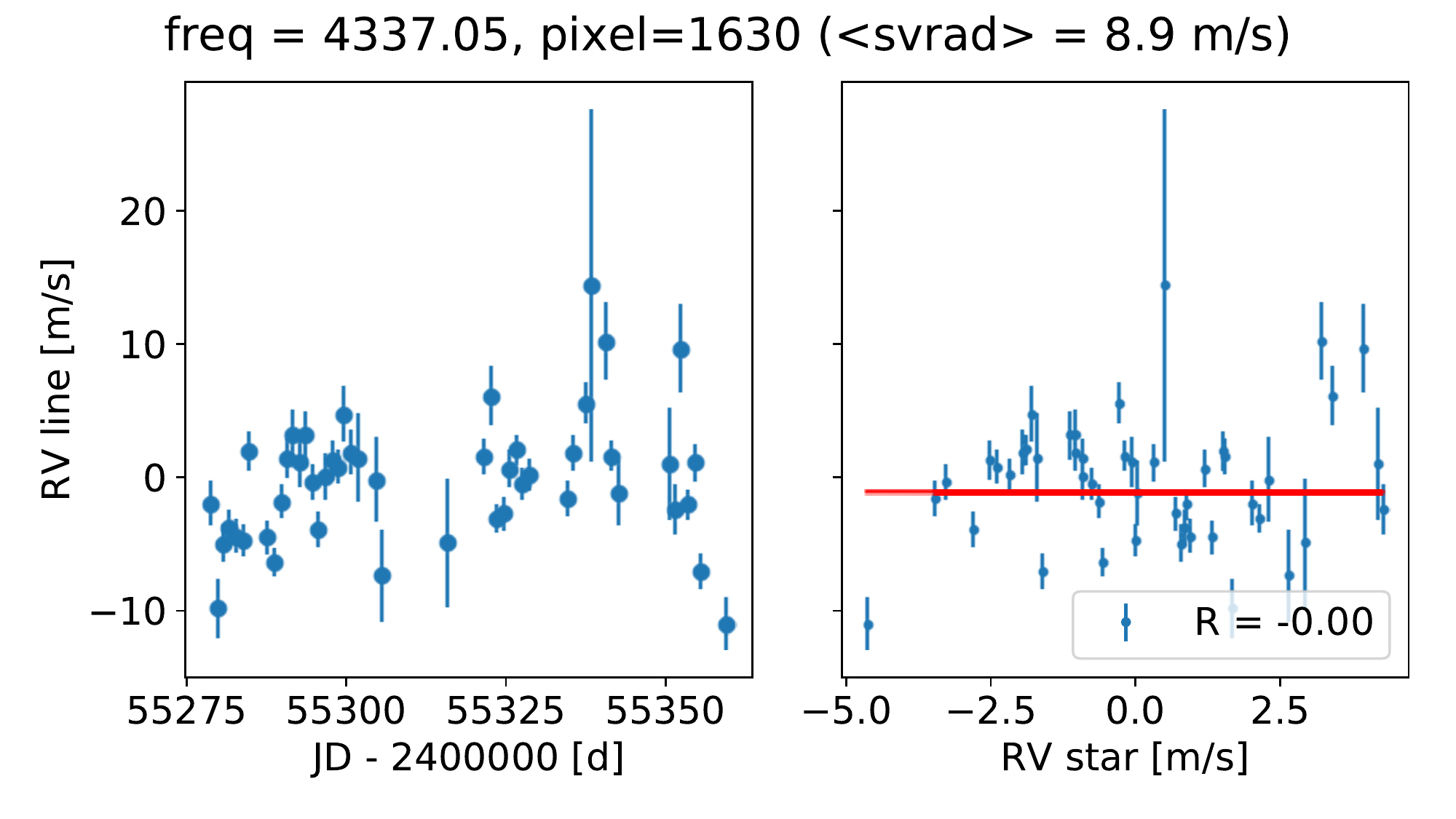}
   \includegraphics[angle=0,width=0.48\textwidth]{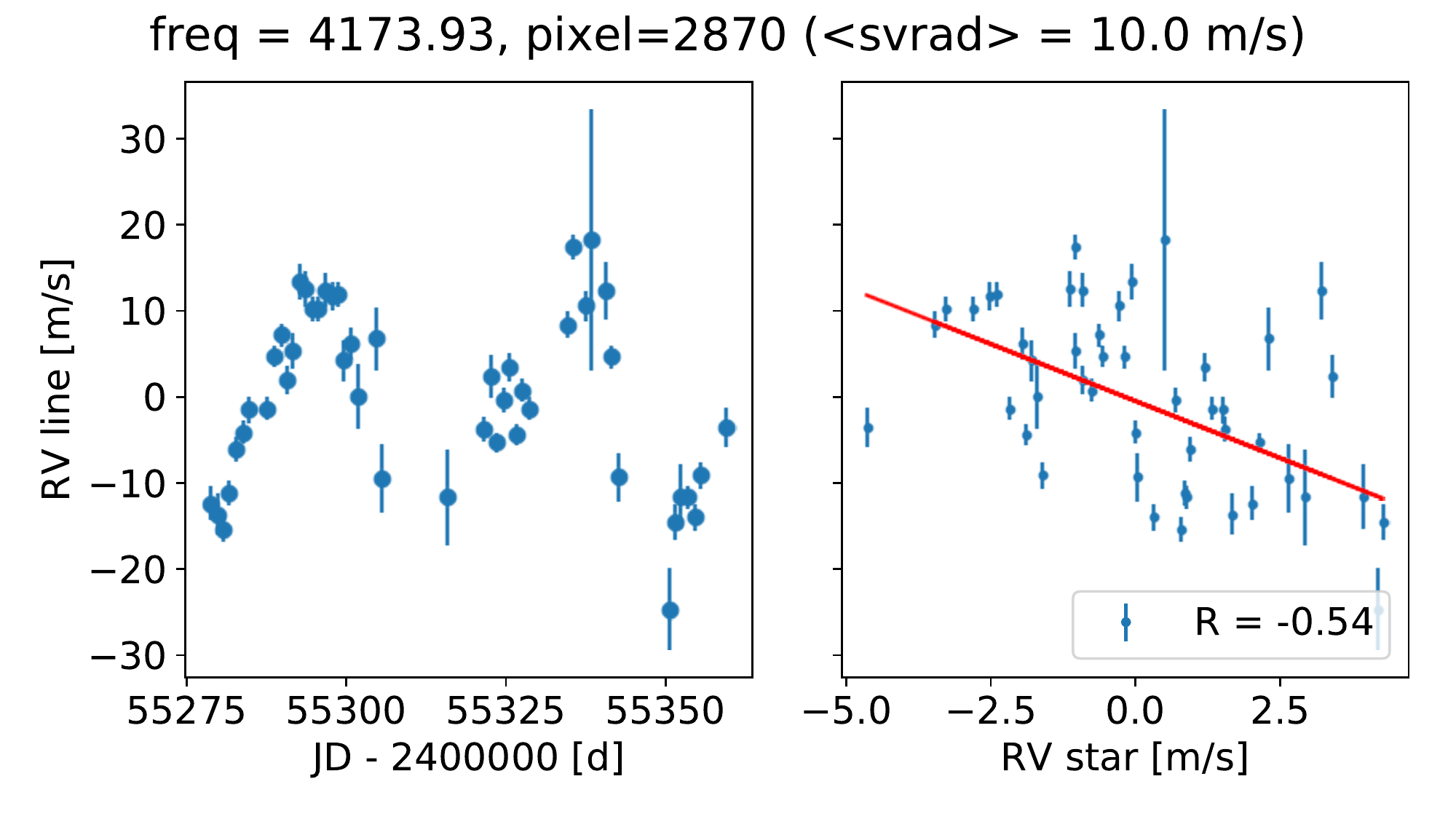}
 \caption[]{The three different panels show the 2010 RVs of $\alpha$\,Cen\,B as measured on three different spectral lines. On each of the panel we show in the title the wavelength of the line in the air, its position on the HARPS CCD in pixel, and the median RV error of this spectral line $<$svrad$>$. Each panel is divided in two, on the left we show the RV of the line as a function of time, and on the right, the correlation of the RV of this line with the RV derived using all the spectral lines. In 2010, $\alpha$\,Cen\,B is activity and the signal seen in the RV of all the lines is mainly due to activity. Therefore, a strong correlation between the RV of a spectral line with those RVs implies that the line is strongly affect by activity signals.}
\label{fig:RV_correlation_with_activity}
\end{figure*}

\begin{figure}[!h]
\center
 \includegraphics[angle=0,width=0.48\textwidth]{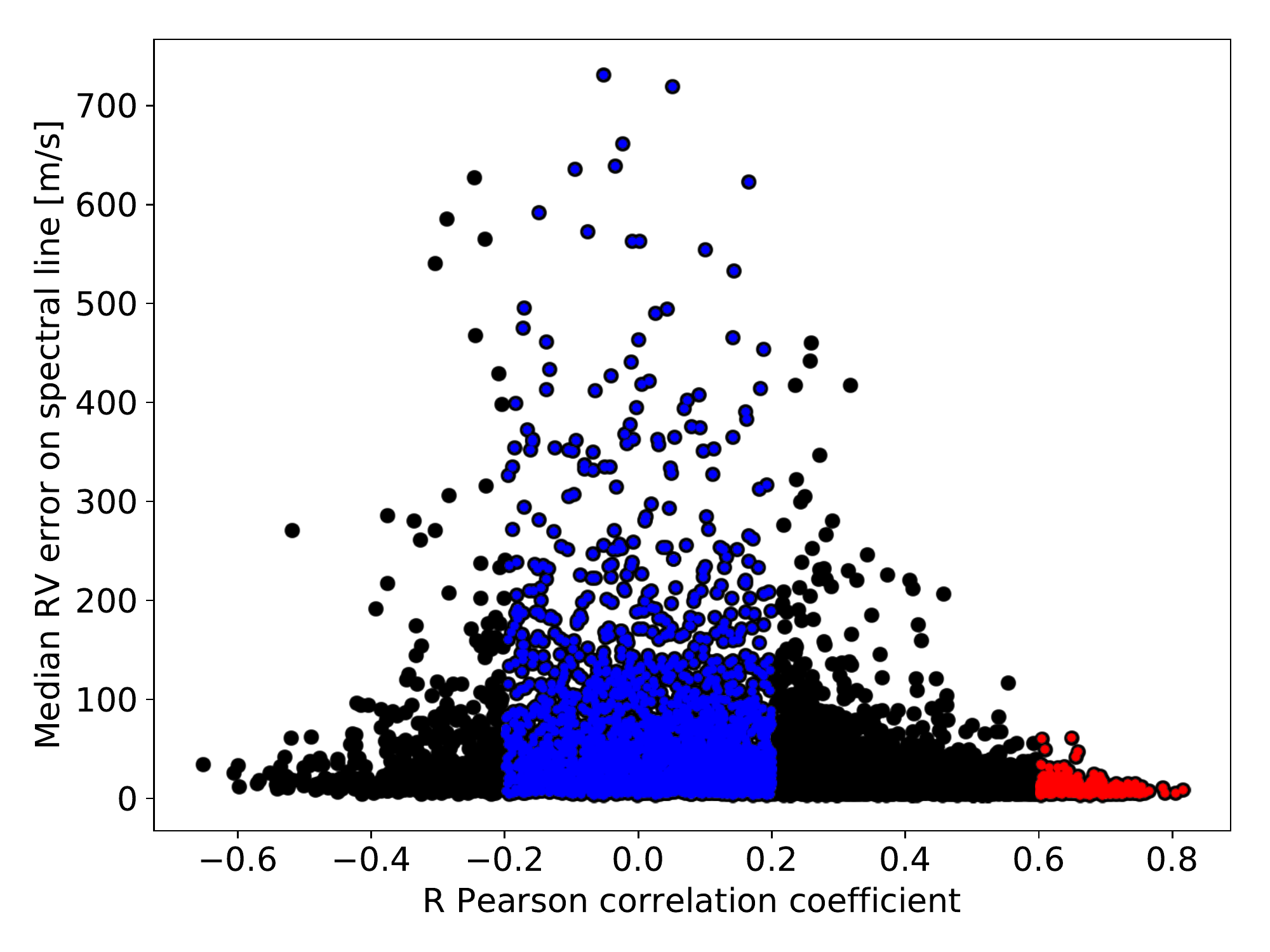}
 \caption[]{RV precision on the RVs of each spectral line as a function of the R Pearson correlation coefficient between the RVs of each individual line with the RVs measured on all the lines, as shown in Fig.~\ref{fig:RV_correlation_with_activity}. The blue selection corresponds to spectral line for which the absolute R correlation coefficient is below 0.2, and therefore to lines that does not seem to be strongly affected by stellar activity. The red select includes lines for which the correlation is stronger than 0.6, and therefore lines that are sensitive to activity.}
\label{fig:R_all_lines}
\end{figure}

In Fig.~\ref{fig:RV_correlation_with_activity}, we compare the RVs of three different spectral lines with respect to the activity signal in $\alpha$\,Cen\,B, which is in our case the RVs measured on all the lines (see preceding paragraph).
As we can see, the spectral line at 5602.91\,\AA\,is strongly correlated with the stellar activity signal, and shows a RV peak-to-peak amplitude of 60\,\ms. On the opposite, we see that the line at 4173.93\,\AA\,is anti-correlated with
the activity signal. We also observe a lot of lines for which the correlation is null, like for the line at 4337.05\,\AA. In Fig.~\ref{fig:R_all_lines}, we show the median RV error of each spectral line with respect to the R Pearson correlation coefficient between the RVs of those lines and stellar activity. As we can see, we observe 2951 spectral lines that have an absolute correlation weaker than 0.2, in blue, and 489 spectral lines in red that present strong correlation with activity (R$>$0.6). 

\begin{figure*}[!h]
\center
 \includegraphics[angle=0,width=0.98\textwidth]{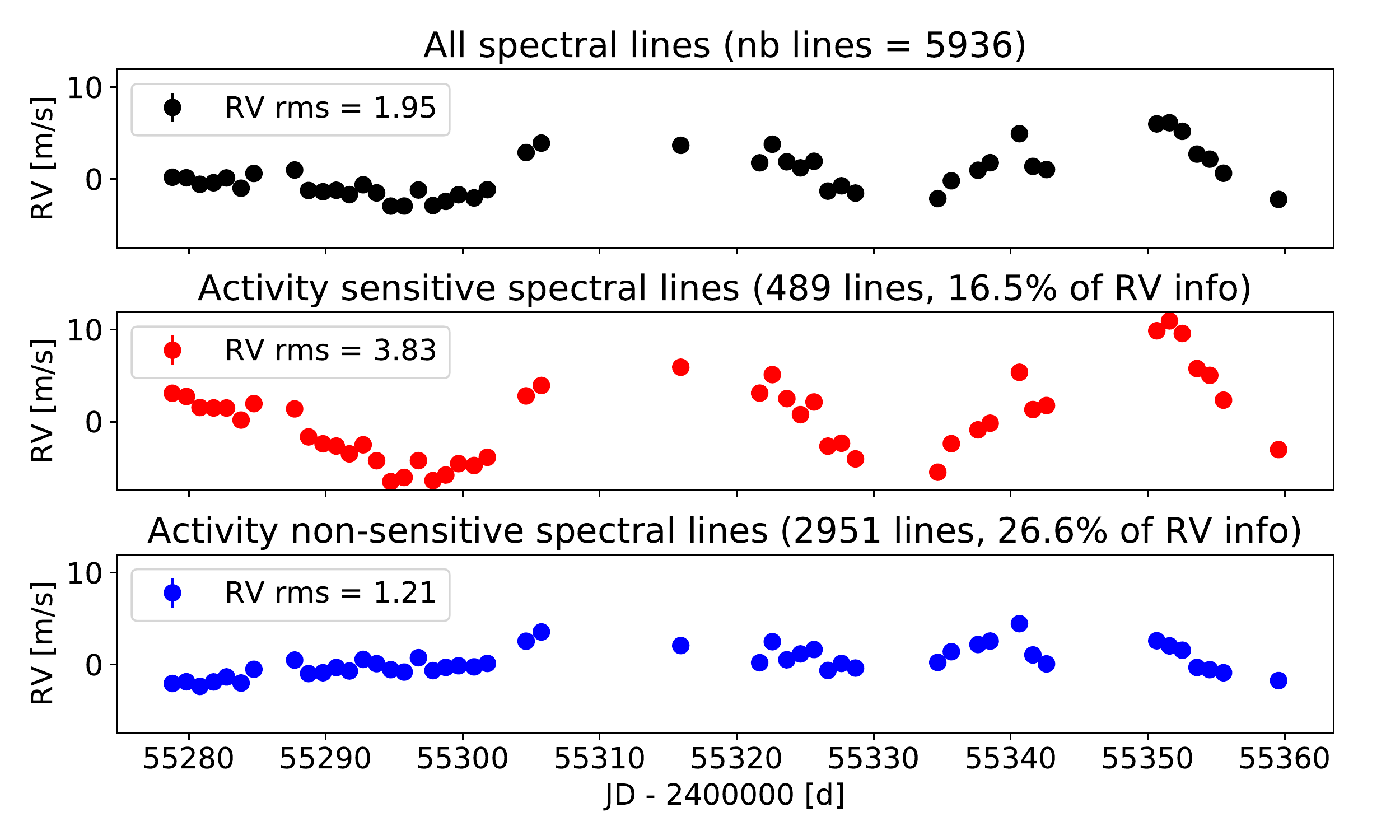}
 \caption[]{Comparison of the 2010 RVs of $\alpha$\,Cen\,B as derived when using all the spectral line (in black), when using lines for which the correlation between their RVs and the RV of all the lines is strong (in red) and when using lines for which the correlation is weak (in blue). The blue and red selections of lines can be see in Fig.~\ref{fig:R_all_lines}. We highlight as well, in the title of each subplot, the number of lines used in each selection, and by how much those lines contribute to the RVs derived using all the spectral lines available. We clearly see that by selecting lines that are strongly correlated or not with the activity signal, we are able to boost the activity signal seen in the RVs by a factor of 2, or mitigate it by a factor of 1.6.}
\label{fig:RV_with_different_line_selection}
\end{figure*}

Now that we observe that the RV of each spectral seems to be affected differently by stellar activity, we want to test if measuring the RV using only affected or non-affected spectral lines can modify the stellar activity signal seen in RV.
We therefore calculat the RVs using all the lines available, using the affected red lines shown in Fig.~\ref{fig:R_all_lines}, and the non-affected blue lines shown in the same figure. Using the same colour code, we show the result in
Fig.~\ref{fig:RV_with_different_line_selection}. 

When combining the RVs of the 489 spectral lines strongly affected by stellar activity, we obtain RVs that present a standard deviation that is twice higher than the one measured on the RVs of all the lines. By using the RV error of each of these lines, which is shown in Fig.~\ref{fig:R_all_lines}, we can compute the RV weight of those 489 spectral lines with respect to the RV weight of all the spectral lines, which by definition is 100\%. Doing so, we find that the RV weight of those 489 lines is 16.5\,\%, which implies that the photon noise will be $1/\sqrt{0.165}=2.5$ times larger compared to the photon noise obtained when measuring the RV on all spectral lines. For $\alpha$\,Cen\,B, the RV photon noise is so low that this increase by a factor 2.5 will not have a significant impact, however 
it could be the case for fainter stars.

When combining the RVs of the 2951 spectral lines that show a weak correlation with activity, we obtain the blue RVs in Fig.~\ref{fig:RV_with_different_line_selection}. The standard deviation of those RVs is 1.6 times smaller than the standard deviation measured on the RVs derived using all the spectral lines. Therefore, the activity signal is significantly mitigated when using this specific selection of spectral line. Those 2951 lines contain 26.6\,\% of the total RV information, implying that the photon noise on the RVs obtained from this line selection is $1/\sqrt{0.266}=1.9$ times larger than the photon noise obtained when measuring the RV on all spectral lines. By measuring the RVs only on spectral lines not significantly affected by activity, we are able to mitigate the red noise induced by stellar activity by a factor of 1.6 while increasing the white photon-noise by a factor of 1.9. Mitigating the red noise induced by stellar activity has more weight than increasing the white photon noise when searching for exoplanet signal in RV data. 

We are confident that deriving the RVs using sub-samples of spectral lines is the key to mitigate the stellar activity signal. When looking at the RV derived using only unaffected lines, it seems that we still see some signal from stellar activity. What we could do is looking at the correlation between the RV of the affected and unaffected lines, and removing it to mitigate even better the activity signal in the unaffected RVs. We however have to be careful, because if a planet is present in the data, its signal will be the same in the RV of the affected or unaffected lines. By removing the correlation, we could modify the planetary signal in the RV residuals and thus obtain incorrect orbital parameters. In conclusion, we have to find what is the best technique to correct for stellar activity using the RV measured on sub-samples of spectral lines, and this is what we will investigate in a forthcoming paper (Cretignier et al. 2018, in prep.).

\section{Discussion}
\label{sec:discussion}

In this paper, we present a new approach to derive RVs from high-resolution spectra. First, we measure the RV of each individual spectral line
and then we combine the RV information of all the spectral lines together to obtain a precise stellar RV.

In Sec.~\ref{sec:RV_comparison} we compare the RVs derived by our new RV extraction procedure, with the RVs obtained from the HARPS DRS, which is the gold standard for precise RV estimate from G and K dwarfs HARPS spectra.
For the three stars studied, $\tau$\,Ceti, $\alpha$\,Cen\,B and HD10180, we find very similar results (see Figs.~\ref{fig:HD10700_comparison_with_DRS}, \ref{fig:HD128621_comparison_with_DRS} and \ref{fig:HD10180_comparison_with_DRS}). In the case of $\tau$\,Ceti (see Sec.~\ref{subsec:RV_HD10700}), although the RVs
are extremely similar when looking at the RV difference between the two reduction and at the distribution of the derived RVs, it seems that ou new RV extraction procedure is slightly
more sensitive to systematics that create spurious RV signals between 100 and 1000 days in period. Those signals are however not significant when considering a false alarm probably of 1\%. 
We still try to investigate the origin of those extra signals in Sec.~\ref{Systematics in HD10700 RVs} of the Appendix, were we show that they might be due to 
the fact that our reduction is more sensitive to systematics induced by spectral line crossing CCD stitchings, and by the effect of tellurics. For HD10180 (see Sec.~\ref{subsec:RV_HD10180}), besides
checking that the RVs from both reductions are extremely similar, we also compared the orbital parameters derived for the planets present in the data. Using the two different RV reductions, we found exactly
the same orbital parameters considering their 1--$\sigma$ uncertainty. Therefore, the RVs derived with our new RV extraction procedure gives RVs that are as good as the HARPS DRS to search for exoplanets.

Using our new RV extraction procedure, we also estimated the drift of the spectrograph overnight, by measuring the RV as a function of time for each of the Th-Ar or FP \'etalon emission spectral lines
that appear on the simultaneous reference spectra that are taken contemporaneously with stellar observations. As we can see in Fig.~\ref{fig:drift_comparison_with_DRS}, our new RV extraction procedure gives also extremely similar results than the HARPS DRS when estimating the drift of the instrument overnight.

Looking at the RV of each individual spectral line, we saw in Sec.~\ref{sec:night_to_night_offsets} and in Sec.~\ref{sec:correcting_night_to_night_offsets} of the Appendix that the RVs of lines falling on the edges of the HARPS detector can be strongly affected by night-to-night offsets. 
We found that these RV offsets are induced by using a different wavelength solution every night. 

To correct for these night-to-night RV offsets, we choose in this paper to build a master Th-Ar spectrum
and to apply to each Th-Ar calibrations the unique wavelength solution of this Th-Ar master. This prevent us of using non-stable wavelength solutions every night, however this requires to correct for the drift of the instrument over time.
Although this approach was significantly complicating the data reduction process (see Sec.~\ref{sec:correcting_night_to_night_offsets} of the Appendix), we were able to demonstrate that this solution can mitigate the night-to-night RV offsets seen on spectral lines falling on the edges of each spectral order.
By deriving the RV of $\alpha$\,Cen\,B from 2008 to 2012 using a unique wavelength 
solution, we could estimate that the standard deviation of the night-to-night instrumental systematics is 0.4\ms. The obtained RVs are therefore less sensitive to the night-to-night RV offsets, and the activity signal of the star, seen at the rotational period, is better characterised. We however discuss in Sec.~\ref{sec:night_to_night_offsets}
and Appendix \ref{HARPS_systematics} that due to a slow modification of the spectrograph focus over time, the RVs derived using a unique wavelength solution will be affected by a long-term drift, that can however be correct for.

Although we could show that our new RV extraction procedure using a unique wavelength solution reduces the night-to-night RV variations, the final gain in terms of RV precision is rather small.
Therefore, what is the gain of doing this complicated reduction, while a simple CCF approach gives extremely similar results. 

First, the main advantage, and this is the most relevant astrophysical phenomena that our new RV extraction demonstrate, is the observation that some lines are much more affected by stellar activity effects than others (see Sec.~\ref{sec:mitigating_activity}). This is expected from stellar physics, however it was never demonstrated that using current spectrographs used for exoplanet characterisation, we had the precision to measure solar-like stellar activity in the RVs of individual spectral lines. 
By measuring the RV of $\alpha$\,Cen\,B using only lines strongly affected or unaffected by stellar activity, we were able to boost the activity signal in RVs by a factor of 2, or mitigate it by a factor of 1.6, respectively. Deriving the RVs on a subset of spectral lines has for consequence the increase in photon noise of the final RVs. Using the line selection that doubles or mitigate by a factor of 1.6 stellar activity worsen the RV precision by a factor of 2.5 or 1.9, respectively.
On $\alpha$\,Cen\,B, this does not have any consequences as the photon noise for this target is extremely low, however, it might be a problem for fainter targets. This is something
that we will investigate soon, as we believe that for planet detection and characterisation, mitigating the red noise induced by stellar activity 
has much more weight than increasing the white photon noise of the derived RVs.

In addition, our new RV extraction approach can be used to probe how the RV varies locally on the detector, and not only from orders to orders, what the HARPS DRS and other data reduction package do, but also within orders. From Figs.~\ref{fig:DRIFT_A_per_region_all} and \ref{fig:drift_A_53400} it is clear that the drift measured on the instrument 
over time is not the same at every location on the HARPS CCD, and we see significant differences between the blue and red detectors, but also between the left and right sides of the orders. 
Our new RV extraction technique is therefore interesting to probe and understand instrumental systematics. 

Moreover, it is important to note that once the RV of each line is computed, it takes nearly no time to measure the RV for any given selection of line, as this process only requires a weighted average. This is of course extremely interesting if we want to compare the RV derived from a huge number of different line selections.

\section{Conclusion}
\label{sec:conclusion}

In this paper, we present a new approach to derive precise RVs from HARPS spectra. This new method tries to use the wealth of information contained in a high-resolution spectrum rather than averaging all the information into the few parameters of the CCF.
We demonstrate that this new RV extraction procedure can be used to probe stellar spectral lines that are sensitive or not to activity, and that by using a smart selection of them, we can either boost or mitigate the stellar activity signal observed in RV. This is of course 
very interesting in the light of new spectrographs like ESPRESSO that will reach a precision of 0.1\,\ms and for which stellar activity signal will dominate the data. We will therefore investigate in more details the effect of activity on each spectral line and try to understand the underlying physics in a forthcoming paper (Cretignier et al. 2018, in prep.). This forthcoming work will present in detail the lines that are strongly affected or unaffected by stellar activity, with their physical properties. We hope that this will motivate other people in the community to join us in this effort for solving the problem of stellar activity in RV measurements, which would be beneficial for the entire exoplanet community.

This new RV extraction procedure can also be used to study the impact of tellurics on the RVs \citep[see Sec.~\ref{fig:HD10700_comparison_subsample_lines} and][]{Cunha-2014}, but also to understand better instrumental systematics (see for example Figs.~\ref{fig:DRIFT_A_per_region_all} and \ref{fig:drift_A_53400}). 

As a ending note, we want to emphasise that our new RV extraction procedure demonstrated that deriving the RV on a spectral line basis brings a huge amount of information, mainly to understand better and correct for stellar activity, and thus improving the methods for extracting locally the RV in a stellar spectra should be investigated further. The results shown here are not specific to HARPS, and can be obtained for the data of any stabilised high-resolution spectrographs.

\begin{acknowledgements}
XD would like to thank F. Pepe, C. Lovis, F. Bouchy and D. Segransan for their help in understanding all the details of the HARPS spectrograph and its data reduction software.
XD is grateful to The Branco Weiss Fellowship--Society in Science for its financial support. 
This work has been carried out in the framework of the National Centre for Competence in Research \emph{PlanetS} supported by the Swiss National Science Foundation (SNSF).
This research has made use of the DACE platform developed in the framework of \emph{PlanetS} (\url{https://dace.unige.ch}).
\end{acknowledgements}

\bibliographystyle{aa}
\bibliography{dumusque_bibliography}

\begin{thebibliography}{62}
\expandafter\ifx\csname natexlab\endcsname\relax\def\natexlab#1{#1}\fi

\bibitem[{{Anglada-Escud{\'e}} {et~al.}(2016){Anglada-Escud{\'e}}, {Amado},
  {Barnes}, {Berdi{\~n}as}, {Butler}, {Coleman}, {de La Cueva}, {Dreizler},
  {Endl}, {Giesers}, {Jeffers}, {Jenkins}, {Jones}, {Kiraga}, {K{\"u}rster},
  {L{\'o}pez-Gonz{\'a}lez}, {Marvin}, {Morales}, {Morin}, {Nelson}, {Ortiz},
  {Ofir}, {Paardekooper}, {Reiners}, {Rodr{\'{\i}}guez},
  {Rodr{\'{\i}}guez-L{\'o}pez}, {Sarmiento}, {Strachan}, {Tsapras}, {Tuomi}, \&
  {Zechmeister}}]{Anglada-Escude-2016}
{Anglada-Escud{\'e}}, G., {Amado}, P.~J., {Barnes}, J., {et~al.} 2016, \nat,
  536, 437

\bibitem[{{Anglada-Escud{\'e}} \& {Butler}(2012)}]{Anglada-Escude-2012}
{Anglada-Escud{\'e}}, G. \& {Butler}, R.~P. 2012, \apjs, 200, 15

\bibitem[{{Arentoft} {et~al.}(2008){Arentoft}, {Kjeldsen}, {Bedding}, {Bazot},
  {Christensen-Dalsgaard}, {Dall}, {Karoff}, {Carrier}, {Eggenberger},
  {Sosnowska}, {Wittenmyer}, {Endl}, {Metcalfe}, {Hekker}, {Reffert}, {Butler},
  {Bruntt}, {Kiss}, {O'Toole}, {Kambe}, {Ando}, {Izumiura}, {Sato}, {Hartmann},
  {Hatzes}, {Bouchy}, {Mosser}, {Appourchaux}, {Barban}, {Berthomieu},
  {Garcia}, {Michel}, {Provost}, {Turck-Chi{\`e}ze}, {Marti{\'c}}, {Lebrun},
  {Schmitt}, {Bertaux}, {Bonanno}, {Benatti}, {Claudi}, {Cosentino}, {Leccia},
  {Frandsen}, {Brogaard}, {Glowienka}, {Grundahl}, \&
  {Stempels}}]{Arentoft-2008}
{Arentoft}, T., {Kjeldsen}, H., {Bedding}, T.~R., {et~al.} 2008, \apj, 687,
  1180

\bibitem[{{Astudillo-Defru} {et~al.}(2015){Astudillo-Defru}, {Bonfils},
  {Delfosse}, {S{\'e}gransan}, {Forveille}, {Bouchy}, {Gillon}, {Lovis},
  {Mayor}, {Neves}, {Pepe}, {Perrier}, {Queloz}, {Rojo}, {Santos}, \&
  {Udry}}]{Astudillo-Defru:2015aa}
{Astudillo-Defru}, N., {Bonfils}, X., {Delfosse}, X., {et~al.} 2015, \aap, 575,
  A119

\bibitem[{{Baranne} {et~al.}(1979){Baranne}, {Mayor}, \&
  {Poncet}}]{Baranne:1979aa}
{Baranne}, A., {Mayor}, M., \& {Poncet}, J.~L. 1979, Vistas in Astronomy, 23,
  279

\bibitem[{{Baranne} {et~al.}(1996){Baranne}, {Queloz}, {Mayor}, {Adrianzyk},
  {Knispel}, {Kohler}, {Lacroix}, {Meunier}, {Rimbaud}, \&
  {Vin}}]{Baranne-1996}
{Baranne}, A., {Queloz}, D., {Mayor}, M., {et~al.} 1996, \aaps, 119, 373

\bibitem[{{Bauer} {et~al.}(2015){Bauer}, {Zechmeister}, \&
  {Reiners}}]{Bauer:2015aa}
{Bauer}, F.~F., {Zechmeister}, M., \& {Reiners}, A. 2015, \aap, 581, A117

\bibitem[{{Bazot} {et~al.}(2012){Bazot}, {Campante}, {Chaplin}, {Carfantan},
  {Bedding}, {Dumusque}, {Broomhall}, {Petit}, {Th{\'e}ado}, {Van Grootel},
  {Arentoft}, {Castro}, {Christensen-Dalsgaard}, {do Nascimento}, {Dintrans},
  {Kjeldsen}, {Monteiro}, {Santos}, {Sousa}, \& {Vauclair}}]{Bazot-2012}
{Bazot}, M., {Campante}, T.~L., {Chaplin}, W.~J., {et~al.} 2012, \aap, 544,
  A106

\bibitem[{{Bonfils} {et~al.}(2007){Bonfils}, {Mayor}, {Delfosse}, {Forveille},
  {Gillon}, {Perrier}, {Udry}, {Bouchy}, {Lovis}, {Pepe}, {Queloz}, {Santos},
  \& {Bertaux}}]{Bonfils-2007}
{Bonfils}, X., {Mayor}, M., {Delfosse}, X., {et~al.} 2007, \aap, 474, 293

\bibitem[{{Borgniet} {et~al.}(2015){Borgniet}, {Meunier}, \&
  {Lagrange}}]{Borgniet-2015}
{Borgniet}, S., {Meunier}, N., \& {Lagrange}, A.-M. 2015, \aap, 581, A133

\bibitem[{{Bouchy} \& {Carrier}(2002)}]{Bouchy-2002}
{Bouchy}, F. \& {Carrier}, F. 2002, \aap, 390, 205

\bibitem[{{Bouchy} {et~al.}(2001){Bouchy}, {Pepe}, \& {Queloz}}]{Bouchy-2001b}
{Bouchy}, F., {Pepe}, F., \& {Queloz}, D. 2001, \aap, 374, 733

\bibitem[{{Bourrier} \& {H{\'e}brard}(2014)}]{Bourrier-2014}
{Bourrier}, V. \& {H{\'e}brard}, G. 2014, ArXiv e-prints
  [\eprint[arXiv]{1406.6813}]

\bibitem[{{Butler} {et~al.}(1999){Butler}, {Marcy}, {Fischer}, {Brown},
  {Contos}, {Korzennik}, {Nisenson}, \& {Noyes}}]{Butler-1999}
{Butler}, R.~P., {Marcy}, G.~W., {Fischer}, D.~A., {et~al.} 1999, \apj, 526,
  916

\bibitem[{{Butler} {et~al.}(1996){Butler}, {Marcy}, {Williams}, {McCarthy},
  {Dosanjh}, \& {Vogt}}]{Butler-1996}
{Butler}, R.~P., {Marcy}, G.~W., {Williams}, E., {et~al.} 1996, PASP, 108, 500

\bibitem[{{Cegla} {et~al.}(2013){Cegla}, {Shelyag}, {Watson}, \&
  {Mathioudakis}}]{Cegla-2013}
{Cegla}, H.~M., {Shelyag}, S., {Watson}, C.~A., \& {Mathioudakis}, M. 2013,
  \apj, 763, 95

\bibitem[{{Cosentino} {et~al.}(2012){Cosentino}, {Lovis}, {Pepe}, {Collier
  Cameron}, {Latham}, {Molinari}, {Udry}, {Bezawada}, {Black}, {Born},
  {Buchschacher}, {Charbonneau}, {Figueira}, {Fleury}, {Galli}, {Gallie},
  {Gao}, {Ghedina}, {Gonzalez}, {Gonzalez}, {Guerra}, {Henry}, {Horne},
  {Hughes}, {Kelly}, {Lodi}, {Lunney}, {Maire}, {Mayor}, {Micela}, {Ordway},
  {Peacock}, {Phillips}, {Piotto}, {Pollacco}, {Queloz}, {Rice}, {Riverol},
  {Riverol}, {San Juan}, {Sasselov}, {Segransan}, {Sozzetti}, {Sosnowska},
  {Stobie}, {Szentgyorgyi}, {Vick}, \& {Weber}}]{Cosentino-2012}
{Cosentino}, R., {Lovis}, C., {Pepe}, F., {et~al.} 2012, in Society of
  Photo-Optical Instrumentation Engineers (SPIE) Conference Series, Vol. 8446,
  Society of Photo-Optical Instrumentation Engineers (SPIE) Conference Series

\bibitem[{{Cunha} {et~al.}(2014){Cunha}, {Santos}, {Figueira}, {Santerne},
  {Bertaux}, \& {Lovis}}]{Cunha-2014}
{Cunha}, D., {Santos}, N.~C., {Figueira}, P., {et~al.} 2014, ArXiv e-prints
  [\eprint[arXiv]{1407.0181}]

\bibitem[{{Davis} {et~al.}(2017){Davis}, {Cisewski}, {Dumusque}, {Fischer}, \&
  {Ford}}]{Davis:2017aa}
{Davis}, A.~B., {Cisewski}, J., {Dumusque}, X., {Fischer}, D.~A., \& {Ford},
  E.~B. 2017, \apj, 846, 59

\bibitem[{{Del Moro}(2004)}]{Del-Moro-2004b}
{Del Moro}, D. 2004, \aap, 428, 1007

\bibitem[{{Del Moro} {et~al.}(2004){Del Moro}, {Berrilli}, {Duvall}, \&
  {Kosovichev}}]{Del-Moro-2004a}
{Del Moro}, D., {Berrilli}, F., {Duvall}, Jr., T.~L., \& {Kosovichev}, A.~G.
  2004, \solphys, 221, 23

\bibitem[{{Delisle} {et~al.}(2018){Delisle}, {S{\'e}gransan}, {Dumusque},
  {Diaz}, {Bouchy}, {Lovis}, {Pepe}, {Udry}, {Alonso}, {Benz}, {Coffinet},
  {Cameron}, {Deleuil}, {Figueira}, {Gillon}, {Curto}, {Mayor}, {Mordasini},
  {Motalebi}, {Moutou}, {Pollacco}, {Pompei}, {Queloz}, {Santos}, \&
  {Wyttenbach}}]{Delisle:2018aa}
{Delisle}, J.-B., {S{\'e}gransan}, D., {Dumusque}, X., {et~al.} 2018, \aap,
  614, A133

\bibitem[{{Desort} {et~al.}(2007){Desort}, {Lagrange}, {Galland}, {Udry}, \&
  {Mayor}}]{Desort-2007}
{Desort}, M., {Lagrange}, A.-M., {Galland}, F., {Udry}, S., \& {Mayor}, M.
  2007, \aap, 473, 983

\bibitem[{{Donati} {et~al.}(1997){Donati}, {Semel}, {Carter}, {Rees}, \&
  {Collier Cameron}}]{Donati:1997aa}
{Donati}, J.-F., {Semel}, M., {Carter}, B.~D., {Rees}, D.~E., \& {Collier
  Cameron}, A. 1997, \mnras, 291, 658

\bibitem[{{Dumusque}(2014)}]{Dumusque-2014c}
{Dumusque}, X. 2014, \apj, 796, 133

\bibitem[{{Dumusque} {et~al.}(2014){Dumusque}, {Boisse}, \&
  {Santos}}]{Dumusque-2014b}
{Dumusque}, X., {Boisse}, I., \& {Santos}, N.~C. 2014, \apj, 796, 132

\bibitem[{{Dumusque} {et~al.}(2011{\natexlab{a}}){Dumusque}, {Lovis},
  {S{\'e}gransan}, {Mayor}, {Udry}, {Benz}, {Bouchy}, {Lo Curto}, {Mordasini},
  {Pepe}, {Queloz}, {Santos}, \& {Naef}}]{Dumusque-2011c}
{Dumusque}, X., {Lovis}, C., {S{\'e}gransan}, D., {et~al.} 2011{\natexlab{a}},
  \aap, 535, A55

\bibitem[{{Dumusque} {et~al.}(2015){Dumusque}, {Pepe}, {Lovis}, \&
  {Latham}}]{Dumusque-2015a}
{Dumusque}, X., {Pepe}, F., {Lovis}, C., \& {Latham}, D.~W. 2015, \apj, 808,
  171

\bibitem[{{Dumusque} {et~al.}(2012){Dumusque}, {Pepe}, {Lovis}, {Segransan},
  {Sahlmann}, {Benz}, {Bouchy}, {Mayor}, {Queloz}, {Santos}, \&
  {Udry}}]{Dumusque-2012}
{Dumusque}, X., {Pepe}, F., {Lovis}, C., {et~al.} 2012, \nat, 491, 207

\bibitem[{{Dumusque} {et~al.}(2011{\natexlab{b}}){Dumusque}, {Santos}, {Udry},
  {Lovis}, \& {Bonfils}}]{Dumusque-2011b}
{Dumusque}, X., {Santos}, N.~C., {Udry}, S., {Lovis}, C., \& {Bonfils}, X.
  2011{\natexlab{b}}, \aap, 527, A82

\bibitem[{{Dumusque} {et~al.}(2011{\natexlab{c}}){Dumusque}, {Udry}, {Lovis},
  {Santos}, \& {Monteiro}}]{Dumusque-2011a}
{Dumusque}, X., {Udry}, S., {Lovis}, C., {Santos}, N.~C., \& {Monteiro},
  M.~J.~P.~F.~G. 2011{\natexlab{c}}, \aap, 525, A140

\bibitem[{{Feng} {et~al.}(2017){Feng}, {Tuomi}, {Jones}, {Barnes},
  {Anglada-Escud{\'e}}, {Vogt}, \& {Butler}}]{Feng:2017ac}
{Feng}, F., {Tuomi}, M., {Jones}, H.~R.~A., {et~al.} 2017, \aj, 154, 135

\bibitem[{{Gray}(2009)}]{Gray-2009}
{Gray}, D.~F. 2009, \apj, 697, 1032

\bibitem[{{Haywood} {et~al.}(2014){Haywood}, {Collier Cameron}, {Queloz},
  {Barros}, {Deleuil}, {Fares}, {Gillon}, {Lanza}, {Lovis}, {Moutou}, {Pepe},
  {Pollacco}, {Santerne}, {S{\'e}gransan}, \& {Unruh}}]{Haywood-2014}
{Haywood}, R.~D., {Collier Cameron}, A., {Queloz}, D., {et~al.} 2014, \mnras,
  443, 2517

\bibitem[{{Jones} {et~al.}(2017){Jones}, {Stenning}, {Ford}, {Wolpert},
  {Loredo}, \& {Dumusque}}]{Jones:2017aa}
{Jones}, D.~E., {Stenning}, D.~C., {Ford}, E.~B., {et~al.} 2017, ArXiv e-prints
  [\eprint[arXiv]{1711.01318}]

\bibitem[{{Kjeldsen} \& {Bedding}(1995)}]{Kjeldsen-1995}
{Kjeldsen}, H. \& {Bedding}, T.~R. 1995, \aap, 293, 87

\bibitem[{{Lefebvre} {et~al.}(2008){Lefebvre}, {Garc{\'{\i}}a},
  {Jim{\'e}nez-Reyes}, {Turck-Chi{\`e}ze}, \& {Mathur}}]{Lefebvre-2008}
{Lefebvre}, S., {Garc{\'{\i}}a}, R.~A., {Jim{\'e}nez-Reyes}, S.~J.,
  {Turck-Chi{\`e}ze}, S., \& {Mathur}, S. 2008, \aap, 490, 1143

\bibitem[{{Lindegren} \& {Dravins}(2003)}]{Lindegren-2003}
{Lindegren}, L. \& {Dravins}, D. 2003, \aap, 401, 1185

\bibitem[{{Lovis} {et~al.}(2011{\natexlab{a}}){Lovis}, {Dumusque}, {Santos},
  {Bouchy}, {Mayor}, {Pepe}, {Queloz}, {S{\'e}gransan}, \&
  {Udry}}]{Lovis-2011b}
{Lovis}, C., {Dumusque}, X., {Santos}, N.~C., {et~al.} 2011{\natexlab{a}},
  ArXiv e-prints [\eprint[arXiv]{1107.5325}]

\bibitem[{{Lovis} {et~al.}(2006){Lovis}, {Mayor}, {Pepe}, {Alibert}, {Benz},
  {Bouchy}, {Correia}, {Laskar}, {Mordasini}, {Queloz}, {Santos}, {Udry},
  {Bertaux}, \& {Sivan}}]{Lovis-2006}
{Lovis}, C., {Mayor}, M., {Pepe}, F., {et~al.} 2006, Nature, 441, 305

\bibitem[{{Lovis} {et~al.}(2011{\natexlab{b}}){Lovis}, {S{\'e}gransan},
  {Mayor}, {Udry}, {Benz}, {Bertaux}, {Bouchy}, {Correia}, {Laskar}, {Lo
  Curto}, {Mordasini}, {Pepe}, {Queloz}, \& {Santos}}]{Lovis-2011a}
{Lovis}, C., {S{\'e}gransan}, D., {Mayor}, M., {et~al.} 2011{\natexlab{b}},
  \aap, 528, A112

\bibitem[{{Mayor} {et~al.}(2003){Mayor}, {Pepe}, {Queloz}, {Bouchy},
  {Rupprecht}, {Lo Curto}, {Avila}, {Benz}, {Bertaux}, {Bonfils}, {Dall},
  {Dekker}, {Delabre}, {Eckert}, {Fleury}, {Gilliotte}, {Gojak}, {Guzman},
  {Kohler}, {Lizon}, {Longinotti}, {Lovis}, {Megevand}, {Pasquini}, {Reyes},
  {Sivan}, {Sosnowska}, {Soto}, {Udry}, {van Kesteren}, {Weber}, \&
  {Weilenmann}}]{Mayor-2003}
{Mayor}, M., {Pepe}, F., {Queloz}, D., {et~al.} 2003, The Messenger, 114, 20

\bibitem[{{Mayor} \& {Queloz}(1995)}]{Mayor-1995}
{Mayor}, M. \& {Queloz}, D. 1995, Nature, 378, 355

\bibitem[{{Meunier} {et~al.}(2010){Meunier}, {Desort}, \&
  {Lagrange}}]{Meunier-2010a}
{Meunier}, N., {Desort}, M., \& {Lagrange}, A.-M. 2010, \aap, 512, A39

\bibitem[{{Meunier} {et~al.}(2016){Meunier}, {Lagrange}, {Mbemba Kabuiku},
  {Alex}, {Mignon}, \& {Borgniet}}]{Meunier-2016}
{Meunier}, N., {Lagrange}, A.-M., {Mbemba Kabuiku}, L., {et~al.} 2016, ArXiv
  e-prints [\eprint[arXiv]{1610.02168}]

\bibitem[{{Noyes} {et~al.}(1984){Noyes}, {Hartmann}, {Baliunas}, {Duncan}, \&
  {Vaughan}}]{Noyes-1984}
{Noyes}, R.~W., {Hartmann}, L.~W., {Baliunas}, S.~L., {Duncan}, D.~K., \&
  {Vaughan}, A.~H. 1984, \apj, 279, 763

\bibitem[{{Pepe} {et~al.}(2011){Pepe}, {Lovis}, {S{\'e}gransan}, {Benz},
  {Bouchy}, {Dumusque}, {Mayor}, {Queloz}, {Santos}, \& {Udry}}]{Pepe-2011}
{Pepe}, F., {Lovis}, C., {S{\'e}gransan}, D., {et~al.} 2011, \aap, 534, A58

\bibitem[{Pepe {et~al.}(2002)Pepe, Mayor, Galland, Naef, Queloz, Santos, Udry,
  \& Burnet}]{Pepe-2002a}
Pepe, F., Mayor, M., Galland, F., {et~al.} 2002, \aap, 388, 632

\bibitem[{{Pepe} {et~al.}(2002){Pepe}, {Mayor}, {Rupprecht}, {Avila},
  {Ballester}, {Beckers}, {Benz}, {Bertaux}, {Bouchy}, {Buzzoni}, {Cavadore},
  {Deiries}, {Dekker}, {Delabre}, {D'Odorico}, {Eckert}, {Fischer}, {Fleury},
  {George}, {Gilliotte}, {Gojak}, {Guzman}, {Koch}, {Kohler}, {Kotzlowski},
  {Lacroix}, {Le Merrer}, {Lizon}, {Lo Curto}, {Longinotti}, {Megevand},
  {Pasquini}, {Petitpas}, {Pichard}, {Queloz}, {Reyes}, {Richaud}, {Sivan},
  {Sosnowska}, {Soto}, {Udry}, {Ureta}, {van Kesteren}, {Weber}, {Weilenmann},
  {Wicenec}, {Wieland}, {Christensen-Dalsgaard}, {Dravins}, {Hatzes},
  {K{\"u}rster}, {Paresce}, \& {Penny}}]{Pepe-2002}
{Pepe}, F., {Mayor}, M., {Rupprecht}, G., {et~al.} 2002, The Messenger, 110, 9

\bibitem[{{Pepe} {et~al.}(2014){Pepe}, {Molaro}, {Cristiani}, {Rebolo},
  {Santos}, {Dekker}, {M{\'e}gevand}, {Zerbi}, {Cabral}, {Di Marcantonio},
  {Abreu}, {Affolter}, {Aliverti}, {Allende Prieto}, {Amate}, {Avila},
  {Baldini}, {Bristow}, {Broeg}, {Cirami}, {Coelho}, {Conconi}, {Coretti},
  {Cupani}, {D'Odorico}, {De Caprio}, {Delabre}, {Dorn}, {Figueira}, {Fragoso},
  {Galeotta}, {Genolet}, {Gomes}, {Gonz{\'a}lez Hern{\'a}ndez}, {Hughes},
  {Iwert}, {Kerber}, {Landoni}, {Lizon}, {Lovis}, {Maire}, {Mannetta},
  {Martins}, {Monteiro}, {Oliveira}, {Poretti}, {Rasilla}, {Riva}, {Santana
  Tschudi}, {Santos}, {Sosnowska}, {Sousa}, {Span{\'o}}, {Tenegi}, {Toso},
  {Vanzella}, {Viel}, \& {Zapatero Osorio}}]{Pepe-2014}
{Pepe}, F., {Molaro}, P., {Cristiani}, S., {et~al.} 2014, ArXiv e-prints
  [\eprint[arXiv]{1401.5918}]

\bibitem[{{Queloz} {et~al.}(2009){Queloz}, {Bouchy}, {Moutou}, {Hatzes},
  {H{\'e}brard}, {Alonso}, {Auvergne}, {Baglin}, {Barbieri}, {Barge}, {Benz},
  {Bord{\'e}}, {Deeg}, {Deleuil}, {Dvorak}, {Erikson}, {Ferraz Mello},
  {Fridlund}, {Gandolfi}, {Gillon}, {Guenther}, {Guillot}, {Jorda}, {Hartmann},
  {Lammer}, {L{\'e}ger}, {Llebaria}, {Lovis}, {Magain}, {Mayor}, {Mazeh},
  {Ollivier}, {P{\"a}tzold}, {Pepe}, {Rauer}, {Rouan}, {Schneider},
  {Segransan}, {Udry}, \& {Wuchterl}}]{Queloz-2009}
{Queloz}, D., {Bouchy}, F., {Moutou}, C., {et~al.} 2009, \aap, 506, 303

\bibitem[{{Queloz} {et~al.}(2001){Queloz}, {Henry}, {Sivan}, {Baliunas},
  {Beuzit}, {Donahue}, {Mayor}, {Naef}, {Perrier}, \& {Udry}}]{Queloz-2001}
{Queloz}, D., {Henry}, G.~W., {Sivan}, J.~P., {et~al.} 2001, \aap, 379, 279

\bibitem[{{Rajpaul} {et~al.}(2015){Rajpaul}, {Aigrain}, {Osborne}, {Reece}, \&
  {Roberts}}]{Rajpaul-2015}
{Rajpaul}, V., {Aigrain}, S., {Osborne}, M.~A., {Reece}, S., \& {Roberts}, S.
  2015, \mnras, 452, 2269

\bibitem[{{Redman} {et~al.}(2014){Redman}, {Nave}, \&
  {Sansonetti}}]{Redman:2014aa}
{Redman}, S.~L., {Nave}, G., \& {Sansonetti}, C.~J. 2014, \apjs, 211, 4

\bibitem[{{Reiners} {et~al.}(2015){Reiners}, {Mrotzek}, {Lemke}, {Hinrichs}, \&
  {Reinsch}}]{Reiners-2015}
{Reiners}, A., {Mrotzek}, N., {Lemke}, U., {Hinrichs}, J., \& {Reinsch}, K.
  2015, ArXiv e-prints [\eprint[arXiv]{1511.03014}]

\bibitem[{{Robertson} {et~al.}(2014){Robertson}, {Mahadevan}, {Endl}, \&
  {Roy}}]{Robertson-2014}
{Robertson}, P., {Mahadevan}, S., {Endl}, M., \& {Roy}, A. 2014, Science, 345,
  440

\bibitem[{{Saar} \& {Donahue}(1997)}]{Saar-1997b}
{Saar}, S.~H. \& {Donahue}, R.~A. 1997, \apj, 485, 319

\bibitem[{{Santos} {et~al.}(2004){Santos}, {Bouchy}, {Mayor}, {Pepe}, {Queloz},
  {Udry}, {Lovis}, {Bazot}, {Benz}, {Bertaux}, {Lo Curto}, {Delfosse},
  {Mordasini}, {Naef}, {Sivan}, \& {Vauclair}}]{Santos-2004e}
{Santos}, N.~C., {Bouchy}, F., {Mayor}, M., {et~al.} 2004, \aap, 426, L19

\bibitem[{{Teixeira} {et~al.}(2009){Teixeira}, {Kjeldsen}, {Bedding}, {Bouchy},
  {Christensen-Dalsgaard}, {Cunha}, {Dall}, {Frandsen}, {Karoff}, {Monteiro},
  \& {Pijpers}}]{Teixeira-2009}
{Teixeira}, T.~C., {Kjeldsen}, H., {Bedding}, T.~R., {et~al.} 2009, \aap, 494,
  237

\bibitem[{{Thompson} {et~al.}(2017){Thompson}, {Watson}, {de Mooij}, \&
  {Jess}}]{Thompson-2017}
{Thompson}, A.~P.~G., {Watson}, C.~A., {de Mooij}, E.~J.~W., \& {Jess}, D.~B.
  2017, ArXiv e-prints [\eprint[arXiv]{1702.01647}]

\bibitem[{{Tuomi} {et~al.}(2013){Tuomi}, {Anglada-Escud{\'e}}, {Gerlach},
  {Jones}, {Reiners}, {Rivera}, {Vogt}, \& {Butler}}]{Tuomi-2013a}
{Tuomi}, M., {Anglada-Escud{\'e}}, G., {Gerlach}, E., {et~al.} 2013, \aap, 549,
  A48

\bibitem[{{Wise} {et~al.}(2018){Wise}, {Dodson-Robinson}, {Bevenour}, \&
  {Provini}}]{Wise:2018aa}
{Wise}, A.~W., {Dodson-Robinson}, S.~E., {Bevenour}, K., \& {Provini}, A. 2018,
  ArXiv e-prints [\eprint[arXiv]{1808.09009}]

\end{thebibliography}

\begin{appendix}

\section{Correcting for the HARPS instrumental drift during the night}
\label{instrumental_drift}

Before the beginning of each night, the HARPS science fibre (fibre A) is illuminated with a Th-Ar lamp to perform 
a wavelength solution and thus fix the zero point of the instrument. At the same time as the Th-Ar spectrum is recorded on the science fibre, the spectrum of another lamp, 
either Th-Ar or FP \'etalon, is recorded on the HARPS reference fibre (fibre B). During the night, this second lamp continues to illuminate the reference fibre, while stars are observed on the science fibre. This allows to probe the instrumental drift within a night by measuring the drift between this simultaneous reference spectrum and the calibration made before the night. In the HARPS DRS, 
this drift is measured using the technique described in Sec.~\ref{subsec:spectral_line_RVs} comparing each order of the Th-Ar or FP simultaneous spectra with the respective order
in the Th-Ar or FP calibration. At the end, a weighted average is performed on all orders to get a measure of the instrumental RV drift for each observation. Here, for consistency 
with the technique used to derive the RV on each individual line, we build a Th-Ar and/or FP master spectrum from all the simultaneous spectra taken during one night and then measure, on each emission line of the Th-Ar or FP simultaneous spectra, the RV drift relative to the Th-Ar or FP master, respectively. Once this is done, we can either perform a weighted average on all the spectral lines to get a single RV drift for each observation, like it is done in the HARPS DRS, or we can measure locally the drift on the HARPS detector.

\begin{figure*}[!h]
\center
\includegraphics[angle=0,width=0.80\textwidth,origin=br]{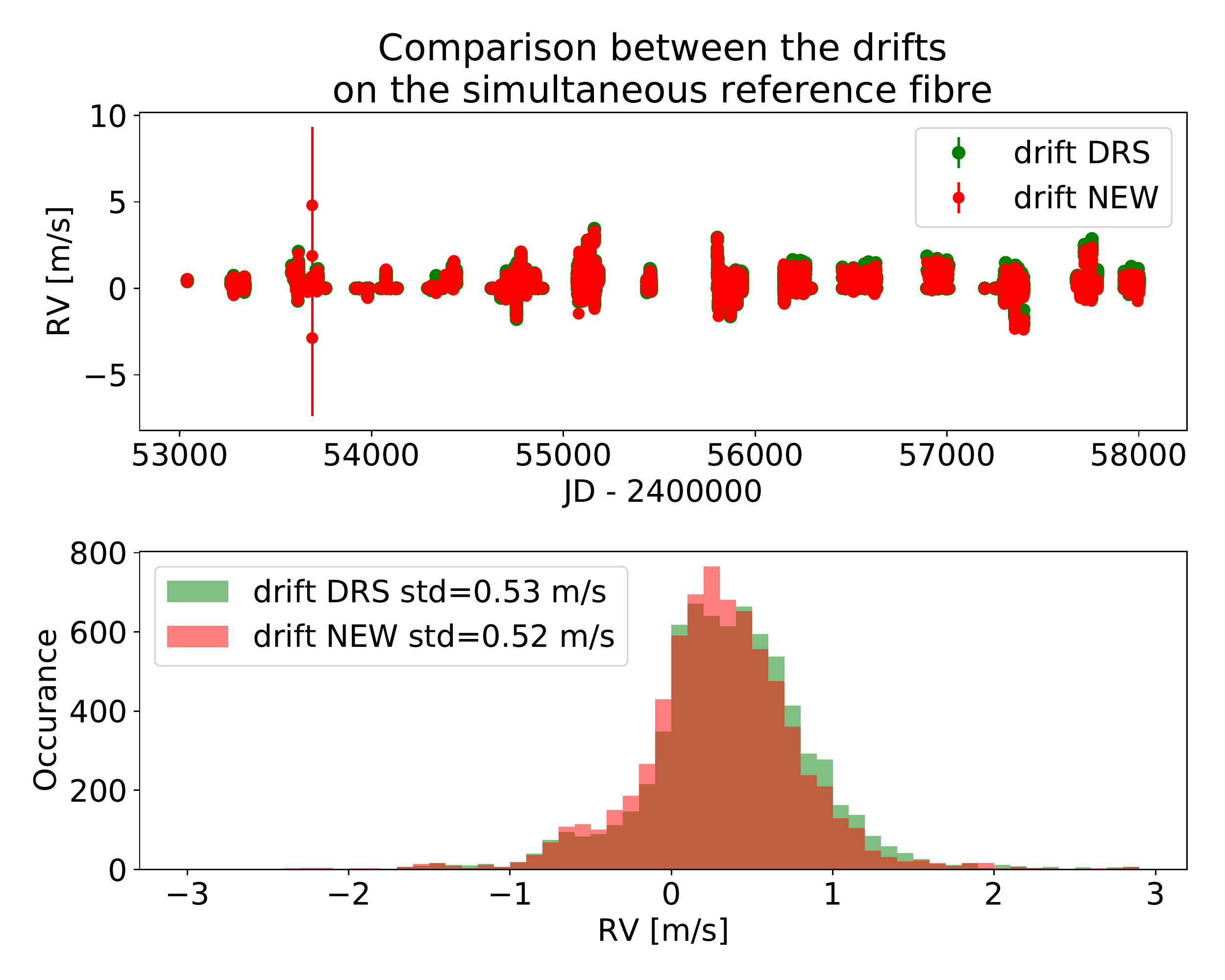}
 \caption[]{\emph{Top:} Comparison of the drift of the instrument measured on the simultaneous reference fibre illuminated either by the Th-Ar or the FP \'etalon lamp and estimated by our new RV extraction procedure (in red) and by the HARPS DRS (in green).
 \emph{Bottom:} Histogram of the drifts measured. The legend shows the standard deviation obtained for both drift data sets.}
\label{fig:drift_comparison_with_DRS}
\end{figure*}

In Fig.~\ref{fig:drift_comparison_with_DRS}, we compare the drifts estimated by the HARPS DRS and the drifts derived by our new RV extraction procedure using all the simultaneous spectra obtained for the $\tau$\,Ceti RV data set that we analyse in Sec.~\ref{subsec:RV_HD10700}. As we can see, our new RV extraction procedure gives extremely similar results than the HARPS DRS and for consistency we use those newly derived drifts to correct for the HARPS systematics when we estimate the RV of each individual spectral line.

\section{Signals between 100 and 1000 days in $\tau$\,Ceti RVs}
\label{Systematics in HD10700 RVs}

\begin{figure*}[!t]
\center
 \includegraphics[angle=0,width=0.60\textwidth,origin=br]{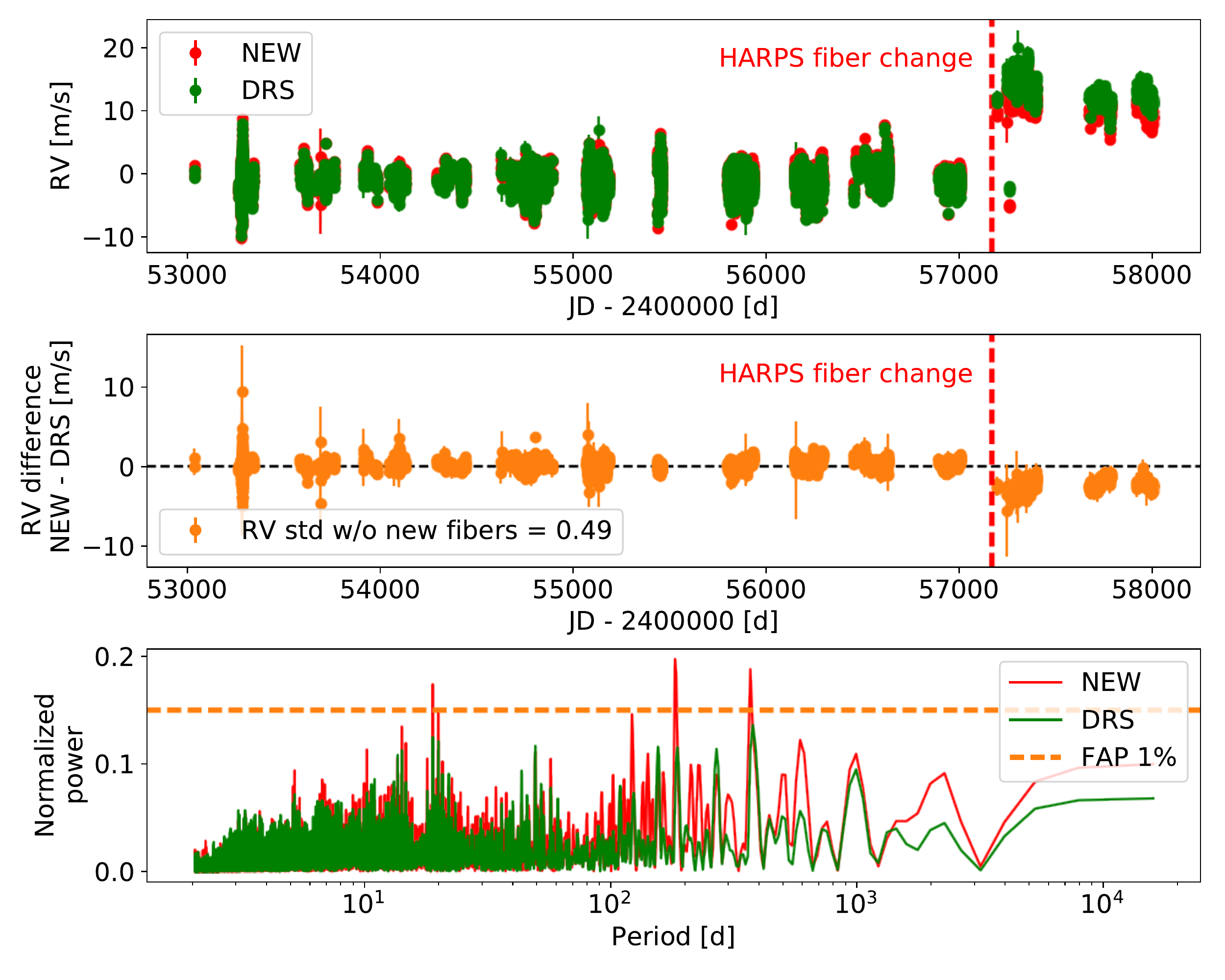}
 \includegraphics[angle=0,width=0.38\textwidth,origin=br]{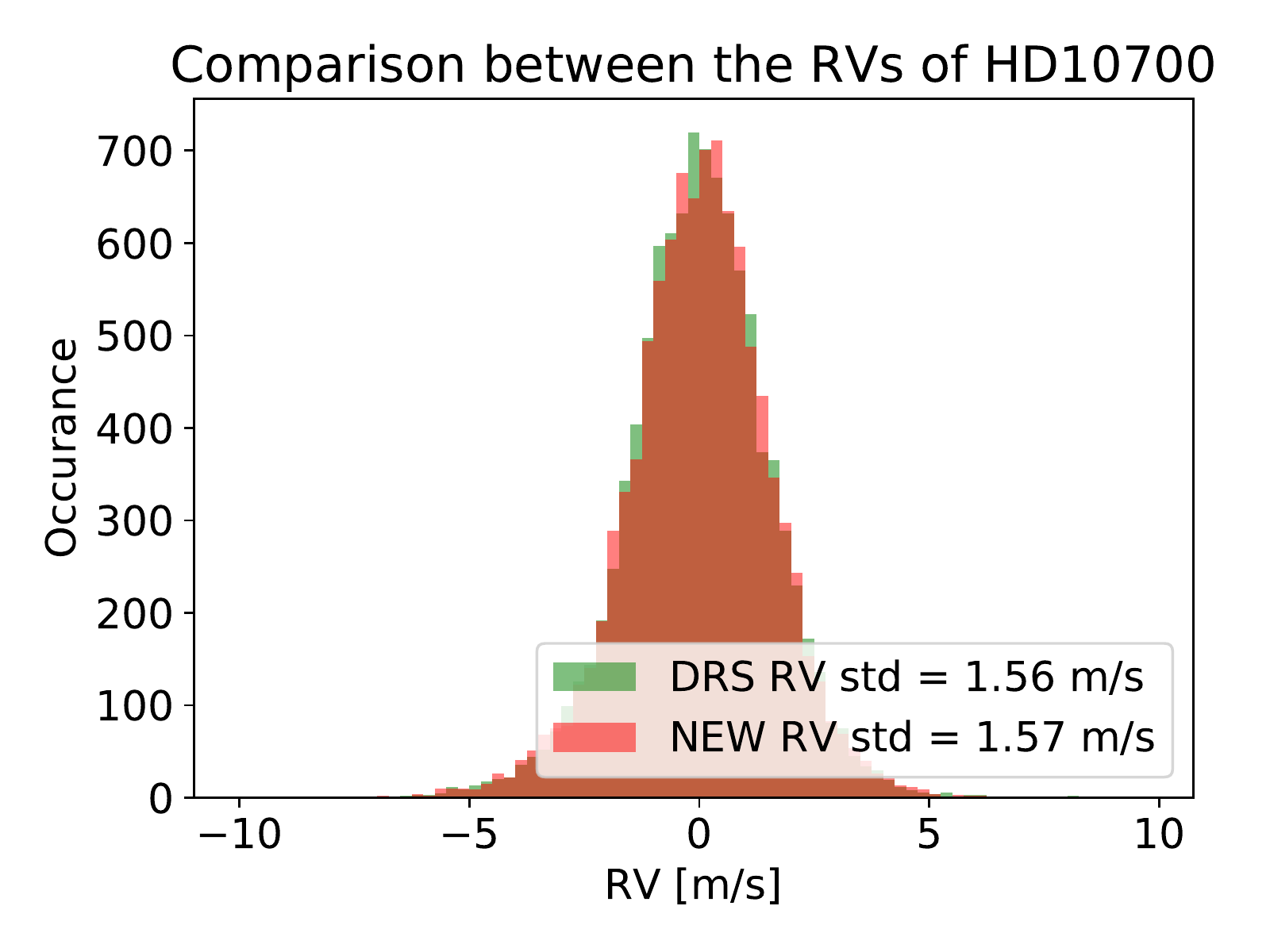}
 \caption[]{Same as Fig.~\ref{fig:HD10700_comparison_with_DRS} but for the RVs of $\tau$\,Ceti derived without considering spectral lines that falls close to CCD stitchings.}
\label{fig:HD10700_comparison_with_DRS_all_lines_stitching_removed}
\end{figure*}

As we can see in Fig.~\ref{fig:HD10700_comparison_with_DRS} of Sec.~\ref{subsec:RV_HD10700}, the RVs derived by our new RV extraction procedure seems to be more affected by spurious RV signals 
with periods between 100 and 1000 days. Our first guess was that our reduction was more sensitive to the systematics created by spectral lines falling close to CCD stitchings, as it was shown by
\citet{Dumusque-2015a} that those lines are responsible for inducing spurious RV signal with periods of one-year and its different harmonics. We therefore removed 1234 spectral lines that were too close to CCD stitchings
and recomputed, using our new RV extraction procedure and the HARPS DRS, the RVs for $\tau$\,Ceti. The results are shown in Fig.~\ref{fig:HD10700_comparison_with_DRS_all_lines_stitching_removed}. We see that some of the strong signals 
between 100 and 1000 days disappear in this new analysis, however, not all of them. The behaviour of spectral lines close to CCD stitchings can therefore not explain all the observed systematics.

\begin{figure*}[!t]
\center
 \includegraphics[angle=0,width=0.33\textwidth,origin=br]{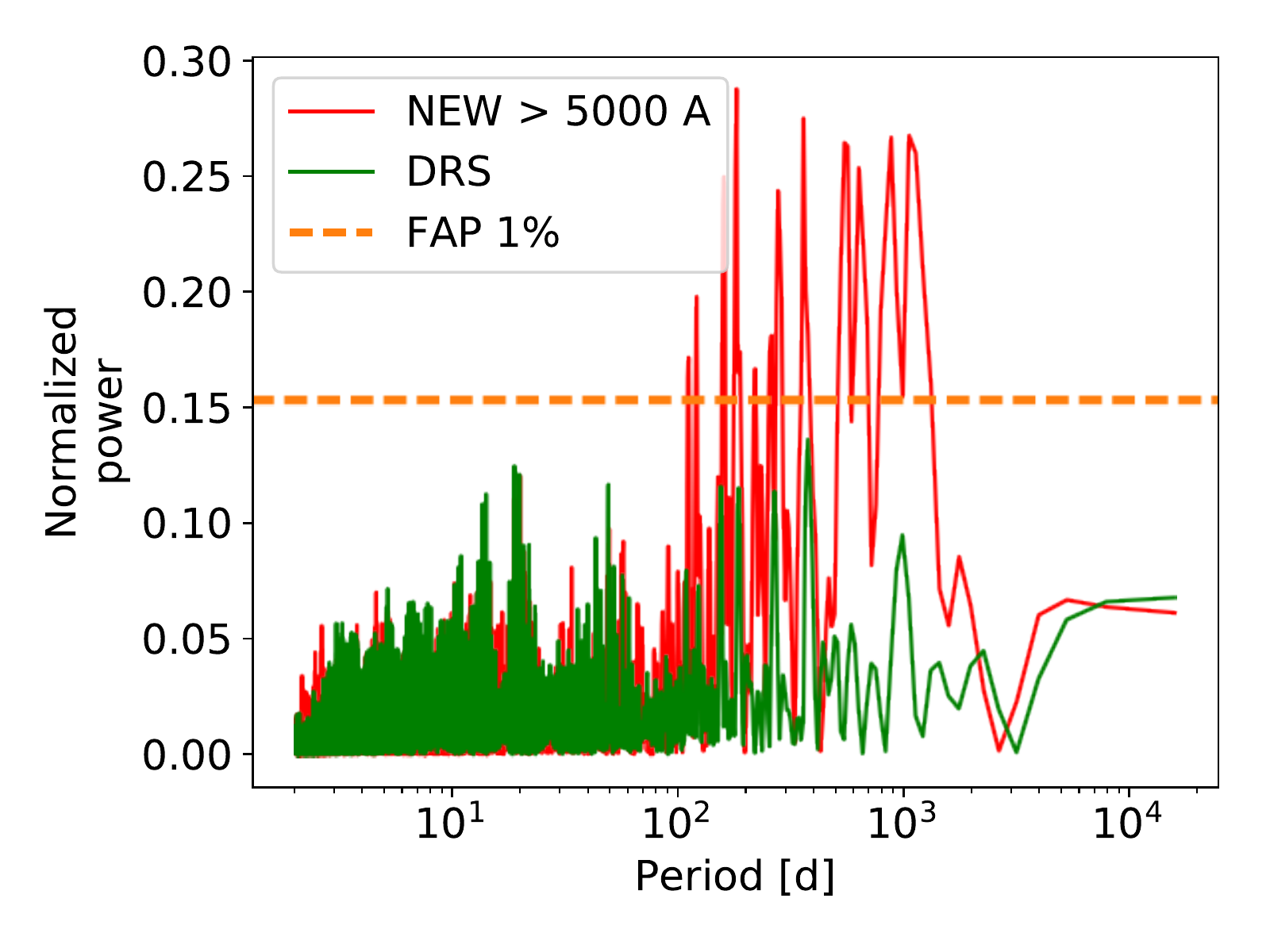}
 \includegraphics[angle=0,width=0.33\textwidth,origin=br]{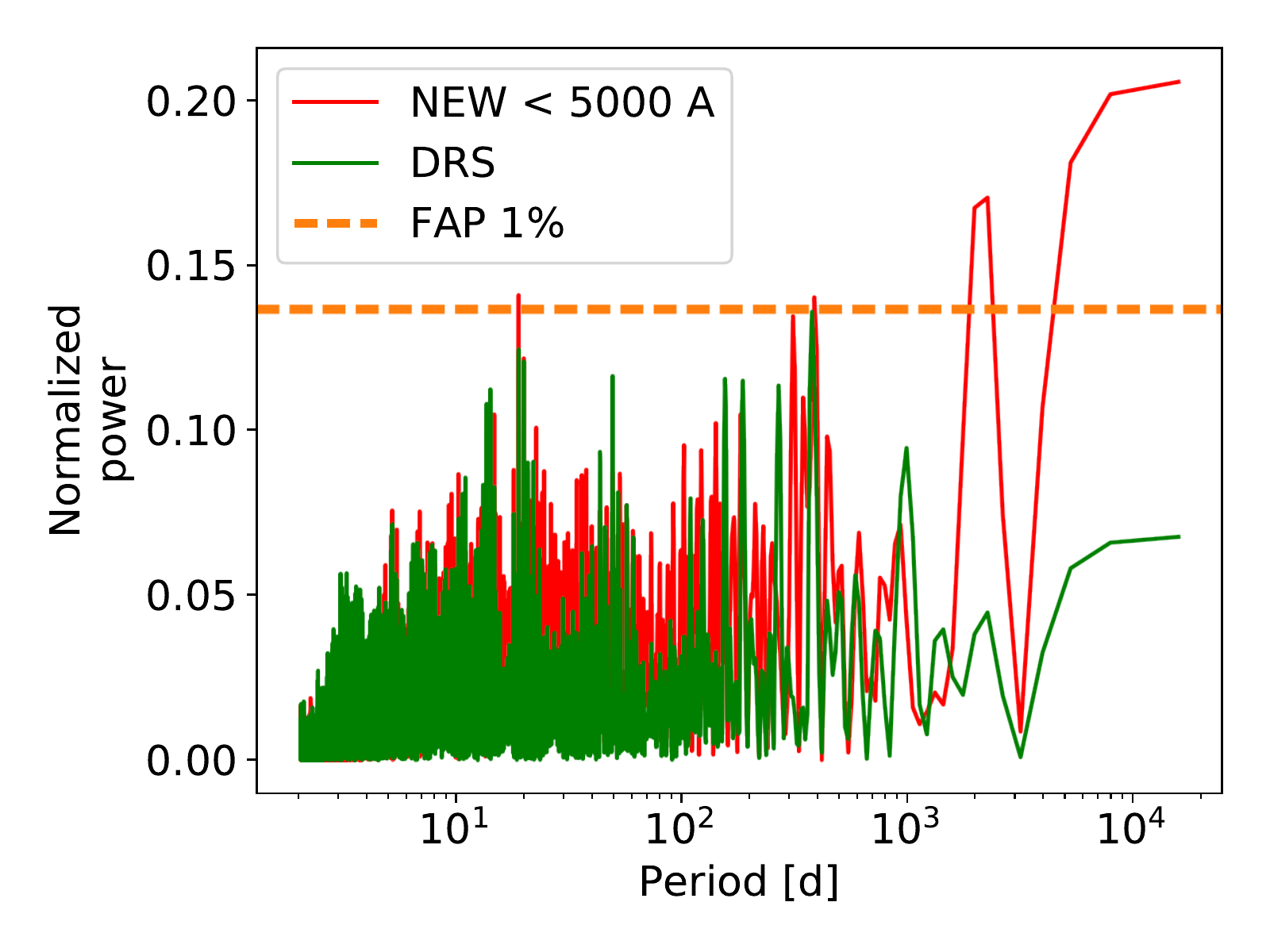}
 \includegraphics[angle=0,width=0.33\textwidth,origin=br]{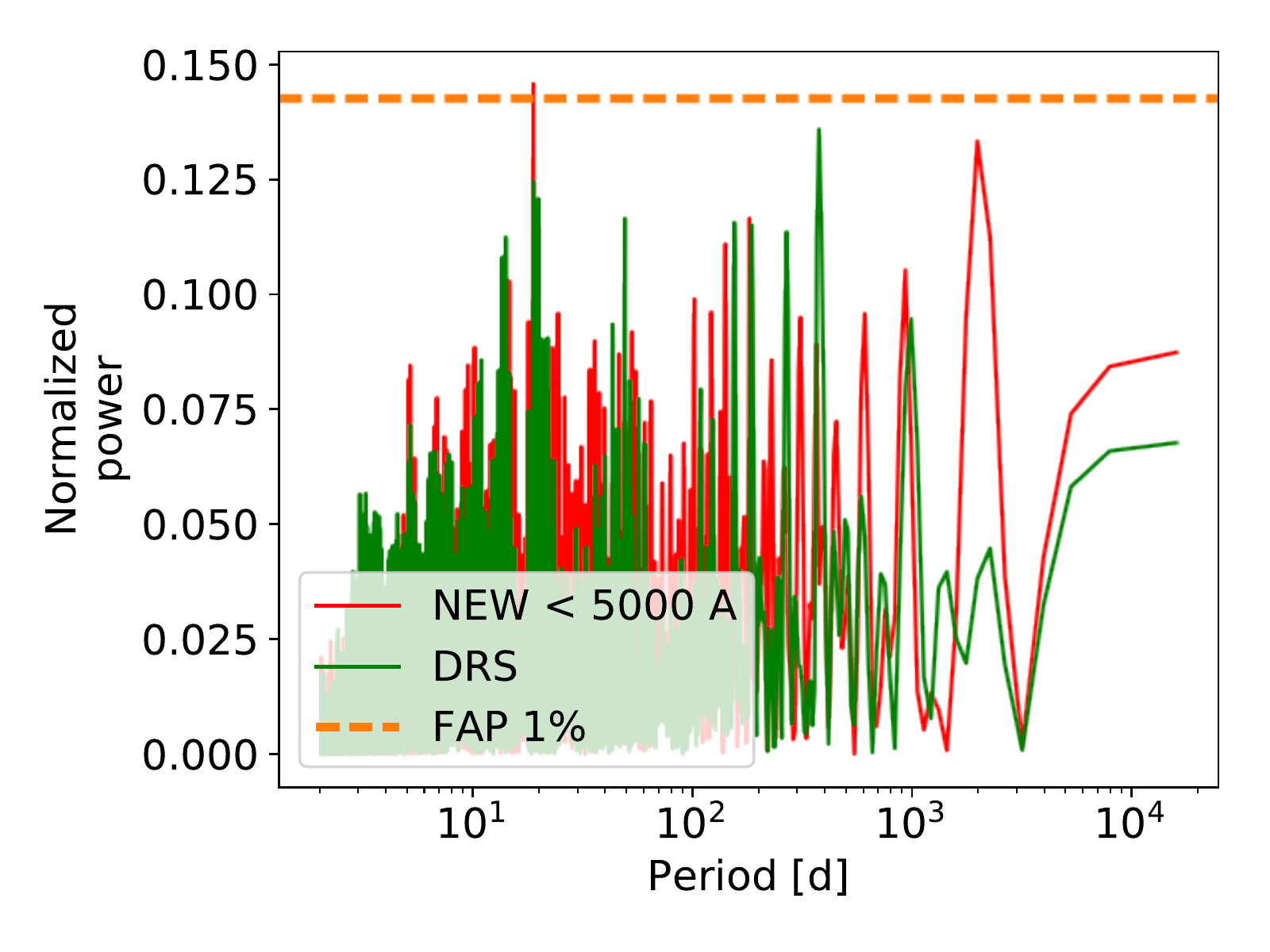}
 \caption[]{RVs of $\tau$\,Ceti derived only using the spectral lines with wavelength longer than 5000\,\AA\,(\emph{left}), shorter than 5000\,\AA\,(\emph{middle}) and shorter than 5000\,\AA\,
 plus a quadratic drift fitted to the residual RVs to remove low frequencies (\emph{right}). This quadratic drift corresponds to a small variation of the RVs of less than 1.5\,\ms\, over 11 years.
 It is clear that spectral lines with wavelengths longer than 5000\,\AA\,induce significant signals at periods between 100 and 1000 days. This is certainly due to telluric and micro-telluric lines
 affecting the red part of the spectrum.}
\label{fig:HD10700_comparison_subsample_lines}
\end{figure*}

Another phenomenon that can create signals with a one-year period is the one caused by tellurics. Tellurics do not move on the HARPS CCD, while stellar spectra are red and blueshifted over a year due to the revolution of Earth around the Sun, 
which modifies the RV of the Earth in the direction of the observed star. Tellurics can therefore induce signals with periods of a year and its different harmonics, depending on the sampling
of the data. To investigate if tellurics could be at the origin of the yearly signal observed, we compared the RVs of all the lines with a wavelength smaller than 5000\,\AA\,with the RVs of all the lines at longer wavelengths.
We selected a cutoff at 5000\,\AA\, here as there is no significant tellurics below this wavelength. In Fig.~\ref{fig:HD10700_comparison_subsample_lines}, we can see that for the lines redder than 5000\,\AA, 
significant signals appear between 100 and 1000 days, while for the lines bluer than 5000\,\AA, no significant signal is observed, except a long-term trend. 
It seems therefore that the extra signals that we see in our new RV extraction procedure RVs at periods around 100 and 1000 days compared to the HARPS DRS RVs is induced by spectral lines redder than 5000\,\AA. Because this is where 
telluric contamination is strong, it is likely that those extra signals are induced by tellurics. This is however something that we 
will investigate with more precision in the future.

\section{Correcting night-to-night RV offsets due to variation in wavelength solutions}
\label{sec:correcting_night_to_night_offsets}

As explained in Sec.~\ref{sec:night_to_night_offsets}, a wavelength solution is derived every night to fix the zero velocity point of the HARPS spectrograph.
To derive a wavelength solution on a Th-Ar calibration, a 3rd order polynomial is fitted on the Th-Ar emission lines present in each order. Because
the flux is low on the edges of the HARPS orders due to the blaze, the polynomial will not be constrained at those locations
and the wavelength solution can vary strongly from night-to-night. 

\begin{figure*}[]
\center
 \includegraphics[angle=0,width=0.48\textwidth]{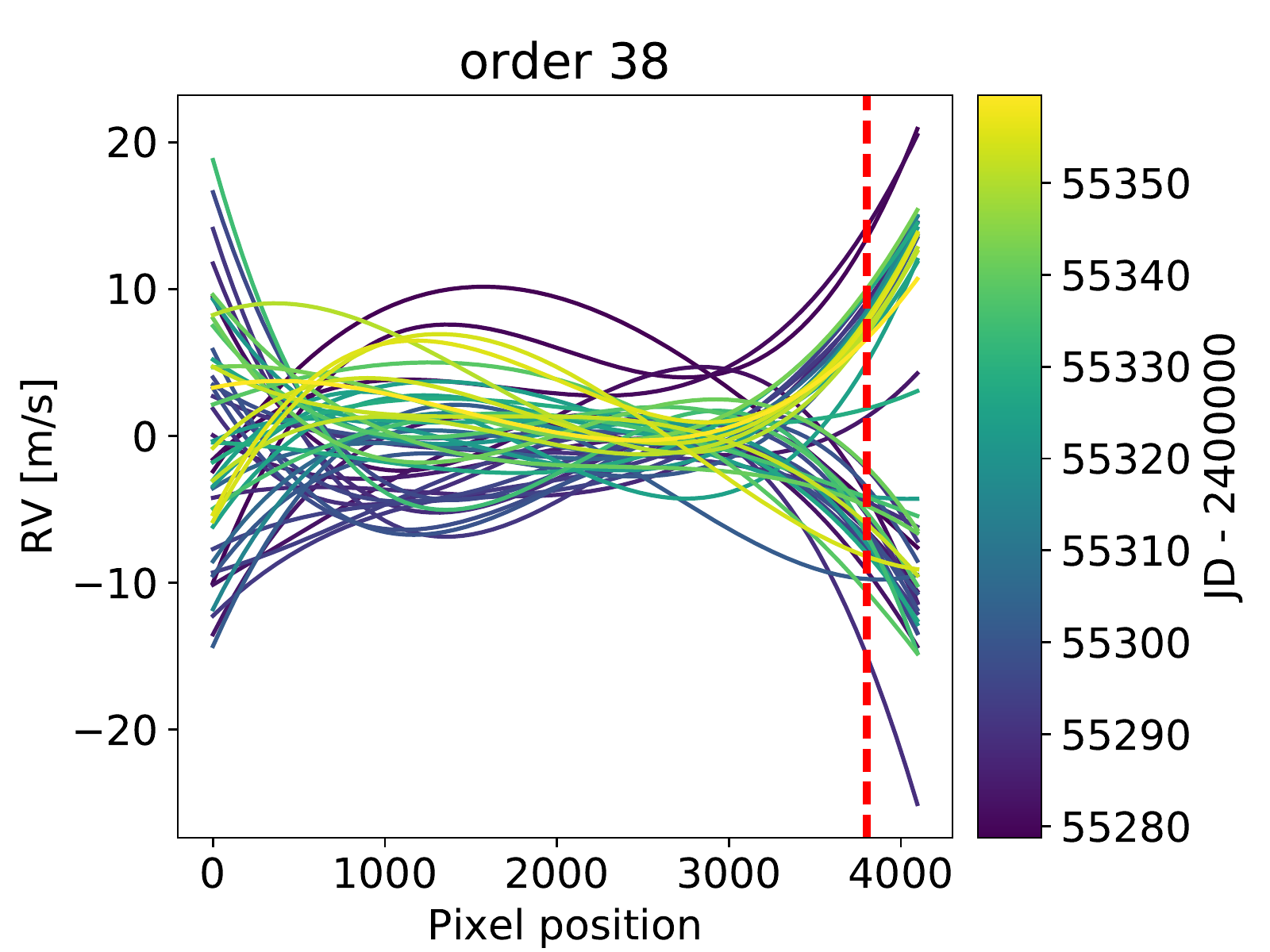}
  \includegraphics[angle=0,width=0.48\textwidth]{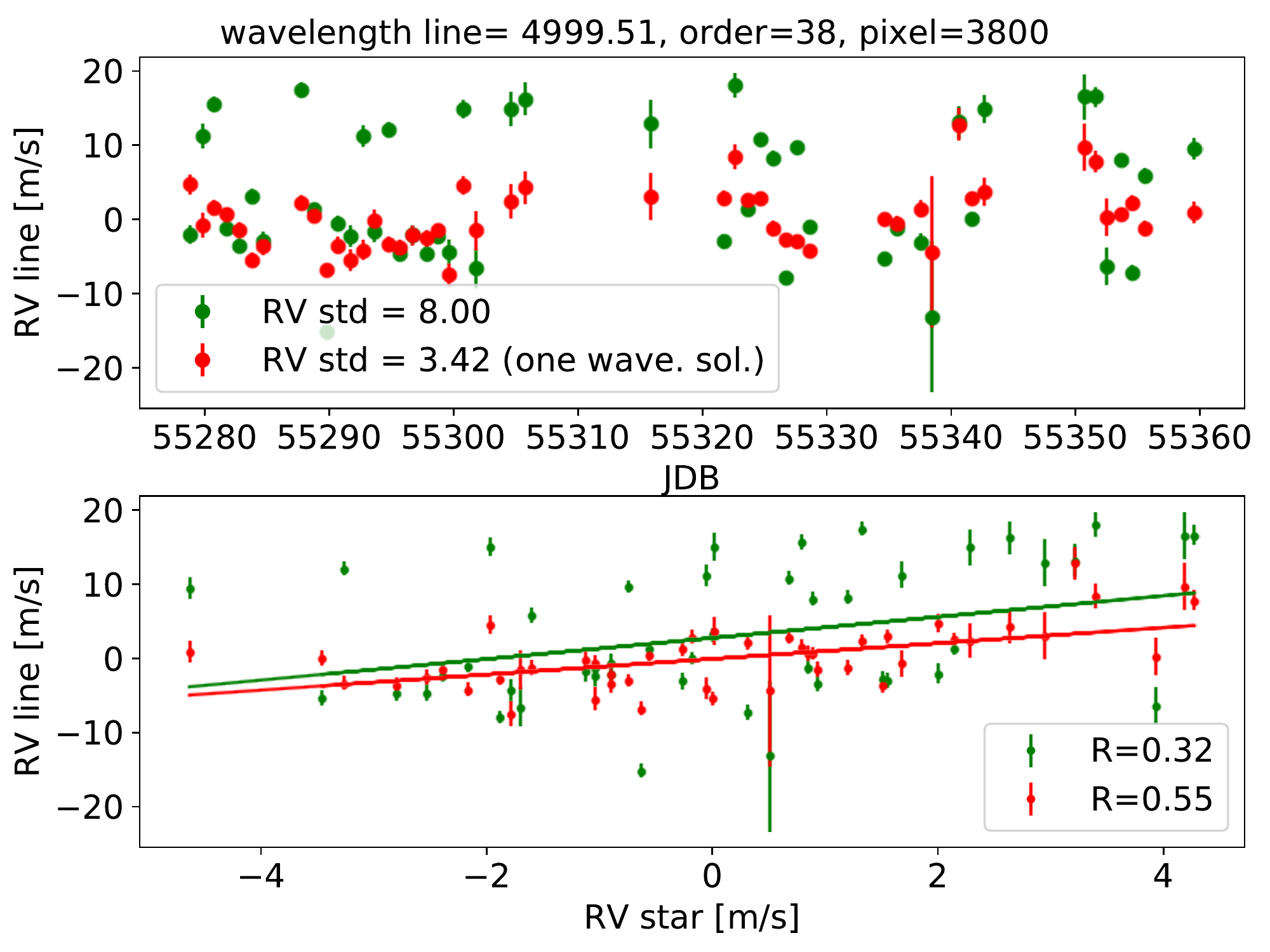}
 \caption[]{\emph{Left:} Wavelength solutions used for the 2010 data of $\alpha$\,Cen\,B, for spectral order 38. As we can see, those wavelength solutions are not stable on the edges of order 38.
 \emph{Top right:} RV measured on the 4999.51\,\AA\,Ti line that falls on pixel 3800 of order 38 (red vertical dashed line in the left pannel). In green, we show the RVs as measured when using the wavelength solutions shown on the left panel. In red we show the RVs derived when using a unique wavelength solution. As we can see, the non-stability of the wavelength solutions on the right edge of spectral order 38 induces night-to-night RV offsets of $\sim$15\,\ms, which is mitigated when using a unique wavelength solution. \emph{Bottom right:} Correlation between the RV of the 4999.51\,\AA\,Ti line and the RV derived using all the spectral lines. As we can see, the correlation is significantly stronger when we use the RVs derived using a unique wavelength solution.}
\label{fig:compa_wavesol}
\end{figure*}

To show the non-stability of the wavelength solution on the edges of the HARPS orders, we plot in the left panel of Fig.~\ref{fig:compa_wavesol} the differences in RV between consecutive wavelength solutions for HARPS spectral order 38 obtained in 2010.
For reasons that will become clear in Sec.~\ref{sec:mitigating_activity}, we analysed only the wavelength solutions used for the RV data of $\alpha$\,Cen\,B in 2010.
As we can see, for pixel values larger than 3500, we observe two categories of wavelength solutions:
one category that presents negative RVs, and another one that exhibits positive RVs. The difference in RV between those two categories of wavelength solution for pixels values larger than 3500 is $\sim$15\,\ms.
With our new RV extraction procedure, we can study the behaviour of a spectral line falling in this region of the detector. We therefore analysed the variation of the 
4999.5\,\AA\,Ti spectral line, which is localised at pixel 3800 in order 38 (red vertical dashed line in the left panel of Fig.~\ref{fig:compa_wavesol}). 
In the top right plot of Fig.~\ref{fig:compa_wavesol}, we show in green the RVs of this spectral line as a function of time as measured with our new RV extraction procedure. Like for the wavelength solutions, we observe that the RVs can also
be classified in two categories: one with positives RVs and another one with negative RVs. The difference between those two categories being also $\sim$15\ms. 
Our new RV extraction procedure therefore allows us to see directly the impact of the non-stability of the wavelength solution on the RVs derived for the 4999.5\,\AA\,Ti spectral line.

To mitigate this observed night-to-night RV offsets induced by the non-stability of the wavelength solutions, a possible solution is to use in combinaison to the Th-Ar spectrum, the much denser spectrum of the FP \'etalon to stabilise the wavelength solution on the edges of the HARPS CCD \citep[][]{Bauer:2015aa}. This work will be soon implemented in the HARPS DRS (Cersullo et al. 2018, submittted to A\&A) and will be available to the community. Performing a wavelength solution every night enables to fix the zero velocity point of the instrument, which simplify greatly the data reduction as it is not necessary to estimate the drift of the instrument from night-to-night. If we want to avoid using a wavelength solution every night,
we have to measure the drift of the instrument from night-to-night. This can be done by building a master Th-Ar spectrum from all the Th-Ar calibrations, then applying the wavelength solution of the master to all Th-Ar spectra and measure the drift between them and the master. This is the method that we explore here and for simplicity, we will refer to this technique in the rest of the paper as the one using a unique wavelength solution.

This approach complicates somehow the data analysis. First, within the 15 years of the HARPS spectrograph, the Th-Ar lamp used for deriving wavelength solutions was changed five times, and thus we need to create a Th-Ar master spectrum for each
of these lamps. In addition, several interventions on the instrument modified the way the Th-Ar spectrum is recorded on the detector. On JD=24571745 (1st of June 2015), the fibres to guide the light from the ESO 3.6m telescope to HARPS where changed 
from circular to octagonal, to mitigate the RV effect induced by telescope guiding errors. This drastically changed the point spread function (PSF) of the spectrograph, which induced a RV offset of $\sim$2\kms (see Fig.~\ref{fig:DRIFT_A_per_region_all} in the Appendix).
Therefore, a new Th-Ar master spectrum needed to be created for analysing the data taken after the fibre change, even though the Th-Ar lamp did not change. On JD=2453399 (19th of January 2005), an intervention on the instrument modified the relative 
drift between the blue and red CCD of HARPS, which can be seen in Fig.~\ref{fig:drift_A_53400} in the Appendix. Besides a general offset and a relative difference between the blue and red detectors of HARPS, we can also see in Fig.~\ref{fig:DRIFT_A_per_region_all}
and Fig.~\ref{fig:drift_A_53400} that the RV offset induced by those interventions is larger on the left side than the right side of the CCD. All those local differences prevent us of measuring a single RV drift for the entire detector to correct for the drift of the spectrograph over time. To solve for this problem, we divided the HARPS CCD in 24 regions, 16 corresponding to the blue CCD (pixels 0 to 1024, 1024 to 2048, 2048 to 3072, 3072 to 4096 and orders 0 to 12, 12 to 24, 24 to 36 and 36 to 47) and 8 corresponding to the red CCD (same separation in pixels but for orders 47 to 60 and 60 to 72). We then measured the drift of the instrument in each of these regions by averaging the RV of all the lines falling in these regions. This allows to correct locally for the drift of the spectrograph over time.

By measuring the drift of the instrument using a unique wavelength solution, we strongly reduce the night-to-night RV offsets observed on spectral lines falling on the edge of the HARPS CCD. In the top right plot of Fig.~\ref{fig:compa_wavesol}, we show in red the RVs of the 4999.5\,\AA\,Ti spectral line derived with this new approach. As we can see, the night-to-night RV offsets are strongly mitigated, which is shown by the reduction of the RV standard deviation from 8 to 3.4\,\ms. On the bottom right plot, we see how the correlation between the RVs of the Ti line and the RVs of all the lines become stronger when we use a unique wavelength solution. This will be very important for the analysis performed in Sec.~\ref{sec:mitigating_activity}.

\begin{figure}[]
\center
 \includegraphics[angle=0,width=0.48\textwidth]{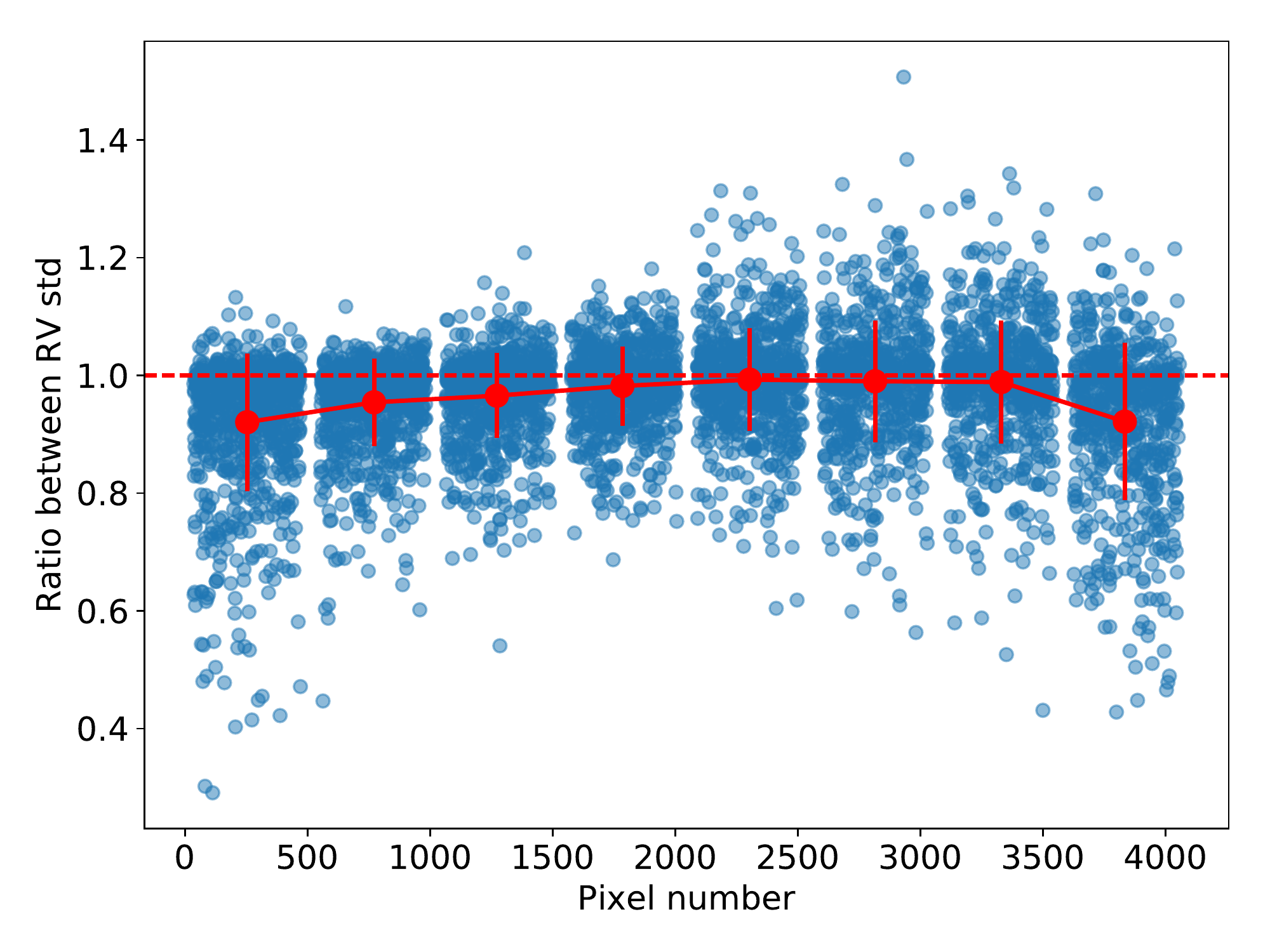}
 \caption[]{Ratio between the standard deviations of the RVs of each spectral line measured using one wavelength solution and using a wavelength solution every night. As we can see, on the edges of the HARPS CCD, i.e. for pixel smaller than 1000 and larger than 3000, our new RV extraction procedure using only one wavelength solution give RVs for which the standard deviation is smaller than when deriving the RVs using a wavelength solution every night.}
\label{fig:compa_wavesol_all_lines}
\end{figure}

To illustrate that using a unique wavelength solution helps in reducing the night-to-night RV offsets of all the lines falling on the edges of the HARPS CCD, we calculated the ratio between the standard deviations of the RVs measured using a unique wavelength solution and the RVs derived using a wavelength solution every night. The result for all the lines present in the 2010 spectra of $\alpha$\,Cen\,B is shown in Fig.~\ref{fig:compa_wavesol_all_lines}. As we can see, this ratio is on average always smaller than one, and the deviation from one becomes more and more important towards the edges of the HARPS CCD. This implies that by measuring the RV drift of the instrument using a unique wavelength solution, we can mitigate significantly the night-to-night RV offsets induced by the non-stability of the wavelength solution computed every night.

\begin{figure*}[!t]
\center
 \includegraphics[angle=0,width=0.60\textwidth,origin=br]{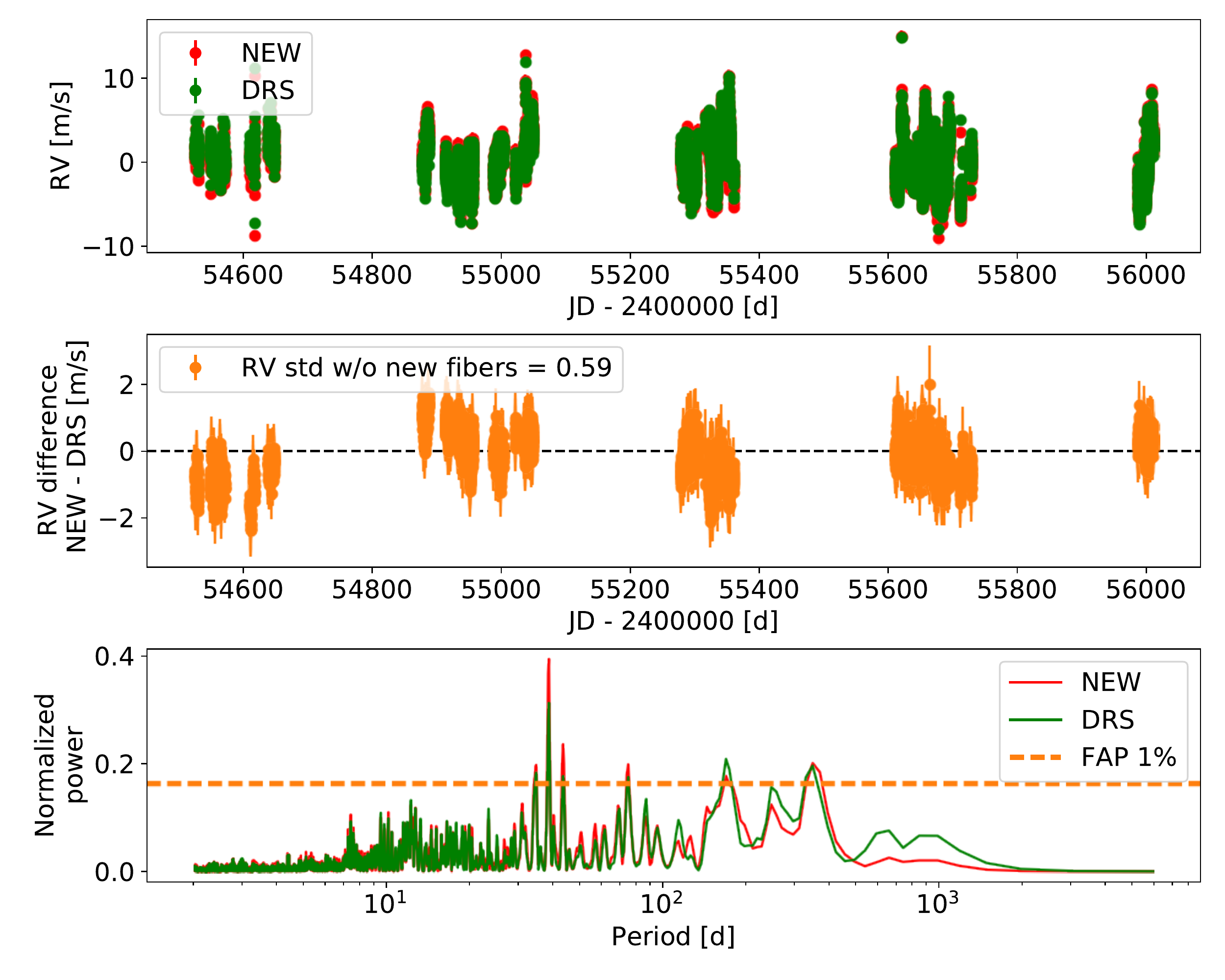}
 \includegraphics[angle=0,width=0.38\textwidth,origin=br]{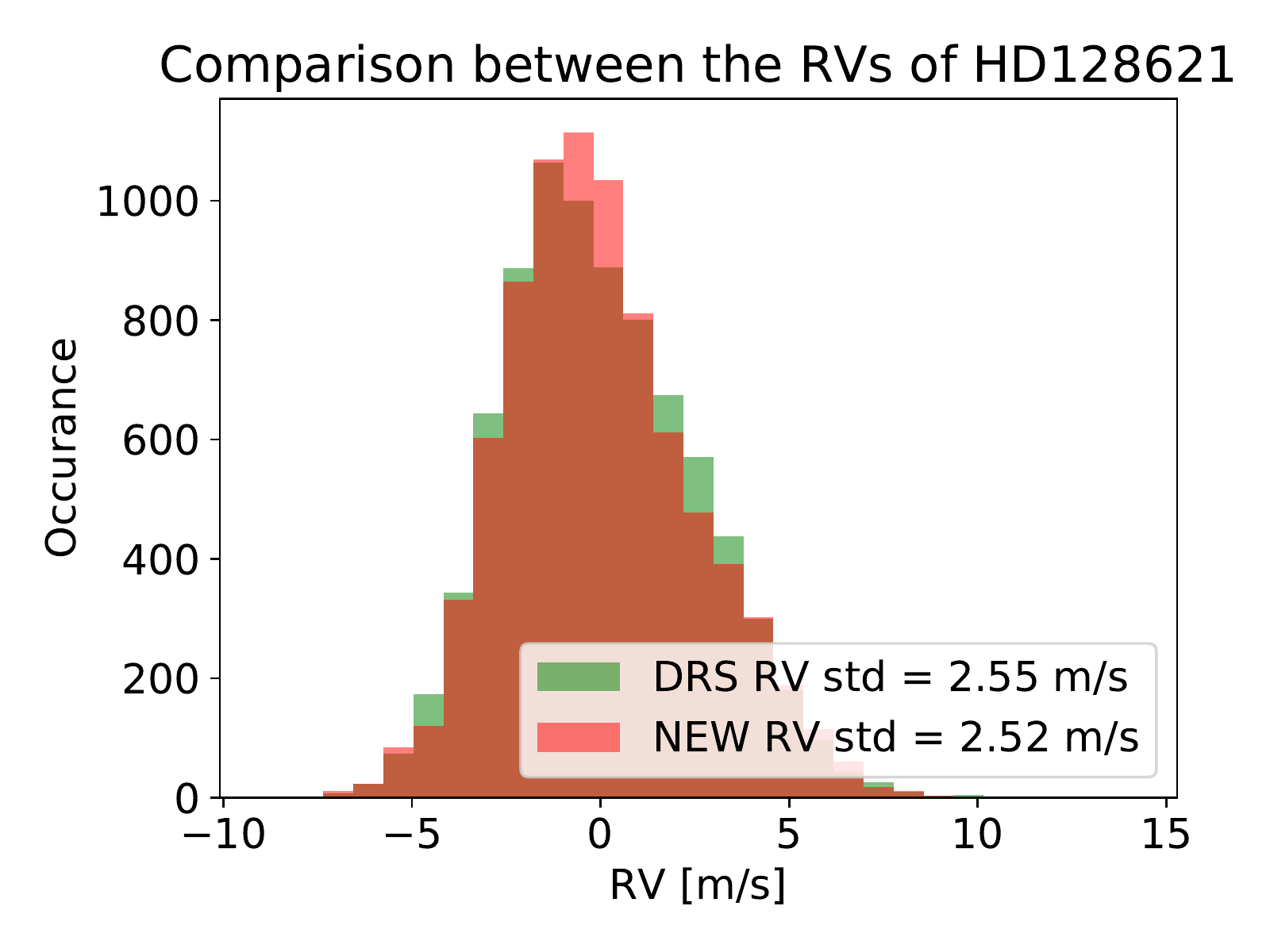}
\caption[]{Same as Fig.~\ref{fig:HD10700_comparison_with_DRS} but for the RVs of $\alpha$\,Cen\,B using only one wavelength solution.}
\label{fig:HD128621_comparison_with_DRS_one_wavesol}
\end{figure*}

\begin{figure}[]
\center
 \includegraphics[angle=0,width=0.48\textwidth]{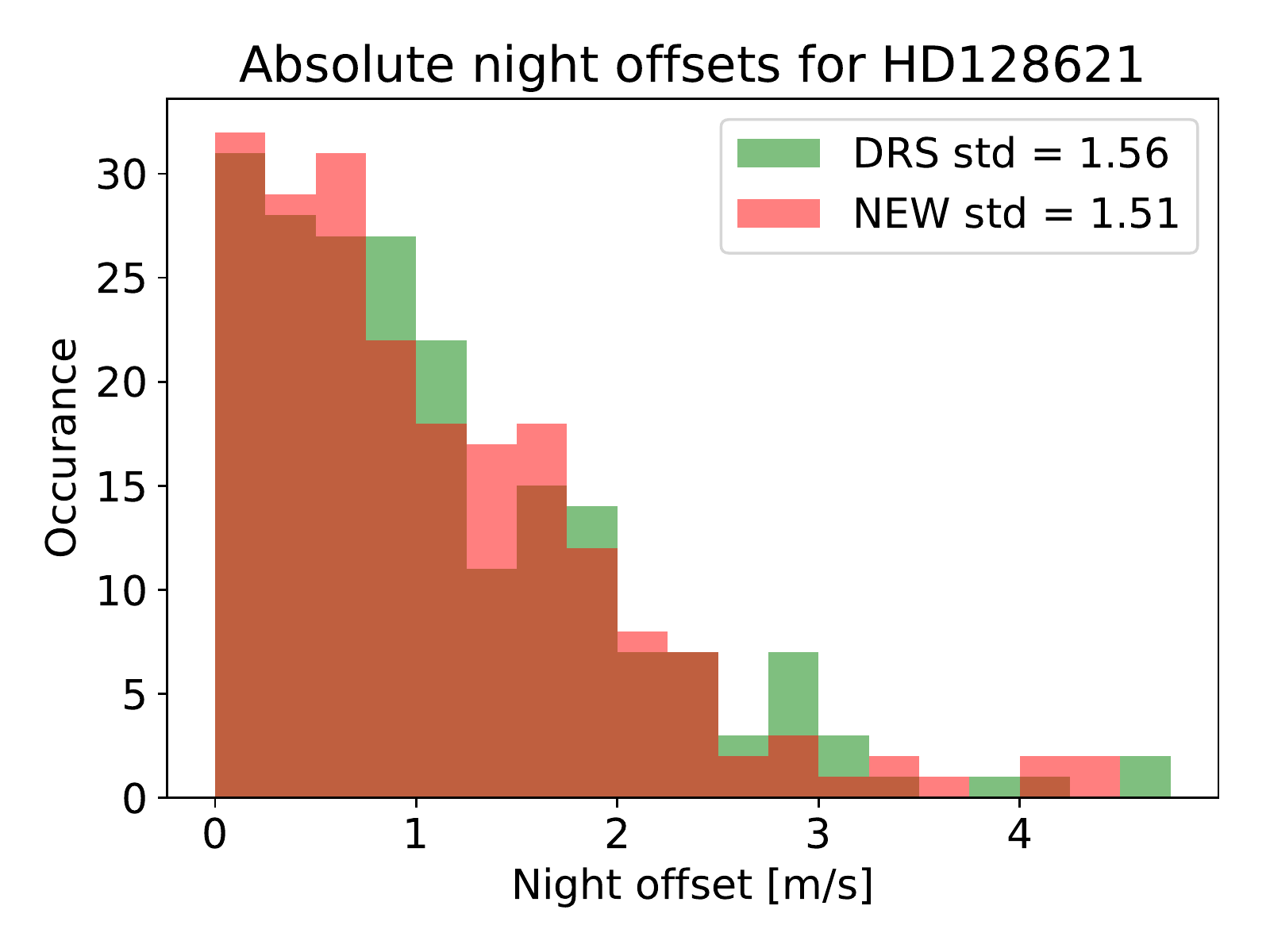}
 \caption[]{Histogram of the absolute night-to-night RV offsets measured on the 2008 to 2012 data of $\alpha$\,Cen\,B. In green, we represent the night-to-night offsets as measured in the RVs from the HARPS DRS, and in red we show the same offsets obtained from the RVs derived using a unique wavelength solution. As we can see, the RV standard distribution of the red histogram is smaller, proving that the night-to-night offsets are reduced by our analysis.}
\label{fig:HD128621_night_offsets}
\end{figure}

In this section, we show that measuring the instrument drift over time using a unique wavelength solution gives RVs that are less affected by night-to-night systematics compared to using a wavelength solution every night. In Fig. ~\ref{fig:HD128621_comparison_with_DRS_one_wavesol}, we compare the RVs of $\alpha$\,Cen\,B estimated from the HARPS DRS and derived with our new RV extraction procedure using a unique wavelength solution. As we can see, those two sets of RVs are extremely similar, with a standard deviation of the RV differences of 0.59\,\ms. Looking at the periodogram of those RVs, we see that the activity signal seen at the stellar rotation period, near 40 days, is stronger in our new RV extraction procedure. This comes from the fact that by mitigating the night-to-night RV offsets using a unique wavelength solution, the activity signal becomes better characterised. The mitigation of this systematics can also be seen in Fig.~\ref{fig:HD128621_night_offsets}, were we plot the histogram of the night-to-night RV offsets as measured on the RV data of $\alpha$\,Cen\,B. As we can see, the standard deviation of the absolute night-to-night RV offsets is smaller when using a unique wavelength solution. Noises adds up quadratically, therefore this analysis shows that for the RV data of $\alpha$\,Cen\,B, the noise induced by the non-stability of the wavelength solutions is 0.4\,\ms ($=\sqrt{1.56^2-1.51^2}$).

In the top plot of Fig. ~\ref{fig:HD128621_comparison_with_DRS_one_wavesol}, we corrected the RVs from the long-term trend induced by the presence of $\alpha$\,Cen\,A using a second order polynomial. The final RVs using a unique wavelength solution are extremely similar to the HARPS DRS RVs after this drift correction, however, the fitted drift is not the same. As this drift is induced by $\alpha$\,Cen\,A, it should be the same. Therefore, the only explanation for this difference is that deriving the RVs using a unique wavelength solution introduce a long-term drift in the RVs. This extra drift is due to the change in focus of the HARPS spectrograph with time. We are able to measure this effect in Sec.~\ref{HARPS_systematics} of the Appendix, and correct for it.

As a note of caution, if a FP \'etalon is available at a RV facility, we strongly encourage the use of a wavelength solution every night if the FP spectrum can be used in combination with the Th-AR spectrum to derive a stable wavelength solution \citep[][Cersullo et al. 2018, submittted to A\&A]{Bauer:2015aa}. We demonstrated here that using a single wavelength solution works as well, however it complicates quite significantly the data analysis process.

\section{HARPS systematics}
\label{HARPS_systematics}

\begin{figure*}[]
\center
 \includegraphics[angle=0,width=0.98\textwidth]{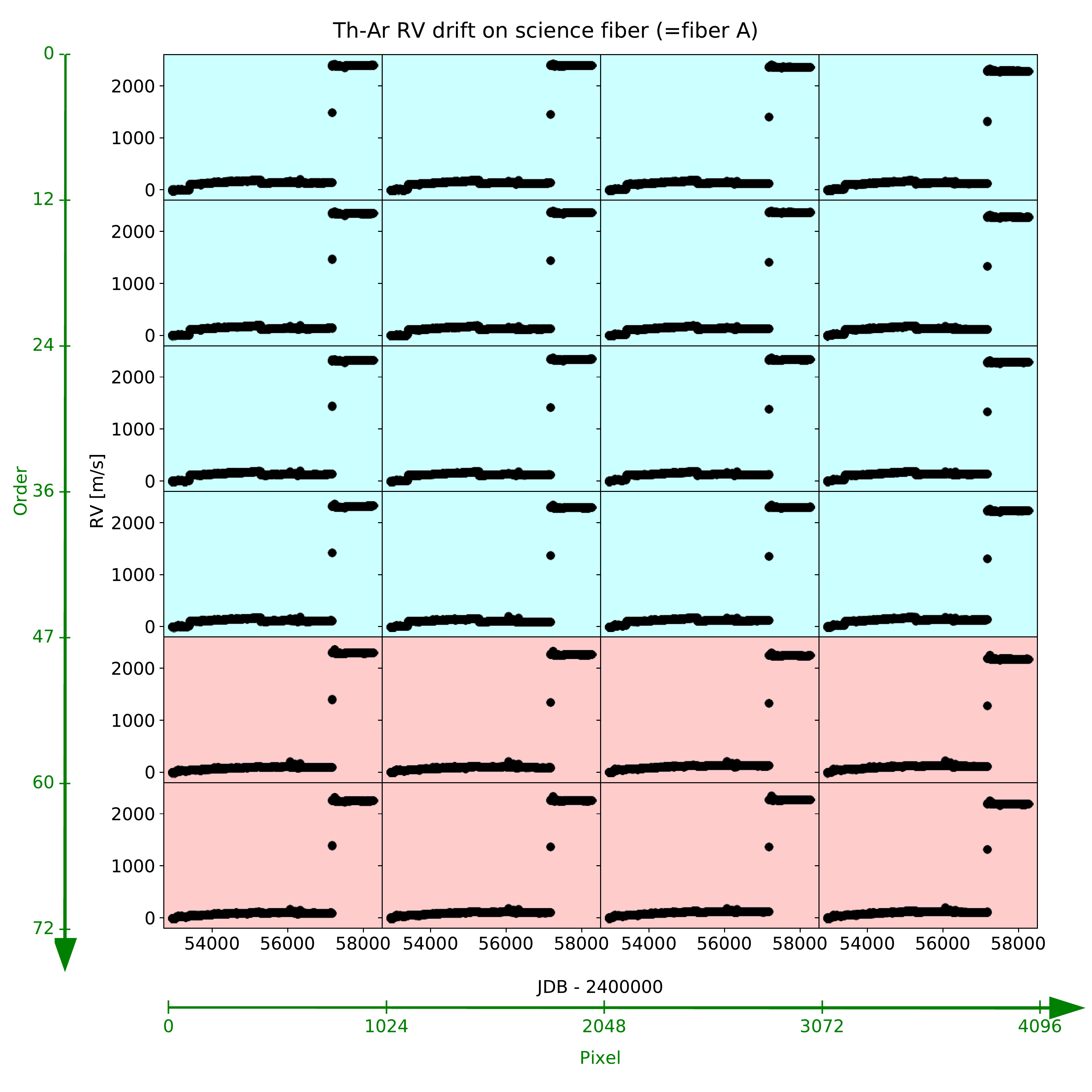}
 \caption[]{Drift of HARPS as a function of time measured on the Th-Ar spectra taken each night to derive the wavelength solution of the detector. Each subplot represents the drift in RVs measured on the Th-Ar emission lines falling in different regions of the HARPS detector.
 The HARPS detector is cut into 24 regions, 16 corresponding to the blue CCD shown with a blue background (pixel 0 to 1024, 1024 to 2048, 2048 to 3072, 3072 to 4096 and orders 0 to 12, 12 to 24, 24 to 36 and 36 to 47) and 8 corresponding to the red CCD shown with a red background (same separation in pixel but for order 47 to 60 and 60 to 72). The main feature that we see here is the huge RV offset on JD=2457174 (1st of June 2015), which is induced by the change of the HARPS fibres from circular to octagonal.}
\label{fig:DRIFT_A_per_region_all}
\end{figure*}

By measuring with our new RV extraction procedure the RVs on each emission spectral line in the Th-Ar spectra used to perform the wavelength solution every night, we can
study the drift of the HARPS spectrograph over time. In Fig.~\ref{fig:DRIFT_A_per_region_all}, we show the drift observed when the fibre from HARPS were changed
from circular to octagonal on the 1st of June 2015. As we can see, this intervention on the instrument induced a RV offset of more than 2\,\kms. In addition, 
by looking carefully, we can also see that the offset is not the same on the right side and on the left side of the detector, with a difference of about 50 to 100\,\ms.

\begin{figure*}[]
\center
 \includegraphics[angle=0,width=0.98\textwidth]{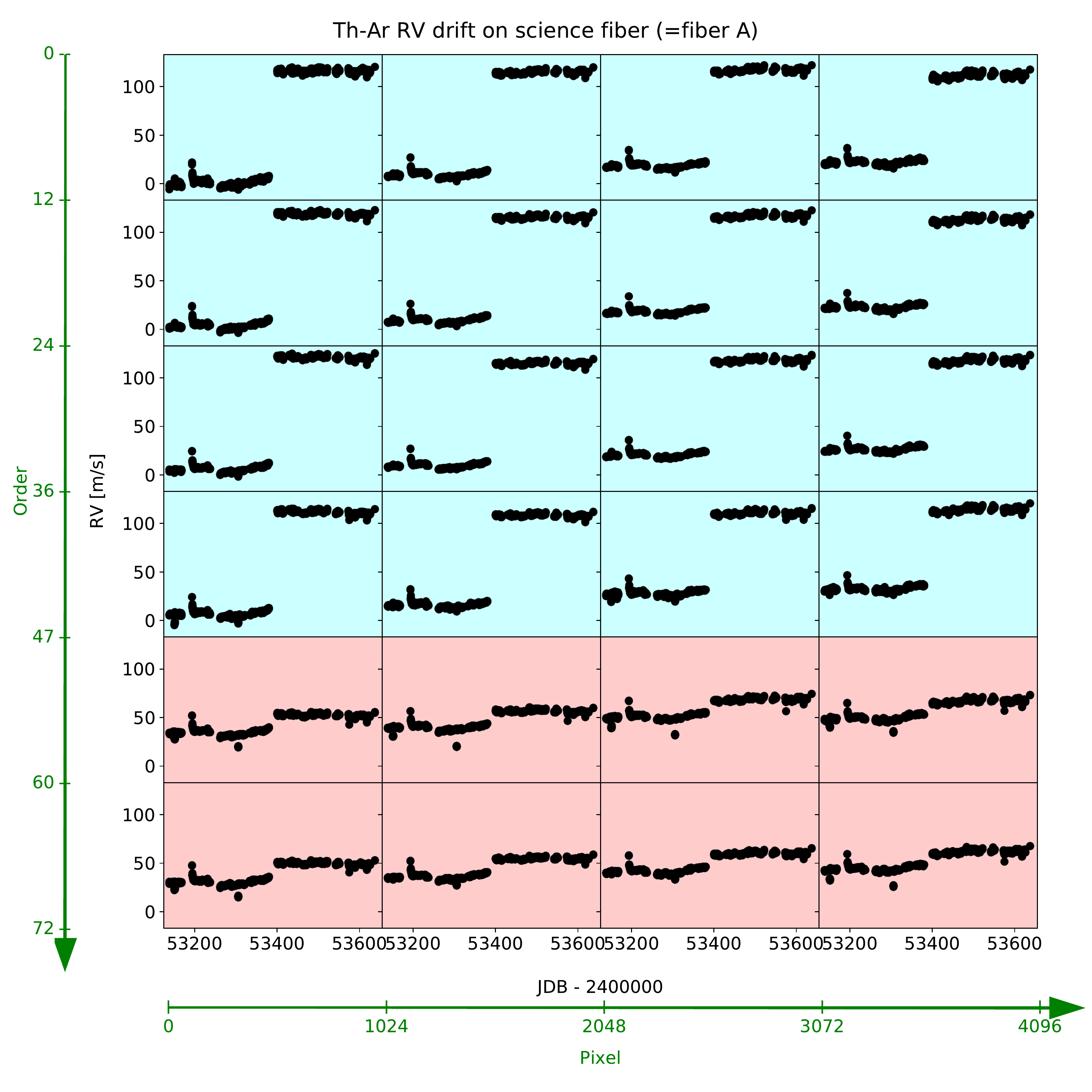}
 \caption[]{Same as Fig.~\ref{fig:DRIFT_A_per_region_all} but here we zoom on the instrument offset happening on JD=2453390 (19th of January 2005).}
\label{fig:drift_A_53400}
\end{figure*}

In Fig.~\ref{fig:drift_A_53400}, we show the drift of the instrument induced by an intervention on HARPS on the 19th of January 2005 (JD=2453390). In this case, we see a different RV offset for the
blue and the red CCD, $\sim$100 and $\sim$10\,\ms, respectively. In this case, we also see a difference in RV offset between the left and right side of the HARPS CCD.

\begin{figure*}[]
\center
 \includegraphics[angle=0,width=0.98\textwidth]{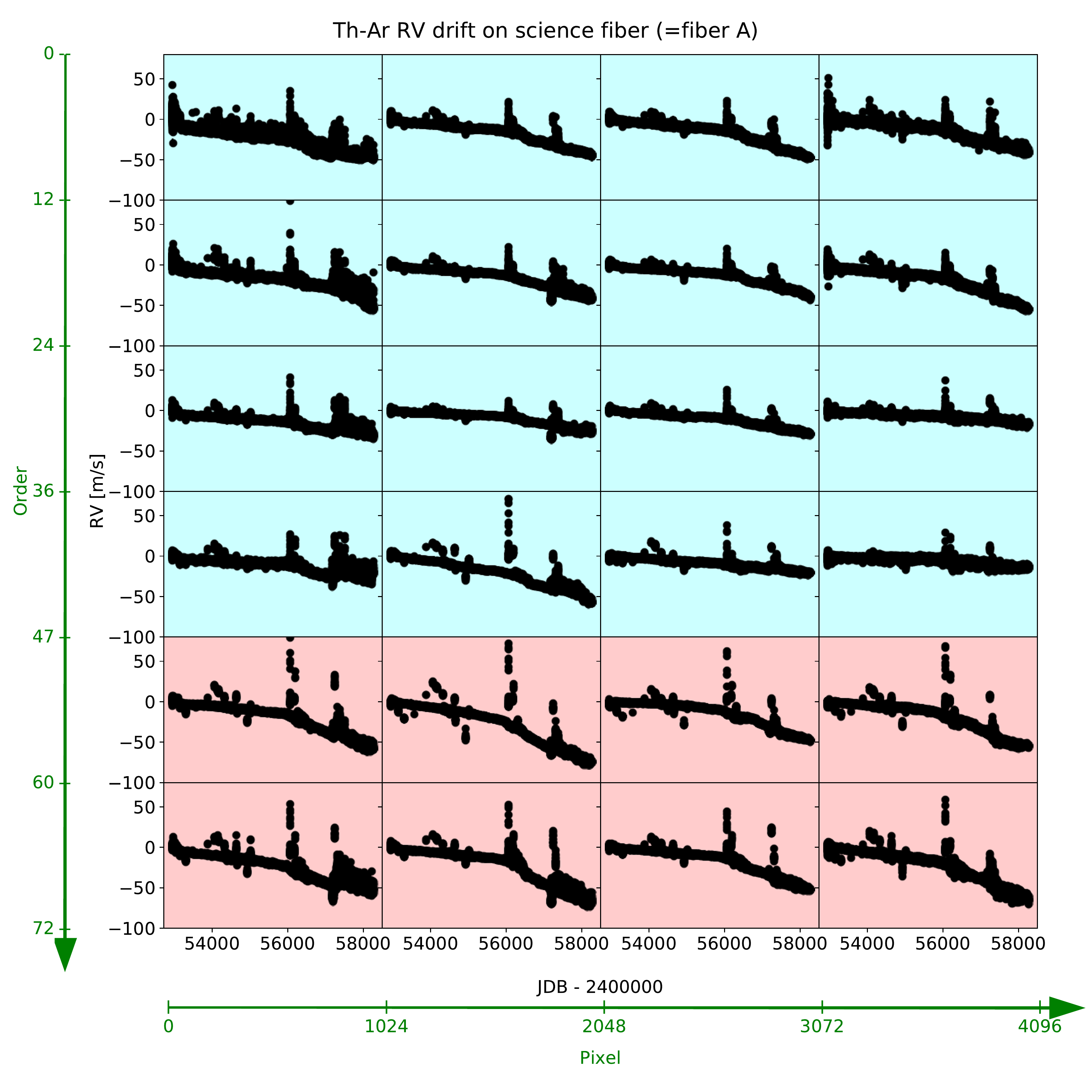}
 \caption[]{Figure with the same configuration as Fig.~\ref{fig:DRIFT_A_per_region_all}. This time however, we show the drift between the Th-Ar spectra with their own wavelength solution with respect to Th-Ar master spectra. Note that we have a master Th-Ar spectrum for each of the 5 lamps used over the 15 years of HARPS, plus an additional one to compensate for the systematics induced by the change of the HARPS fibres. In this plot, we corrected for the offset that exists between each consecutive Th-Ar master spectra. The drift observed over time is due to a slow variation of the focus of the instrument, which changes the PSF and therefore the shape of the spectral lines over time in the Th-Ar spectra recorded on the CCD. The master Th-Ar spectra is by construction an average of all the Th-Ar spectra, and therefore do not account for the change in the focus of HARPS.}
\label{fig:DRIFT_A_per_region_wavelenght_solution_every_night}
\end{figure*}

To correct for the night-to-night RV offsets due to the non-stability of the wavelength solution on the edges of HARPS orders, we first built a master Th-Ar spectrum from all the Th-Ar calibration spectra.
Then, we applied for all Th-Ar calibrations the wavelength solution of the master Th-Ar spectrum and measured the drift between each of these calibrations and the master. Note that because we used five Th-Ar calibrations 
lamps over the 15 years of HARPS, we needed to build a master Th-Ar spectrum for each of them. In addition, an extra master was also considered to account for the drastic change induced by the modification of the HARPS fibres. Besides changing the calibration lamp that induces a RV offset but which is easy to correct for, this method requires that the spectrum
of the Th-Ar lamp recorded on the CCD does not change over time. On HARPS, we know that this is not true as that the focus of the instrument slowly varies with time. This implies that the PSF of the instruments 
changes, and therefore the shape of the lines in the Th-Ar spectrum recorded on the detector changes as well. To measure this effect, we estimated the drift between each Th-Ar calibration with their own wavelength solution and the master Th-Ar
spectra used. The result of this analysis is shown in Fig.~\ref{fig:DRIFT_A_per_region_wavelenght_solution_every_night}. If the shape of the Th-Ar spectral lines varies with time, this analysis should reveal a long-term drift. This is exactly what is seen, with a drift of $\sim$50\,\ms\,over 15 years. This proves that the focus of the instrument is slightly changing over time. From this result, we can conclude that using a unique wavelength solution will induce a drift in the derived RVs. This drift can
be corrected for by using the results shown in Fig.~\ref{fig:DRIFT_A_per_region_wavelenght_solution_every_night}, or simply by fitting a polynomial as we did for the RVs of $\alpha$\,Cen\,B in Sec.~\ref{sec:night_to_night_offsets}.

\end{appendix}

\end{document}